\documentclass[aps,pra,reprint,amsmath,amssymb,groupedaddress,showpacs]{revtex4-1}
\usepackage{graphicx}                  
\usepackage{float}
\usepackage{dcolumn}                   
\usepackage{bm}                        
\usepackage[bookmarks=false]{hyperref} 
\usepackage[font={small}]{caption}     
\usepackage{enumitem}                  
\usepackage{textcomp}                  
\usepackage{mathtools}                 
\hypersetup{
    unicode=false,                     
    pdftoolbar=true,                   
    pdfmenubar=true,                   
    pdffitwindow=true,                 
    pdfstartview={},                   
    pdftitle={Entanglement Universality of TGX States in Qubit-Qutrit Systems},                              
    pdfauthor={Samuel R. Hedemann},    
    pdfsubject={Quantum Entanglement}, 
    pdfcreator={},                     
    pdfproducer={},                    
    pdfkeywords={Entanglement,} {2x3,} {TGX States,} {True-Generalized X States,} {TGX,} {Entanglement-Preserving Unitary,} {EPU,} {Minimal TGX States,} {EPU-Minimal TGX States,} {Super-Generalized X States,} {SGX States,} {SGX,} {Minimal SGX States},           
    pdfnewwindow=true,                 
    colorlinks=true,                   
    linkcolor=black,                   
    citecolor=black,                   
    filecolor=black,                   
    urlcolor=black                     
}
\newcommand{\Sec}[1]{\hyperref[sec:#1]{Sec.{\kern 2pt}\ref*{sec:#1}}}
\newcommand{\Section}[1]{\hyperref[sec:#1]{Section~\ref*{sec:#1}}}
\newcommand{\Secs}[2]{\protect\hyperref[sec:#1]{{Secs.{\kern 2pt}\ref*{sec:#1}--\ref*{sec:#2}}}}
\newcommand{\Fig}[2][]{\hyperref[fig:#2]{Fig.{\kern 2pt}\ref*{fig:#2}#1}}
\newcommand{\Figure}[2][]{\hyperref[fig:#2]{Figure~\ref*{fig:#2}#1}}
\newcommand{\App}[1]{\hyperref[sec:App.#1]{App.{\kern 2pt}\ref*{sec:App.#1}}}
\newcommand{\Appendix}[1]{\hyperref[sec:App.#1]{Appendix~\ref*{sec:App.#1}}}
\newcommand{\EqLab}[1]{\\\noindent\smash{\raisebox{6pt}[0pt][0pt]{\hypertarget{eq:#1}{}}}\vspace{-11pt}}
\newcommand{\Eq}[1]{\protect\hyperlink{eq:#1}{(\ref*{eq.#1})}}
\newcommand{\Eqs}[2]{\protect\hyperlink{eq:#1}{(\ref*{eq.#1}--\ref*{eq.#2})}}
\newenvironment{Equation}[1]{\EqLab{#1}\begin{equation}\label{eq.#1}}{\end{equation}\par\noindent\ignorespacesafterend}
\newcommand{\Table}[2][]{\hyperref[tab:#2]{Table~\ref*{tab:#2}#1}}
\newcommand{\Tables}[2]{\hyperref[tab:#1]{Tables~\ref*{tab:#1}--\ref*{tab:#2}}}
\newcommand{\ThmLab}[1]{\noindent\smash{\raisebox{11pt}[0pt][0pt]{\hypertarget{Thm:#1}{}}}\textbf{Theorem{\kern 3pt}#1:~~}}
\newcommand{\Thm}[1]{\protect\hyperlink{Thm:#1}{Theorem{\kern 3pt}#1}}
\newcommand{\Thms}[2]{\protect\hyperlink{Thm:#1}{Theorems{\kern 3pt}#1--#2}}
\newcommand{\tr}[0]{\text{tr}}
\newcommand{\shiftmath}[2]{\textnormal{\raisebox{#1}[#1][#1]{$#2$}}}
\newcommand{\scalemath}[2]{\textnormal{\scalebox{#1}{$#2$}}}
\newcommand{\hsp}[1]{{\kern #1pt}}
\renewcommand{\geq}[0]{\geqslant}
\renewcommand{\ge}[0]{\geqslant}
\renewcommand{\leq}[0]{\leqslant}
\renewcommand{\le}[0]{\leqslant}
\newcommand{\redx}[2]{{#1}{\kern -5.3pt}{{\textnormal{\raisebox{-1.2pt}{\scalebox{1.2}{\textasciicaron}}}}}{\kern -4.2pt}{~}^{(#2)}}
\newcommand{\mbar}[0]{\mathop {m}\limits^{{\kern -3.5pt}~_{\overline{{\kern 7pt}}}}}
\newcommand{\mbarsub}[0]{{\kern 0.0pt}\mathop {m}\limits^{{\kern -0.3pt}{\overline{{\kern 5.5pt}}}}{\kern 0.0pt}}
\newcommand{\nmaxnot}{n_{\,\overline{{\kern -1.8pt}\max^{~^{~^{~}}}\!\!\!\!\!\!\!\!\!\!}}\,}
\newcommand{\X}[0]{{\mathop{\rm X}\nolimits}}

\newcommand{\MEMS}[0]{{\mathop{\rm MEMS}\nolimits}}
\newcommand{\TGX}[0]{{\mathop{\rm TGX}\nolimits}}

\newcommand{\LU}[0]{{\mathop{\rm LU}\nolimits}}
\newcommand{\LPU}[0]{{\mathop{\rm LPU}\nolimits}}
\newcommand{\EPU}[0]{{\mathop{\rm EPU}\nolimits}}
\newcommand{\epre}{\overline{\hsp{-0.8}e\rule{0pt}{4.75pt}}}
\newcommand{\SGX}[0]{{\mathop{\rm SGX}\nolimits}}
\newcommand{\ZeroRhoab}[2]{\vphantom{\rho_{#1,#2}}\scalemath{0.5}{\phantom{\rho_{#1,#2}}}\mathclap{\cdot}\scalemath{0.5}{\phantom{\rho_{#1,#2}}}}
    \newcommand{\rhoTGX}[0]{{\mathop{\rho_\TGX}\nolimits}}

    \newcommand{\rhoMinTGX}[0]{{\mathop{\rho_\TGX^{\protect\shiftmath{0.2pt}{\rm min}}}\nolimits}}
    
    \newcommand{\rhoEPUminTGX}[0]{{\mathop{\rho_{\EPU_\TGX^{\rm min}}}\nolimits}}
    
    \newcommand{\rhoSGX}[0]{{\mathop{\rho_\SGX}\nolimits}}
    
    \newcommand{\rhoMinSGX}[0]{{\mathop{\rho_\SGX^{\protect\shiftmath{0.2pt}{\rm min}}}\nolimits}}
\begin{document}
\title{Entanglement Universality of TGX States in Qubit-Qutrit Systems}
\author{Samuel R. Hedemann}
\noaffiliation
\date{August 2, 2022}
\begin{abstract}
We prove that all states (mixed or pure) of qubit-qutrit ($2\times 3$) systems have entanglement-preserving unitary (EPU) equivalence to a compact subset of  true-generalized X (TGX) states called EPU-minimal TGX states which we give explicitly. Thus, for any spectrum-entanglement combination achievable by general states, there exists an EPU-minimal TGX state of the same spectrum and entanglement. We use I-concurrence to measure entanglement and give an  explicit formula for it for all $2\times 3$ minimal TGX states (a more general set than EPU-minimal TGX states) whether mixed or pure, yielding its minimum average value over all decompositions. We also give a computable I-concurrence formula for a more general family called minimal super-generalized X (SGX) states, and give optimal decompositions for minimal SGX states and all of their subsets.
\end{abstract}
\maketitle
\section{\label{sec:I}Introduction and Review of $2\times 2$}
\vspace{-8pt}
Quantum entanglement \cite{Schr,EPR} is a powerful resource in emerging technologies such as quantum computing \cite{Feyn,DiVi,Sho1,Sho2,Grov,Deu1,DeJo,CEMM}, quantum communications \cite{BBCJ,BPM1,BPM2,HedA,HKLo,Pati,BDSS,AgPP}, and quantum machine learning \cite{GuZi,Trug,ScSP,BWPR}. In recent years, many explicit results have been found for simple systems like two qubits ($2\times 2$) which can help us study more complicated systems. While it is well-known that entanglement can be \textit{detected} in qubit-qutrit ($2\times 3$) systems, measures such as \textit{negativity} \cite{Pere,Vida} do \textit{not} give minimum average entanglement over all decompositions. Here, we extend the work of \cite{HedX,MeMG,HeXU} to show that entanglement universality exists in $2\times 3$ in analogy to $2\times 2$, with \textit{computable} optimal values for special families of states.

To review, a measure of entanglement of any two-qubit state $\rho$ (mixed or pure), is the \textit{concurrence} \cite{HiWo,Woot},
\begin{Equation}                      {1}
C(\rho ) \equiv \max \{0,\xi _1  - \xi _2  - \xi _3  - \xi _4 \},
\end{Equation}
where $\xi_{1}\geqslant\cdots\geqslant\xi_{4}$ are the eigenvalues of the Hermitian operator $\sqrt {\sqrt \rho  \widetilde {\rho} \sqrt \rho  } $ (or square roots of the eigenvalues of nonHermitian operator $\rho\widetilde {\rho}$), where $\widetilde{\rho}  \equiv (\sigma_2  \otimes \sigma _2 )\rho ^* (\sigma _2  \otimes \sigma _2 )$ and \smash{$\sigma_2 \equiv\binom{0\hsp{3}-i}{\hsp{-1.2}i\hsp{8}0}$}. If $\rho$ is an X state, defined as
\begin{Equation}                      {2}
\rho_{\X}  =\! \left( {\begin{array}{*{20}c}
   {\rho _{1,1} } & \cdot & \cdot & {\rho _{1,4} }  \\
   \cdot & {\rho _{2,2} } & {\rho _{2,3} } & \cdot  \\
   \cdot & {\rho _{3,2} } & {\rho _{3,3} } & \cdot  \\
   {\rho _{4,1} } & \cdot & \cdot & {\rho _{4,4} }  \\
\end{array}} \right)\!,
\end{Equation}
for \smash{$\rho _{a,b}\!\equiv\!\langle a|\rho_{\X}|b\rangle$} with $a,b\!\in\!\{1,\ldots,4\}$, \Eq{1} simplifies to
\begin{Equation}                      {3}
C(\rho_\X ) = 2\max \{ 0,|\rho _{1,4} | \!-\! \sqrt {\rho _{2,2}\rho _{3,3} },
|\rho _{2,3} | \!-\! \sqrt {\rho _{1,1}\rho _{4,4} } \,\},
\end{Equation}
\cite{YuEb,WBPS}. For \textit{any general} $\rho$ of spectrum \mbox{$\lambda _1  \!\ge  \cdots  \ge\! \lambda _4 $, $C \!\in\!$} $[0,\max\{0,c_{\MEMS}\}]$, and $c_{\MEMS}\equiv\lambda _1  - \lambda _3  - 2\sqrt {\lambda _2 \lambda _4 } $, there is an entanglement-preserving unitary (EPU)-equivalent X state of the same $C$ and spectrum given by \cite{HeXU} as
\begin{Equation}                      {4}
\rho _{\EPU_\X}  = \!\left(\!\! {\begin{array}{*{20}c}
   {\frac{{\lambda _1  + \lambda _3  + \sqrt \Omega  }}{2}} &  \cdot  &  \cdot  & {\frac{{\sqrt {(\lambda _1  - \lambda _3 )^2  - \Omega } }}{2}}  \\
    \cdot  & {\lambda _2 } &  \cdot  &  \cdot   \\
    \cdot  &  \cdot  & {\lambda _4 } &  \cdot   \\
   {\frac{{\sqrt {(\lambda _1  - \lambda _3 )^2  - \Omega } }}{2}} &  \cdot  &  \cdot  & {\frac{{\lambda _1  + \lambda _3  - \sqrt \Omega  }}{2}}  \\
\end{array}} \right)\!,
\end{Equation}
\smash{$\Omega  \equiv \max \{ 0,Q\}$} and \smash{$Q \equiv (\lambda _1  - \lambda _3 )^2  - (C + 2\sqrt {\lambda _2 \lambda _4 } )^2 $}, where \smash{$c_{\MEMS}$} is the $2\times 2$ spectral-MEMS preconcurrence, and MEMS are \textit{maximally entangled mixed states} \cite{IsHi,ZiBu,HoBM,VeAM,WNGK}.
\section{\label{sec:II}Summary of New Results for $2\times 3$}
\vspace{-8pt}
In a bipartite quantum system with density operators $\rho$ in a Hilbert space $\mathcal{H}\equiv\mathcal{H}^{(1)}\otimes\mathcal{H}^{(2)}$ where $\dim[\mathcal{H}^{(m)}]\equiv n_m$ so $\dim(\mathcal{H})\equiv n=n_1 n_2$, a $2\times 3$ system has $\mathbf{n}\equiv(n_1,n_2)=(2,3)$, and the entanglement of any $\rho$ in $2\times 3$ is given by the \textit{I-concurrence} \cite{AuVD,RBCH,Woo2,AlFe,ZZFL}, as (from \App{A})
\begin{Equation}                      {5}
E(\rho ) \equiv \mathop {\min }\limits_{\forall \{ p_j ,\rho _j \}} \sum\nolimits_j {p_j \|\mathbf{C}(\rho _j )\|_2 } ,
\end{Equation}
\rule{0pt}{9pt}where \smash{$\mathbf{C}(\rho ) \equiv [C(\rho ^{\{ 1,2,4,5\} } ),C(\rho ^{\{ 1,3,4,6\} } ),C (\rho ^{\{ 2,3,5,6\} } )]$} is the \textit{subspace concurrence vector}, where \textit{subspaces} of $\rho$ are \smash{$\rho ^{\{ \mathbf{v}\} }\equiv \rho_{\mathbf{v},\mathbf{v}}  \equiv \sum\nolimits_{\shiftmath{0.7pt}{a,\!b\! =\! 1,\!1}}^{\shiftmath{-0.5pt}{d,\!d} } {\rho _{v_a ,v_b } |a^{[d]}\rangle \langle b^{[d]}|}$} for levels {$\mathbf{v}\equiv(v_{1},\ldots,v_{d})$} where \smash{$|\hsp{0.2}a^{\hsp{0.5}\shiftmath{-1.5pt}{[d]}}\rangle$} are $d$-level computational basis states,\hsp{-1.5} subspaces\hsp{-1.5} $\{\mathbf{q}_k\}$\hsp{-1.5} in\hsp{-1.5} $\mathbf{C}$\hsp{-1.5} are\hsp{-1.5} \textit{quartets}\hsp{-1.5} (see\hsp{-2} \App{B}),\hsp{-2} and\hsp{-2} $\rho \hsp{-1} =\hsp{-3} \sum\nolimits_j \hsp{-1}{p_j \rho _j }$\hsp{-2} (see\hsp{-2} \App{C}).\hsp{-2} If\hsp{-1} $\rho$\hsp{-1} is\hsp{-1} a\hsp{-1} \textit{minimal\hsp{-2} true-generalized\hsp{-1} X\hsp{-1} (TGX)\hsp{-1} state}\hsp{-1} \cite{HedX,HedD,MeMG,HedE,MeMH,HedC,HeXU,HCor,HMME},\hsp{-1} which\hsp{-1} in\hsp{-1} $2\times 3$\hsp{-1} is
\begin{Equation}                      {6}
\rhoMinTGX  =\! \left( {\begin{array}{*{20}c}
   {\rho _{1,1} } &  \cdot  &  \cdot  &  \cdot  & \cdot & {\rho _{1,6}}  \\[-0.5pt]
    \cdot  & {\rho _{2,2} } &  \cdot  & \cdot &  \cdot  & \cdot  \\[-0.5pt]
    \cdot  &  \cdot  & {\rho _{3,3} } & {\rho _{3,4}} & \cdot &  \cdot   \\[-0.5pt]
    \cdot  & \cdot & {\rho _{4,3}} & {\rho _{4,4} } &  \cdot  &  \cdot   \\[-0.5pt]
   \cdot &  \cdot  & \cdot &  \cdot  & {\rho _{5,5} } &  \cdot   \\[-0.5pt]
   {\rho _{6,1}} & \cdot &  \cdot  &  \cdot  &  \cdot  & {\rho _{6,6} }  \\[-0.5pt]
\end{array}} \right)\!,
\end{Equation}
for \smash{$\rho _{a,b}\!\equiv\!\langle a|\rhoMinTGX|b\rangle$}, $a,b\!\in\!\{1,\ldots,6\}$, which is a TGX state with nonzero off-diagonals in only \textit{one} of the quartets in \Eq{5} [e.g., $\{1,3,4,6\}$; see \Eq{16}], then \Eq{5} becomes
\begin{Equation}                      {7}
\begin{array}{*{20}l}
   {E(\rhoMinTGX )\! =\! 2\max\hsp{-1.5}\{  0, } &\!\!\! { |\rho _{1,5} | \!- \hsp{-2.0}\sqrt {\rho _{2,2} \rho _{4,4} } ,|\rho _{2,4} | \!- \hsp{-2.0}\sqrt {\rho _{1,1} \rho _{5,5} } ,}  \\
   {} &\!\!\! {|\rho _{1,6} | \!- \hsp{-2.0}\sqrt {\rho _{3,3} \rho _{4,4} } ,|\rho _{3,4} | \!- \hsp{-2.0}\sqrt {\rho _{1,1} \rho _{6,6} } ,}  \\
   {} &\!\!\! {|\rho _{2,6} | \!- \hsp{-2.0}\sqrt {\rho _{3,3} \rho _{5,5} } ,|\rho _{3,5} | \!- \hsp{-2.0}\sqrt {\rho _{2,2} \rho _{6,6} }  \}. }  \\
\end{array}\!
\end{Equation}
For \textit{any} $\rho$ of spectrum $\lambda _1  \ge  \cdots  \ge \lambda _6 $, entanglement $E \!\in\! [0,\max\{0,e_{\MEMS}\}]$, $e_{\MEMS}\equiv\lambda _1  - \lambda _5  - 2\sqrt {\lambda _4 \lambda _6 }$, there exists an \textit{EPU-equivalent minimal TGX state} given by
\begin{Equation}                      {8}
\rhoEPUminTGX \! =\! \!\left(\!\! {\begin{array}{*{20}c}
   {\frac{{\lambda _1  + \lambda _5  + \sqrt \Omega  }}{2}} &\!  \cdot  &\!  \cdot  &\!  \cdot  &\!  \cdot  &\! {\frac{{\sqrt {(\lambda _1  - \lambda _5 )^2  - \Omega } }}{2}}  \\[-0.5mm]%
    \cdot  &\! {\lambda _2 } &\!  \cdot  &\!  \cdot  &\!  \cdot  &\!  \cdot   \\[-0.5mm]%
    \cdot  &\!  \cdot  &\! {\lambda _4 } &\!  \cdot  &\!  \cdot  &\!  \cdot   \\[-0.5mm]%
    \cdot  &\!  \cdot  &\!  \cdot  &\! {\lambda _6 } &\!  \cdot  &\!  \cdot   \\[-0.5mm]%
    \cdot  &\!  \cdot  &\!  \cdot  &\!  \cdot  &\! {\lambda _3 } &\!  \cdot   \\[-0.5mm]%
   {\frac{{\sqrt {(\lambda _1  - \lambda _5 )^2  - \Omega } }}{2}} &\!  \cdot  &\!  \cdot  &\!  \cdot  &\!  \cdot  &\! {\frac{{\lambda _1  + \lambda _5  - \sqrt \Omega  }}{2}}  \\
\end{array}} \right)\!,
\end{Equation}
$\Omega  \equiv \max \{ 0,Q\} $ and $Q \equiv (\lambda _1  - \lambda _5 )^2  - (E + 2\sqrt {\lambda _4 \lambda _6 } )^2 $.%

Details and proofs of these results are in \Sec{III}, but first we summarize a few other important results.
\begin{itemize}[leftmargin=*,labelindent=4pt]\setlength\itemsep{0pt}
\item[\textbf{1.}]\hypertarget{ExtraSummaryResults:1}{}The minimal TGX I-concurrence formula of \Eq{7} gives the minimum average entanglement over all decompositions as proved in \Sec{III.A} and visualized in \Fig{1}.
\vspace{-15pt}\\
\begin{figure}[H]
\centering
\includegraphics[width=1.00\linewidth]{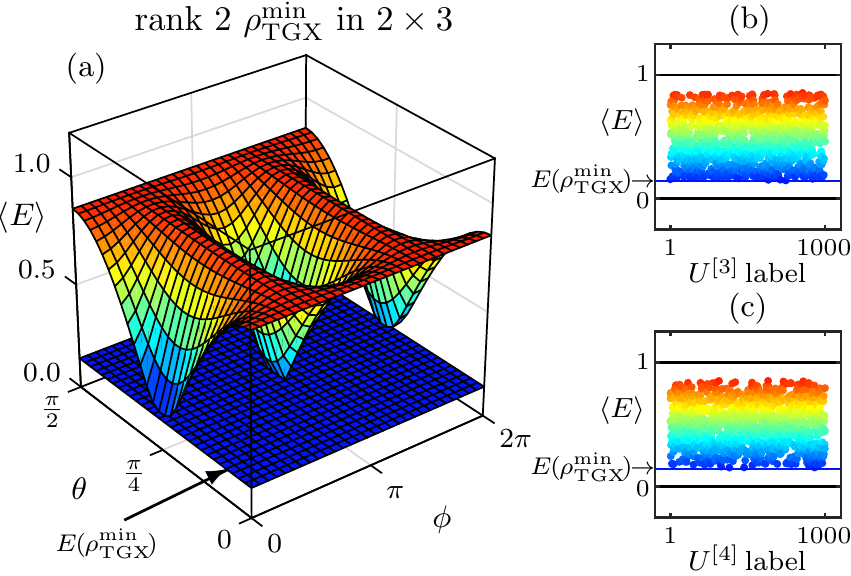}
\vspace{-17pt}
\caption{(color online) Entanglement as minimum average I-concurrence \smash{$\langle E\rangle_{\min}$} over many decompositions built from \smash{$U^{[D]}\!\equiv\! U^{[D]}(\theta,\phi,\ldots)$} [see \App{C}] for numbers of pure decomposition states $D\!\in\!\{r,\ldots,r^2\}\!=\!\{2,3,4\}$  for a rank-$2$ minimal TGX state \smash{$\rhoMinTGX$} from \Eq{6} [or \Eq{16}] with $\lambda_1 \!=\!\lambda_2\!=\!0.5$, and where \smash{$E(\rhoMinTGX)$} is from \Eq{7}. Each grid point or dot is $\langle E\rangle$ for a different decomposition of \smash{$\rhoMinTGX$} [$900$ decompositions for (a), $1000$ for (b) and (c)]. Trials of $10^3$ \textit{other} \smash{$\rhoMinTGX$} \textit{all} succeeded with \smash{$E(\rhoMinTGX)=\langle E\rangle_{\min}$}, using $10^4$ decompositions each.}
\label{fig:1}
\end{figure}

$\phantom{\text{in}}$\Figure{1} visually demonstrates that \Eq{7} is the correct minimal TGX I-concurrence formula, because the full I-concurrence \Eq{5} was used for each pure decomposition state to get $\langle E\rangle$, and \smash{$\langle E\rangle_{\min}$} matches \smash{$E(\rhoMinTGX)$} from \Eq{7}. These tests are not proofs; they are merely necessary tests of the proofs in \Sec{III}, which prove these results for \textit{all} ranks, not just rank $2$.

$\phantom{\text{in}}$Note that I-concurrence $E(\rho)$ from \Eq{5} is a necessary and sufficient (N\&S) entanglement measure (EM) for $2\times 3$, whereas \textit{generalized concurrence} \cite{Uhl1} is a verified N\&S EM \textit{only} in $2\times 2$, and that although the \textit{negativity} \cite{Pere,Vida} \textit{is} an N\&S EM in $2\times 2$ and $2\times 3$, it is \textit{not} a minimum average value over all decompositions unless it is also convex-roof extended.
\item[\textbf{2.}]\hypertarget{ExtraSummaryResults:2}{}An explicit \textit{Lewenstein-Sanpera} (LS) \cite{LS,LSD} decomposition (derived in \Sec{III.D}) of \Eq{8} is
\begin{Equation}                      {9}
\rhoEPUminTGX  = p_E \rho _E  + (1 - p_E )\rho _S ,
\end{Equation}
\vspace{-14pt}\\
with entangled-part probability $p_E$ and state $\rho _E$,
\begin{Equation}                      {10}
p_E  = \frac{{E\langle x_1 |x_1 \rangle }}{{\xi _1 +\delta_{{\xi_1},0}}},\;\;\;\rho _E  = \frac{{|x_1 \rangle \langle x_1 |}}{{\langle x_1 |x_1 \rangle }},
\end{Equation}
\vspace{-14pt}\\
such that the entanglement is
\begin{Equation}                      {11}
\begin{array}{*{20}l}
   {E(\rhoEPUminTGX)} &\!\! {= p_E E(\rho _E) = E}  \\
   {} &\!\! {=\text{max}\{0,\xi_{1}-\xi_{2}-\xi_{3}-\xi_{4}\},}  \\
\end{array}
\end{Equation}
\vspace{-14pt}\\
and its separable part $\rho_S$ is
\begin{widetext}
{~}\vspace{-26pt}
\begin{Equation}                      {12}
\begin{array}{*{20}l}
   {\rho _S} &\!\! {= \frac{1}{{1 - p_E +\delta_{p_{E},1}}}\left( {\lambda _2 |2 \rangle \langle 2 |  + \lambda _3 |5 \rangle \langle 5 |  + \frac{{\min \{ \xi _1 ,\xi _2  + \xi _3  + \xi _4 \}+\delta_{{\xi_1},0} }}{{\xi _1 +\delta_{{\xi_1},0}}}|x_1 \rangle \langle x_1 | + \sum\limits_{a = 2}^{4} {|x_a \rangle \langle x_a |} } \right)\!,}  \\
\end{array}
\end{Equation}
\vspace{-14pt}\\
where \smash{$\lambda _1  \ge  \cdots  \ge \lambda _6 $}, with computational basis states $|k \rangle$ for $k\in 1,\ldots,6$, and Wootters decomposition states,
\begin{Equation}                      {13}
\begin{array}{*{20}c}
   {\begin{array}{*{20}l}
   {|x_1 \rangle  = \frac{i}{{N_1 }}\!\left( {\begin{array}{*{20}c}
   {\frac{{\sqrt {\lambda _1 \lambda _5 \Omega }  + \delta _{\lambda _5 \Omega ,0} \Delta }}{\Delta }\sqrt {\lambda _1 } \sqrt {\frac{{\Delta  + \sqrt \Omega  }}{{2\Delta }}}  - \frac{{\xi _1 \Delta  - \lambda _1 \sqrt {\Delta ^2  - \Omega } }}{\Delta }\sqrt {\lambda _5 } \sqrt {\frac{{\Delta  - \sqrt \Omega  }}{{2\Delta }}} }  \\
    \cdot   \\[-1ex]
    \cdot   \\[-1ex]
    \cdot   \\[-1ex]
    \cdot   \\
   {\frac{{\sqrt {\lambda _1 \lambda _5 \Omega }  + \delta _{\lambda _5 \Omega ,0} \Delta }}{\Delta }\sqrt {\lambda _1 } \sqrt {\frac{{\Delta  - \sqrt \Omega  }}{{2\Delta }}}  + \frac{{\xi _1 \Delta  - \lambda _1 \sqrt {\Delta ^2  - \Omega } }}{\Delta }\sqrt {\lambda _5 } \sqrt {\frac{{\Delta  + \sqrt \Omega  }}{{2\Delta }}} }  \\
\end{array}} \right)\!,} &\!\! {|x_3 \rangle  = \frac{1}{{\sqrt 2 }}\!\left( {\begin{array}{*{20}c}
    \cdot   \\
    \cdot   \\
   {\sqrt {\lambda _4 } }  \\
   {\sqrt {\lambda _6 } }  \\
    \cdot   \\
    \cdot   \\
\end{array}} \right)\!,}  \\[7.7ex]
   {|x_2 \rangle  = \frac{1}{{N_2 }}\!\left( {\begin{array}{*{20}c}
   {\frac{{\xi _2 \Delta  - \lambda _5 \sqrt {\Delta ^2  - \Omega } }}{\Delta }\sqrt {\lambda _1 } \sqrt {\frac{{\Delta  + \sqrt \Omega  }}{{2\Delta }}}  + \frac{{\sqrt {\lambda _1 \lambda _5 \Omega }  + \delta _{\lambda _5 \Omega ,0} \Delta }}{\Delta }\sqrt {\lambda _5 } \sqrt {\frac{{\Delta  - \sqrt \Omega  }}{{2\Delta }}} }  \\
    \cdot   \\[-1ex]
    \cdot   \\[-1ex]
    \cdot   \\[-1ex]
    \cdot   \\
   {\frac{{\xi _2 \Delta  - \lambda _5 \sqrt {\Delta ^2  - \Omega } }}{\Delta }\sqrt {\lambda _1 } \sqrt {\frac{{\Delta  - \sqrt \Omega  }}{{2\Delta }}}  - \frac{{\sqrt {\lambda _1 \lambda _5 \Omega }  + \delta _{\lambda _5 \Omega ,0} \Delta }}{\Delta }\sqrt {\lambda _5 } \sqrt {\frac{{\Delta  + \sqrt \Omega  }}{{2\Delta }}} }  \\
\end{array}} \right)\!,} &\!\! {|x_4 \rangle  = \frac{i}{{\sqrt 2 }}\!\left(\!\!\! {\begin{array}{*{20}c}
    \cdot   \\
    \cdot   \\
   {\phantom{+}\sqrt {\lambda _4 } }  \\
   { - \sqrt {\lambda _6 } }  \\
    \cdot   \\
    \cdot   \\
\end{array}}\! \right)\!,}  \\
\end{array}}  \\
   {\begin{array}{*{20}c}
   {\,\rule{0pt}{18pt}N_1  \equiv \sqrt {(\frac{{\xi _1 \Delta  - \lambda _1 \sqrt {\Delta ^2  - \Omega } }}{\Delta })^2  + (\frac{{\sqrt {\lambda _1 \lambda _5 \Omega }  + \delta _{\lambda _5 \Omega ,0} \Delta }}{\Delta })^2 } ,} & {N_2  \equiv \sqrt {(\frac{{\xi _2 \Delta  - \lambda _5 \sqrt {\Delta ^2  - \Omega } }}{\Delta })^2  + (\frac{{\sqrt {\lambda _1 \lambda _5 \Omega }  + \delta _{\lambda _5 \Omega ,0} \Delta }}{\Delta })^2 },}  \\
\end{array}}  \\
\end{array}
\end{Equation}
\vspace{-24pt}\\
\end{widetext}
where \smash{$\rho _{\EPU_{\TGX}^{\min } }^{\{1,3,4,6\}}\!=\!\sum\nolimits_{a=1}^{4}|x_{a}\rangle\langle x_{a}|$}, and
\begin{Equation}                      {14}
\begin{array}{*{20}l}
   {\xi _1 } &\!\! { = \frac{{\sqrt {4\lambda _1 \lambda _5 \Delta ^2  + (\lambda _1  - \lambda _5 )^2 (\Delta ^2  - \Omega )}  + (\lambda _1  - \lambda _5 )\sqrt {\Delta ^2  - \Omega } }}{{2\Delta }}}  \\[0.5ex]
   {\xi _2 } &\!\! { = \frac{{\sqrt {4\lambda _1 \lambda _5 \Delta ^2  + (\lambda _1  - \lambda _5 )^2 (\Delta ^2  - \Omega )}  - (\lambda _1  - \lambda _5 )\sqrt {\Delta ^2  - \Omega } }}{{2\Delta }}}  \\[0.5ex]
   {\xi _3 } &\!\! { =\xi _4 = \sqrt {\lambda _4 \lambda _6 }, }  \\
\end{array}
\end{Equation}
in \Eqs{10}{13}, and also
\begin{Equation}                      {15}
\begin{array}{*{20}l}
   \Delta  &\!\! { \equiv \lambda _1  - \lambda _5  + \delta _{\lambda _1 ,\lambda _5 } }  \\
   \Omega  &\!\! { \equiv \max \{ 0,Q\} ,}  \\
   Q &\!\! { \equiv (\lambda _1  - \lambda _5 )^2  - (E + 2\sqrt {\lambda _4 \lambda _6 } )^2 .}  \\
\end{array}
\end{Equation}
\vspace{-12pt}\\
See \Sec{III.D} for a full derivation of these results.
\item[\textbf{3.}]\hypertarget{ExtraSummaryResults:3}{}TGX form and entanglement of states such as \Eq{6} or \Eq{8} are both preserved by local-permutation unitary (LPU) operations  $U_{\LPU}\equiv\Pi^{(1)}\otimes\Pi^{(2)}$ where $\Pi^{(m)}$ is a mode-$m$ unitary permutation operator. The full set of minimal TGX states in $2\times 3$ is [from \Eq{6}],
\begin{Equation}                      {16}
\begin{array}{*{20}l}
   {\rhoMinTGX  =\! \left( {\begin{array}{*{20}c}
   {\rho _{1,1} } &  \cdot  &  \cdot  &  \cdot  & {\rho _{1,5} } & \cdot  \\
    \cdot  & {\rho _{2,2} } &  \cdot  & {\rho _{2,4} } &  \cdot  & \cdot  \\
    \cdot  &  \cdot  & {\rho _{3,3} } & \cdot & \cdot &  \cdot   \\
    \cdot  & {\rho _{4,2} } & \cdot & {\rho _{4,4} } &  \cdot  &  \cdot   \\
   {\rho _{5,1} } &  \cdot  & \cdot &  \cdot  & {\rho _{5,5} } &  \cdot   \\
   \cdot & \cdot &  \cdot  &  \cdot  &  \cdot  & {\rho _{6,6} }  \\
\end{array}} \right)\!,}  \\
   {\rhoMinTGX  =\! \left( {\begin{array}{*{20}c}
   {\rho _{1,1} } &  \cdot  &  \cdot  &  \cdot  & \cdot & {\rho _{1,6}}  \\
    \cdot  & {\rho _{2,2} } &  \cdot  & \cdot &  \cdot  & \cdot  \\
    \cdot  &  \cdot  & {\rho _{3,3} } & {\rho _{3,4} } & \cdot &  \cdot   \\
    \cdot  & \cdot & {\rho _{4,3} } & {\rho _{4,4} } &  \cdot  &  \cdot   \\
   \cdot &  \cdot  & \cdot &  \cdot  & {\rho _{5,5} } &  \cdot   \\
   {\rho _{6,1} } & \cdot &  \cdot  &  \cdot  &  \cdot  & {\rho _{6,6} }  \\
\end{array}} \right)\!,}  \\
   {\rhoMinTGX  =\! \left( {\begin{array}{*{20}c}
   {\rho _{1,1} } &  \cdot  &  \cdot  &  \cdot  & \cdot & \cdot  \\
    \cdot  & {\rho _{2,2} } &  \cdot  & \cdot &  \cdot  & {\rho _{2,6} }  \\
    \cdot  &  \cdot  & {\rho _{3,3} } & \cdot & {\rho _{3,5} } &  \cdot   \\
    \cdot  & \cdot & \cdot & {\rho _{4,4} } &  \cdot  &  \cdot   \\
   \cdot &  \cdot  & {\rho _{5,3} } &  \cdot  & {\rho _{5,5} } &  \cdot   \\
   \cdot & {\rho _{6,2} } &  \cdot  &  \cdot  &  \cdot  & {\rho _{6,6} }  \\
\end{array}} \right)\!,}  \\
\end{array}
\end{Equation}
which are LPU variations of each other. All LPU variations of \Eq{8} and \Eq{16} are confined to TGX space:
\begin{Equation}                      {17}
\rhoTGX =\!\left( {\begin{array}{*{20}c}
   {\rho _{1,1} } &  \cdot  &  \cdot  &  \cdot  & {\rho _{1,5} } & {\rho _{1,6} }  \\
    \cdot  & {\rho _{2,2} } &  \cdot  & {\rho _{2,4} } &  \cdot  & {\rho _{2,6} }  \\
    \cdot  &  \cdot  & {\rho _{3,3} } & {\rho _{3,4} } & {\rho _{3,5} } &  \cdot   \\
    \cdot  & {\rho _{4,2} } & {\rho _{4,3} } & {\rho _{4,4} } &  \cdot  &  \cdot   \\
   {\rho _{5,1} } &  \cdot  & {\rho _{5,3} } &  \cdot  & {\rho _{5,5} } &  \cdot   \\
   {\rho _{6,1} } & {\rho _{6,2} } &  \cdot  &  \cdot  &  \cdot  & {\rho _{6,6} }  \\
\end{array}} \right)\!.
\end{Equation}
\item[\textbf{4.}]\hypertarget{ExtraSummaryResults:4}{}By \hyperlink{ExtraSummaryResults:3}{Item 3}, \Eq{7} is invariant under LPU operations. \{The I-concurrence of \Eq{5} is local-unitary (LU) invariant, and while LU operations on any $\rho$ \textit{do} preserve entanglement, \Eq{7} is only valid for \textit{minimal TGX} states [just as \Eq{3} is only valid for X states], which means only states LPU-equivalent to \Eq{16}.\} Thus, the I-concurrence of each state in \Eq{16} is given by \Eq{7} for all spectra, as is that of the simpler EPU-minimal TGX states such as \Eq{8} and its LPU variations.
\item[\textbf{5.}]\hypertarget{ExtraSummaryResults:5}{}Despite having X form, \Eq{8} is a TGX state, since there are X states $\rho^{\{2,5\}}$ which are not EPU equivalent to \Eq{8} in general since $\rho^{\{2,5\}}$ is \textit{always separable}  since $\{|2\rangle,|5\rangle\}=\{|1,2\rangle,|2,2\rangle\}$. [Throughout, we use $\{ |1\rangle , \ldots ,|6\rangle \}  = \{ |1,1\rangle ,|1,2\rangle ,|1,3\rangle ,|2,1\rangle ,|2,2\rangle ,|2,3\rangle \} $ (see App.\hsp{2.5}U of \cite{HCor}) where $|a,b\rangle\equiv|a\rangle|b\rangle\equiv|a\rangle\otimes|b\rangle$.]
\item[\textbf{6.}]\hypertarget{ExtraSummaryResults:6}{}In analogy to \cite{HeXU}, the EPU that transforms general $\rho$ to \smash{$\rhoEPUminTGX$} of \Eq{8} is
\begin{Equation}                      {18}
U_{\EPU_{\TGX}^{\min} }  \equiv \epsilon _{\rhoEPUminTGX} \epsilon _\rho ^\dag ,
\end{Equation}
where $\epsilon _\rho$ is the unitary eigenvector matrix of $\rho$ and \smash{$\epsilon _{\rhoEPUminTGX}$} is that of \Eq{8}, shown explicitly in \Eq{105}.
\item[\textbf{7.}]\hypertarget{ExtraSummaryResults:7}{}One major difference from $2\times 2$ is that a computable formula for the I-concurrence $E(\rho)$ of general mixed states $\rho$ in $2\times 3$ is not yet known.  \textit{If} we had such a formula, we could use the $E$ and spectrum of any $\rho$ to immediately construct \raisebox{1pt}{\smash{$\rhoEPUminTGX$}} of \Eq{8}, and achieve transformations like that of the first figure of \cite{HeXU}. Nevertheless, \Eq{8} is extremely powerful since it allows us to \textit{parameterize} states of all physical spectrum-entanglement combinations in $2\times 3$.
\item[\textbf{8.}]\hypertarget{ExtraSummaryResults:8}{}The entanglement $E$ of \smash{$\rhoEPUminTGX$} also agrees with the minimal\hsp{-0.5} TGX I-concurrence\hsp{-0.5} formula of \Eq{7},\hsp{1} giving the minimum average entanglement over all decompositions as proved in \Sec{III.A} and visualized in \Fig{2}.
\begin{figure}[H]
\centering
\includegraphics[width=1.00\linewidth]{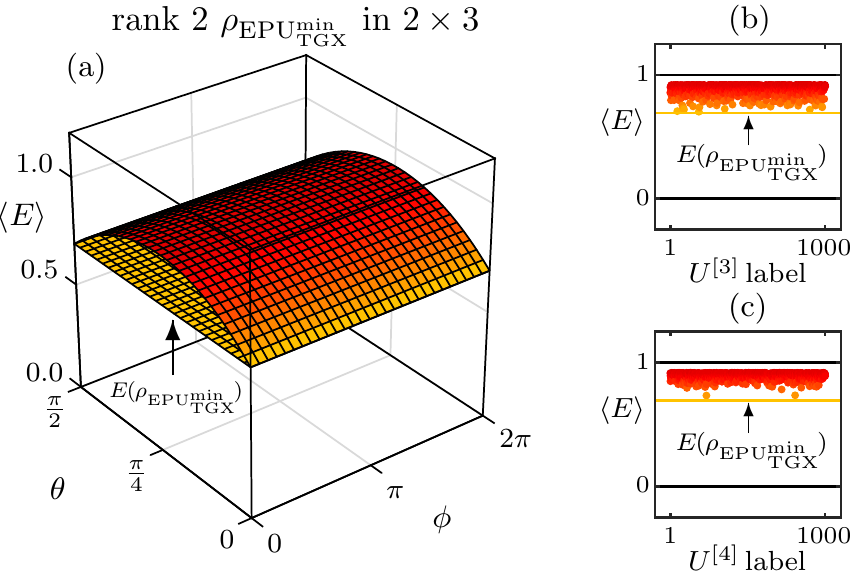}
\vspace{-17pt}
\caption{(color online) Entanglement as minimum average I-concurrence \smash{$\langle E\rangle_{\min}$} over many decompositions built from \smash{$U^{[D]}\!\equiv\! U^{[D]}(\theta,\phi,\ldots)$} [see \App{C}] for numbers of pure decomposition states $D\!\in\!\{r,\ldots,r^2\}\!=\!\{2,3,4\}$ for a rank-$2$ EPU-minimal TGX state \protect\raisebox{1.5pt}{\smash{$\rhoEPUminTGX$}} from \Eq{8} with $\{\lambda_1,\lambda_2\}\!=\!\{0.7,0.3\}$ and $E=0.693$, with \smash{$E(\protect\shiftmath{1.0pt}{\rhoEPUminTGX})$} computed from \Eq{7}. Each grid point or dot is $\langle E\rangle$ for a \protect\rule{0pt}{8.5pt}different decomposition of \smash{$\rhoEPUminTGX$} [$900$ decompositions for (a), $1000$ for (b) and (c)]. This \protect\rule{0pt}{8.5pt}test was repeated for $10^3$ \textit{other} \smash{$\rhoEPUminTGX$}, and \textit{all} succeeded with \smash{$E(\rhoEPUminTGX)=\langle E\rangle_{\min}=E$}.}
\label{fig:2}
\end{figure}
{~\vspace{-22pt}\\}
$\phantom{\text{id}}$\Figure{2} visually demonstrates that \Eq{7} is the correct minimal TGX I-concurrence formula, but also verifies that $E$ in \Eq{8} is correct, because the full I-concurrence of \Eq{5} was used for each pure decomposition state to get $\langle E\rangle$, and \smash{$\langle E\rangle_{\min}$} matches both\hsp{-0.5} \smash{$E(\rhoEPUminTGX)$} from \Eq{7} and $E$ from \Eq{8}. These\hsp{0.5} tests\hsp{0.5} are\hsp{0.5} not\hsp{0.5} proofs;\hsp{-2} they are merely necessary tests of the proofs in \Sec{III}, which prove these results for \textit{all} ranks.
\item[\textbf{9.}]\hypertarget{ExtraSummaryResults:9}{}Nomenclature: The term \textit{EPU-equivalent minimal TGX states} (or simply \textit{EPU-minimal TGX states}) \smash{$\rhoEPUminTGX$} refers to the simplest possible set of TGX states\hsp{1} that has EPU equivalence to general states (where ``simplest'' means it has the most zeros).

$\phantom{\text{id}}$The term \textit{minimal TGX states} \smash{$\rhoMinTGX$} refers to a somewhat broader class of states definable as the most general TGX states whose simplified I-concurrence still has the same special formula as \smash{$\rhoEPUminTGX$}. In $2\times 3$, a \smash{$\rhoMinTGX$} can be formed from a \smash{$\rhoEPUminTGX$} by promoting to dense X form any quartets needed to diagnose entanglement in that particular \raisebox{1.0pt}{\smash{$\rhoEPUminTGX$}}, where \textit{quartets} are the $2\times 2$ product subspaces defined in \App{B}.

$\phantom{\text{id}}$For example, the EPU-minimal TGX state \smash{$\rhoEPUminTGX$} in \Eq{8} only has nonzero off-diagonals in subspace $\{1,6\}$, while its minimal TGX state counterpart \smash{$\rhoMinTGX$} in \Eq{6} [the second state in \Eq{16}] has nonzero off-diagonals in both $\{1,6\}$ and $\{3,4\}$ since those are all the X-state elements in the $\{1,3,4,6\}$ quartet which contained the nonzero off-diagonals of \smash{$\rhoEPUminTGX$}.

$\phantom{\text{id}}$We can form an even more general set of states called \textit{minimal super-generalized X (SGX) states} \smash{$\rhoMinSGX$} by generalizing \smash{$\rhoMinTGX$} just enough for the simple I-concurrence formula to have the same form but with full concurrences rather than X concurrences. In $2\times 3$, this means promoting the entanglement-containing quartet of \smash{$\rhoMinTGX$} to dense form, as well as any remaining subspace still permitting that I-concurrence form.
\item[\textbf{10.}]\hypertarget{ExtraSummaryResults:10}{}As introduced in \hyperlink{ExtraSummaryResults:9}{Result 9}, the full set of minimal SGX states in $2\times 3$ is
{\\\vspace{-14.5pt}}%
\begin{Equation}                      {19}
\begin{array}{*{20}l}
   {\rhoMinSGX  =\! \left( {\begin{array}{*{20}c}
   {\rho _{1,1} } & {\rho _{1,2} } &  \cdot  & {\rho _{1,4} } & {\rho _{1,5} } &  \cdot   \\
   {\rho _{2,1} } & {\rho _{2,2} } &  \cdot  & {\rho _{2,4} } & {\rho _{2,5} } &  \cdot   \\
    \cdot  &  \cdot  & {\rho _{3,3} } &  \cdot  &  \cdot  & {\rho _{3,6} }  \\
   {\rho _{4,1} } & {\rho _{4,2} } &  \cdot  & {\rho _{4,4} } & {\rho _{4,5} } &  \cdot   \\
   {\rho _{5,1} } & {\rho _{5,2} } &  \cdot  & {\rho _{5,4} } & {\rho _{5,5} } &  \cdot   \\
    \cdot  &  \cdot  & {\rho _{6,3} } &  \cdot  &  \cdot  & {\rho _{6,6} }  \\
\end{array}} \right)\!,}  \\
   {\rhoMinSGX  =\! \left( {\begin{array}{*{20}c}
   {\rho _{1,1} } &  \cdot  & {\rho _{1,3} } & {\rho _{1,4} } &  \cdot  & {\rho _{1,6} }  \\
    \cdot  & {\rho _{2,2} } &  \cdot  &  \cdot  & {\rho _{2,5} } &  \cdot   \\
   {\rho _{3,1} } &  \cdot  & {\rho _{3,3} } & {\rho _{3,4} } &  \cdot  & {\rho _{3,6} }  \\
   {\rho _{4,1} } &  \cdot  & {\rho _{4,3} } & {\rho _{4,4} } &  \cdot  & {\rho _{4,6} }  \\
    \cdot  & {\rho _{5,2} } &  \cdot  &  \cdot  & {\rho _{5,5} } &  \cdot   \\
   {\rho _{6,1} } &  \cdot  & {\rho _{6,3} } & {\rho _{6,4} } &  \cdot  & {\rho _{6,6} }  \\
\end{array}} \right)\!,}  \\
   {\rhoMinSGX  =\! \left( {\begin{array}{*{20}c}
   {\rho _{1,1} } &  \cdot  &  \cdot  & {\rho _{1,4} } &  \cdot  &  \cdot   \\
    \cdot  & {\rho _{2,2} } & {\rho _{2,3} } &  \cdot  & {\rho _{2,5} } & {\rho _{2,6} }  \\
    \cdot  & {\rho _{3,2} } & {\rho _{3,3} } &  \cdot  & {\rho _{3,5} } & {\rho _{3,6} }  \\
   {\rho _{4,1} } &  \cdot  &  \cdot  & {\rho _{4,4} } &  \cdot  &  \cdot   \\
    \cdot  & {\rho _{5,2} } & {\rho _{5,3} } &  \cdot  & {\rho _{5,5} } & {\rho _{5,6} }  \\
    \cdot  & {\rho _{6,2} } & {\rho _{6,3} } &  \cdot  & {\rho _{6,5} } & {\rho _{6,6} }  \\
\end{array}} \right)\!,}  \\
\end{array}
\end{Equation}
{\vspace{-10pt}\\}%
which are all LPU variations of each other.  Note that minimal SGX states have \textit{non}TGX form in general, but minimal TGX states are always subsets of minimal SGX states (although the full TGX space is not a subspace of any of the minimal SGX states).  The full SGX space is the union of all minimal SGX states and in $2\times 3$ simply yields the set of \textit{all states},
{\\\vspace{-14.5pt}}%
\begin{Equation}                      {20}
\rhoSGX  =\! \left( {\begin{array}{*{20}c}
   {\rho _{1,1} } & {\rho _{1,2} } & {\rho _{1,3} } & {\rho _{1,4} } & {\rho _{1,5} } & {\rho _{1,6} }  \\
   {\rho _{2,1} } & {\rho _{2,2} } & {\rho _{2,3} } & {\rho _{2,4} } & {\rho _{2,5} } & {\rho _{2,6} }  \\
   {\rho _{3,1} } & {\rho _{3,2} } & {\rho _{3,3} } & {\rho _{3,4} } & {\rho _{3,5} } & {\rho _{3,6} }  \\
   {\rho _{4,1} } & {\rho _{4,2} } & {\rho _{4,3} } & {\rho _{4,4} } & {\rho _{4,5} } & {\rho _{4,6} }  \\
   {\rho _{5,1} } & {\rho _{5,2} } & {\rho _{5,3} } & {\rho _{5,4} } & {\rho _{5,5} } & {\rho _{5,6} }  \\
   {\rho _{6,1} } & {\rho _{6,2} } & {\rho _{6,3} } & {\rho _{6,4} } & {\rho _{6,5} } & {\rho _{6,6} }  \\
\end{array}} \right)\!.
\end{Equation}
{\vspace{-10pt}\\}%
$\phantom{\text{id}}$Minimal SGX states in $2\times 3$ are not TGX states in general for many reasons. TGX states were proposed in \cite{HedX} as a family of states that achieves EPU equivalance and contains particular sets of states that have the same entanglement properties as the Bell states. 

$\phantom{\text{id}}$For instance, the Bell states are maximally entangled, form a complete orthonormal basis [called a maximally entangled basis (MEB)], have balanced superposition while having multiple state coefficients be $0$, and have diagonal reductions. In \cite{HedE}, it was proved that TGX states always admit such families of states in all systems, and it showed how to construct the proper generalization of Bell states in all systems wrt full $N$-partite entanglement.  

$\phantom{\text{id}}$Here, we can use \Eq{19} to show that minimal SGX states cannot contain any MEBs of pure ME TGX states, and therefore minimal SGX states do not achieve the same properites as TGX states in general, even though they are general enough to achieve EPU-equivalence.  But even then, the simplest states that do so are the much-simpler EPU-minimal TGX states which are always confined within TGX space.
\item[\textbf{11.}]\hypertarget{ExtraSummaryResults:11}{}A computable formula for I-concurrence of minimal SGX states (mixed or pure) in $2\times 3$ is
\begin{Equation}                      {21}
E(\rhoMinSGX) \!=\! \|\mathbf{C}(\rhoMinSGX)\|_\infty  \! =\! \max \{ C^{\{ \mathbf{q}_1 \} } ,C^{\{ \mathbf{q}_2 \} } ,C^{\{ \mathbf{q}_3 \} } \} ,
\end{Equation}
where \smash{$C^{\{ \mathbf{q}\} }  \equiv C([\rhoMinSGX]^{\{ \mathbf{q}\} } )$} is the concurrence of the generally mixed $\mathbf{q}$ subspace of \smash{$\rhoMinSGX$}, where $\mathbf{q}_k$ for $k=1,2,3$ are the quartets from \Eq{5}.
\item[\textbf{12.}]\hypertarget{ExtraSummaryResults:12}{}The minimal SGX I-concurrence formula of \Eq{21} gives the minimum average I-concurrence over all decom- positions as proved in \Sec{III.A} and shown in \Fig{3}.
\vspace{-12pt}\\
\begin{figure}[H]
\centering
\includegraphics[width=1.00\linewidth]{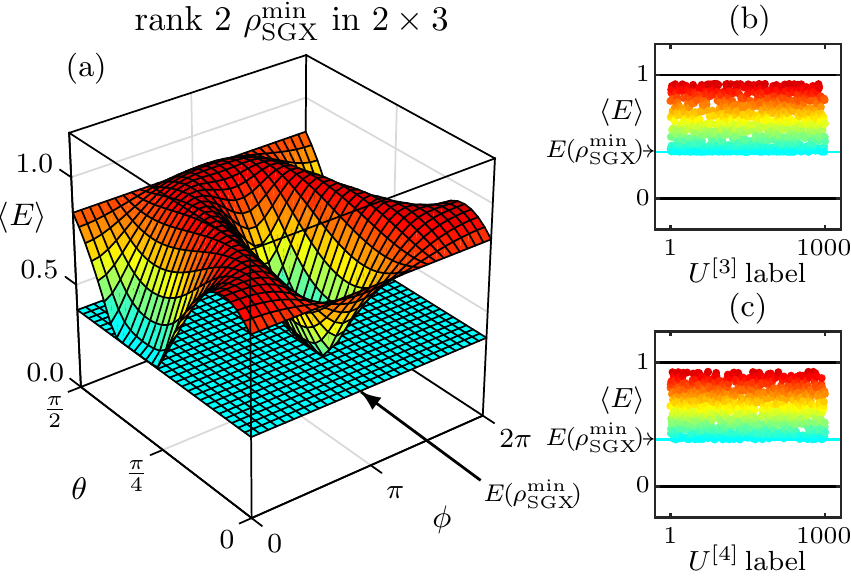}
\vspace{-17pt}
\caption{(color online) Minimum average I-concurrence \smash{$\langle E\rangle_{\min}$} over many decompositions built from \smash{$U^{[D]}\!\equiv\! U^{[D]}(\theta,\phi,\ldots)$} [see \App{C}] for numbers of pure decomposition states $D\!\in\!\{r,\ldots,r^2\}\!=\!\{2,3,4\}$  for a rank-$2$ minimal SGX state \smash{$\rhoMinSGX$} from \Eq{19} with $\{\lambda_1,\lambda_2\}\!=\!\{0.73,0.27\}$, with \smash{$E(\rhoMinSGX)$} from \Eq{21}. Each grid point or dot is $\langle E\rangle$ for a different decomposition of \smash{$\rhoMinSGX$} [$900$ decompositions for (a), $1000$ for (b) and (c)]. Trials of $10^3$ \textit{other} \smash{$\rhoMinSGX$} \textit{all} succeeded with \smash{$E(\rhoMinSGX)\!=\!\langle E\rangle_{\min}$}.}
\label{fig:3}
\end{figure}
$\;$\vspace{-24pt}\\
$\phantom{\text{id}}$\Figure{3} demonstrates that \Eq{21} gives the correct minimum average I-concurrence over all decompositions of minimal SGX states. Again, while this example is merely a necessary test, the \textit{proof} that \Eq{21} works for all \smash{$\rhoMinSGX$} of all ranks is in \Sec{III.A}.
\end{itemize}

We now prove all these new results in \Sec{III}.
\section{\label{sec:III}Proofs and Derivations}
\vspace{-6pt}
Here we prove the results summarized in \Sec{II}.
\vspace{-26pt}\\
\subsection{\label{sec:III.A}Proof of the I-Concurrence Formulas for Minimal SGX States and Minimal TGX States}
\vspace{-6pt}%
We give our proof of \smash{$E(\rhoMinSGX )$} from \Eq{21} and \smash{$E(\rhoMinTGX )$} from \Eq{7} as a series of facts that build upon each other:
\begin{itemize}[leftmargin=*,labelindent=9.5pt]\setlength\itemsep{0pt}
\item[\textbf{1.}]\hypertarget{minTGXIconcurrenceFact:1}{}The I-concurrence of any \textit{pure} state $\rho$ in $2\times 3$ is
\begin{Equation}                      {22}
E(\rho ) \!=\! \sqrt {C^2 (\rho ^{\{ 1,2,4,5\} } )\! +\! C^2 (\rho ^{\{ 1,3,4,6\} } )\! +\! C^2 (\rho ^{\{ 2,3,5,6\} } )} ,
\end{Equation}
derived in \App{A}, using definitions in the text after \Eq{5}. The sets $\{ 1,2,4,5\}$, $\{ 1,3,4,6\}$, and $\{ 2,3,5,6\}$ in \Eq{22} and \Eq{5} are called \textit{quartets} (see \App{B}).
\item[\textbf{2.}]\hypertarget{minTGXIconcurrenceFact:2}{}All pure TGX states in $2\times 3$ have at most $L_*  = 2$ nonzero probability amplitudes. The sets of $L_*$ indices of nonzero levels of each of these possible pure TGX subspaces are called \textit{ME TGX tuples} \cite{HMME}.  From \cite{HedE}, all ME TGX tuples in $2\times 3$ are
\begin{Equation}                      {23}
\{ 1,5\},\{ 1,6\},\{ 2,4\},\{ 2,6\},\{ 3,4\},\{ 3,5\}.
\end{Equation}
{~\vspace{-30pt}\\}
\item[\textbf{3.}]\hypertarget{minTGXIconcurrenceFact:3}{}In $2\times 3$, each ME TGX tuple in \Eq{23} is a \textit{subspace} of a \textit{single quartet} in \Eq{22}. In particular, these subspaces form \textit{inseparable qubits} (the inner or outer pair in each quartet) as defined in \App{B}. Thus, for any \textit{pure} TGX state $\rho _{|\psi _{\TGX} \rangle }  \equiv |\psi _{\TGX} \rangle \langle \psi _{\TGX} |$ in $2\times 3$, its subspace for any quartet has only two nonzero levels at most, given by one of the ME TGX tuples of \Eq{23}, and only in an inseparable qubit.  For example, if the nondiagonal quartet of a given $\rho _{|\psi _{\TGX} \rangle }$ is $\mathbf{q} \equiv \{ 1,3,4,6\} $, then its subspace for this $\mathbf{q}$ might be
\begin{Equation}                      {24}
\rho _{|\psi _{\TGX} \rangle }^{\{ 1,3,4,6\} }  = \left( {\begin{array}{*{20}c}
   {\rho _{1,1} } &  \cdot  &  \cdot  & {\rho _{1,6} }  \\
    \cdot  &  \cdot  &  \cdot  &  \cdot   \\
    \cdot  &  \cdot  &  \cdot  &  \cdot   \\
   {\rho _{6,1} } &  \cdot  &  \cdot  & {\rho _{6,6} }  \\
\end{array}} \right)\!,
\end{Equation}
{~\vspace{-12pt}\\}
where \smash{$\rho _{a,b}  \equiv \langle a|\rho _{|\psi _{\TGX} \rangle } |b\rangle $}, which has only two nonzero levels, while all other subspace quartets of that pure state are diagonal or $0$. (Note: all pure TGX states in $2\times 3$ have two nonzero levels except for computational basis states which have one.)
\item[\textbf{4.}]\hypertarget{minTGXIconcurrenceFact:4}{}By \hyperlink{minTGXIconcurrenceFact:3}{Fact 3}, any pure TGX state in $2\times 3$ has \smash{$E(\rho _{|\psi _{\TGX} \rangle } ) = C(\rho _{|\psi _{\TGX} \rangle }^{\{ \mathbf{q}_k \} } )$} for only \textit{one} quartet $\mathbf{q}_k$, since the others must all be zero since only one subspace quartet can be nondiagonal, and diagonal subspaces have zero subspace concurrence. Therefore, \textit{for all pure TGX states in} $2\times 3$,
\begin{Equation}                      {25}
\begin{array}{*{20}l}
   {E(\rho _{|\psi _{\TGX} \rangle } )} &\!\!\! { =\! \max \{ C(\rho _{|\psi _{\TGX} \rangle }^{\{ \mathbf{q}_1 \} } ),C(\rho _{|\psi _{\TGX} \rangle }^{\{ \mathbf{q}_2 \} } ),C(\rho _{|\psi _{\TGX} \rangle }^{\{ \mathbf{q}_3 \} } )\} }  \\
   {} &\!\!\! { =\! \|\mathbf{C}(\rho _{|\psi _{\TGX} \rangle } )\|_\infty , }  \\
\end{array}
\end{Equation}
{~\vspace{-16pt}\\}
where $\mathbf{C}(\rho)$ is defined after \Eq{5}. [Note: since only one subspace concurrence can be nonzero for pure TGX states in $2\times 3$, $E(\rho _{|\psi _{\TGX} \rangle } )$ simplifies to \textit{any} $p$-norm of $\mathbf{C}(\rho _{|\psi _{\TGX} \rangle } )$ for $p\geq 1$, but the infinity norm offers the simplest final form as we will see.]
\item[\textbf{5.}]\hypertarget{minTGXIconcurrenceFact:5}{}\hyperlink{minTGXIconcurrenceFact:4}{Fact 4} suggests that any state with all of its coherence (nonzero off-diagonals) in just \textit{one} fully dense quartet might also satisfy \Eq{25}. However, since the quartets \textit{overlap} we need to check whether that causes multiple quartets to be nonzero.

$\phantom{\text{id}}$For example, if only $\rho^{\{1,3,4,6\}}$ is dense with zeros everywhere else, then $\rho^{\{1,2,4,5\}}$ has nonzero nondiagonals in $\{1,4\}$ and $\rho^{\{2,3,5,6\}}$ has nonzero nondiagonals in $\{3,6\}$. But since $\{|1\rangle,|4\rangle\}\!=\!\{|1\rangle|1\rangle,|2\rangle|1\rangle\}$ and $\{|3\rangle,|6\rangle\}\!=\!\{|1\rangle|3\rangle,|2\rangle|3\rangle\}$, then $\rho^{\{1,2,4,5\}}$ and $\rho^{\{2,3,5,6\}}$ are \textit{separable} since their only nondiagonality is in a separable subspace, so we can always find an explicit separable decomposition for them, such as $\rho^{\{1,2,4,5\}}\!=\!(\rho_{1,1}|1^{[2]}\rangle\langle 1^{[2]}|\!+\!\rho_{4,1}|2^{[2]}\rangle\langle 1^{[2]}|\!+\!\rho_{1,4}|1^{[2]}\rangle\langle 2^{[2]}|\!+\!\rho_{4,4}|2^{[2]}\rangle\langle 2^{[2]}|)\otimes|1^{[2]}\rangle\langle 1^{[2]}|$, where $\{|1^{[2]}\rangle,|2^{[2]}\rangle\}$ are $2$-dimensional basis states of each mode of the $2\times 2$ subspace state $\rho^{\{1,2,4,5\}}$.

$\phantom{\text{id}}$Therefore, pure states in $2\times 3$ with just one quartet that is fully dense, such as $\{1,3,4,6\}$, can only have \textit{one} nonzero subspace concurrence.  Such pure states qualify as \textit{minimal SGX states}, as seen in \Eq{19}. The example above also applies to \textit{mixed} minimal SGX states, but for now, we focus on \textit{pure} minimal SGX states, meaning that, for example, while $\rho^{\{1,3,4,6\}}$ can be dense, the rest of the elements are all zero to obey the geometric-mean rule for pure states that $|\rho_{a,b}|=\sqrt{\rho_{a,a}\rho_{b,b}}$.  (Alternatively, a pure minimal SGX state could be dense in $\{2,5\}$ with zeros everywhere else, but since $\{|2\rangle,|5\rangle\}=\{|1\rangle|2\rangle,|2\rangle|2\rangle\}$ is separable, such states are automatically separable, so \textit{all} subspace concurrences are $0$ so they are a special case where there is one potentially nonzero subspace concurrence whose value is $0$.)  Furthermore, note that this example generalizes to all pure minimal SGX states by LPU variation.

$\phantom{\text{id}}$Thus, for \textit{pure} minimal SGX states \smash{$\rho_{|\psi_{\SGX}\rangle}$} in $2\times 3$, since only \textit{one} subspace concurrence can be nonzero, their I-concurrence simplies to 
{~\vspace{-16pt}\\}
\begin{Equation}                      {26}
\begin{array}{*{20}l}
   {E(\rho _{|\psi _{\SGX} \rangle } )} &\!\!\! { =\! \max \{ C(\rho _{|\psi _{\SGX} \rangle }^{\{ \mathbf{q}_1 \} } ),C(\rho _{|\psi _{\SGX} \rangle }^{\{ \mathbf{q}_2 \} } ),C(\rho _{|\psi _{\SGX} \rangle }^{\{ \mathbf{q}_3 \} } )\} }  \\
   {} &\!\!\! { =\! \|\mathbf{C}(\rho _{|\psi _{\SGX} \rangle } )\|_\infty . }  \\
\end{array}
\end{Equation}
{~\vspace{-16pt}\\}
$\phantom{\text{id}}$At this point, since all pure TGX states are subsets of pure minimal SGX states [which can be seen by noting that each ME TGX tuple from \Eq{23} is a subset of one 2-level nondiagonal subspace of one of the minimal SGX states of \Eq{19}], we will focus our proof on minimal SGX states, and then derive special results for minimal TGX states at the end.
\item[\textbf{6.}]\hypertarget{minTGXIconcurrenceFact:6}{}In any system, not just in $2\times 3$, we say that two or more matrices are \textit{spacewise orthogonal} if the potentially nonzero elements in each of them are always in regions where all of the others always have zeros.

$\phantom{\text{in}}$We call a matrix \textit{internally spacewise orthogonal} if it can be rearranged through general permutation to be \textit{block-diagonal} with at least one generally nonzero block strictly smaller than the full matrix (to exclude the case of a fully dense matrix being its own block-diagonal form, since then none of its subspaces would be spacewise orthogonal). Here, by ``generally nonzero block'' we mean a square of contiguous matrix elements which \textit{can} all be nonzero but can also have zeros (including all of them being zero).

$\phantom{\text{in}}$A \textit{minimal SGX} state in $2\times 3$, such as the middle of \Eq{19}, is internally spacewise orthogonal since the $\{1,3,4,6\}$ subspace and the $\{2,5\}$ subspace are each generally nonzero blocks that could be permuted to block-diagonal form (note that we do not actually permute the matrix to block-diagonal form). These internal nonzero blocks are spacewise orthogonal to each other because their orthogonality comes from living in different subspaces with no overlaping levels.  

$\phantom{\text{in}}$For example, the minimal SGX state in the middle of \Eq{19} is internally spacewise orthogonal because it has two spacewise-orthogonal subspaces as
\begin{Equation}                      {27}
{\setlength\fboxsep{0pt}   
\rhoMinSGX\!  =\!\! \left(\! {\begin{array}{*{20}c}
   {\begin{array}{*{20}l}
   {{\framebox{$\begin{array}{*{20}l}
   {\rho_{1,1}}  \\[0.3ex]
\end{array}$}}_{\,1}}  \\
   {{\makebox{$\begin{array}{*{20}l}
   {\ZeroRhoab{2}{1} }  \\[0.3ex]
\end{array}$}}_{\,\phantom{2}}}  \\
   {{\framebox{$\begin{array}{*{20}l}
   {\rho _{3,1}}  \\
   {\rho _{4,1}}  \\[0.3ex]
\end{array}$}}_{\,1}}  \\
   {{\makebox{$\begin{array}{*{20}l}
   {\ZeroRhoab{5}{1} }  \\[0.3ex]
\end{array}$}}_{\,\phantom{2}}}  \\
   {{\framebox{$\begin{array}{*{20}l}
   {\rho_{6,1}}  \\[0.3ex]
\end{array}$}}_{\,1}}  \\
\end{array}} &\!\!\!\!\!\!
   {\begin{array}{*{20}l}
   {{\makebox{$\begin{array}{*{20}l}
   {\ZeroRhoab{1}{2} }  \\[0.3ex]
\end{array}$}}_{\,\phantom{2}}}  \\
   {{\framebox{$\begin{array}{*{20}l}
   {\rho_{2,2}}  \\[0.3ex]
\end{array}$}}_{\,2}}  \\
   {{\makebox{$\begin{array}{*{20}l}
   {\ZeroRhoab{3}{2}}  \\
   {\ZeroRhoab{4}{2}}  \\[0.3ex]
\end{array}$}}_{\,\phantom{2}}}  \\
   {{\framebox{$\begin{array}{*{20}l}
   {\rho_{5,2}}  \\[0.3ex]
\end{array}$}}_{\,2}}  \\
   {{\makebox{$\begin{array}{*{20}l}
   {\ZeroRhoab{6}{2} }  \\[0.3ex]
\end{array}$}}_{\,\phantom{2}}}  \\
\end{array}} &\!\!\!\!\!\!
   {\begin{array}{*{20}l}
   {{\framebox{$\begin{array}{*{20}l}
   {\rho _{1,3}} & {\rho _{1,4}}  \\[0.3ex]
\end{array}$}}_{\,1}}  \\
   {{\makebox{$\begin{array}{*{20}l}
   {\ZeroRhoab{2}{3}} & {\ZeroRhoab{2}{4}}  \\
\end{array}$}}_{\,\phantom{1}}}  \\
   {{\framebox{$\begin{array}{*{20}l}
   {\rho _{3,3}} & {\rho _{3,4}}  \\
   {\rho _{4,3}} & {\rho _{4,4}}  \\[0.3ex]
\end{array}$}}_{\,1}}  \\
   {{\makebox{$\begin{array}{*{20}l}
   {\ZeroRhoab{5}{3}} & {\ZeroRhoab{5}{4}}  \\[0.3ex]
\end{array}$}}_{\,\phantom{1}}}  \\
   {{\framebox{$\begin{array}{*{20}l}
   {\rho _{6,3}} & {\rho _{6,4}}  \\[0.3ex]
\end{array}$}}_{\,1}}  \\
\end{array}} &\!\!\!\!\!\!
   {\begin{array}{*{20}l}
   {{\makebox{$\begin{array}{*{20}l}
   {\ZeroRhoab{1}{5} }  \\[0.3ex]
\end{array}$}}_{\,\phantom{2}}}  \\
   {{\framebox{$\begin{array}{*{20}l}
   {\rho_{2,5}}  \\[0.3ex]
\end{array}$}}_{\,2}}  \\
   {{\makebox{$\begin{array}{*{20}l}
   {\ZeroRhoab{3}{5}}  \\
   {\ZeroRhoab{4}{5}}  \\[0.3ex]
\end{array}$}}_{\,\phantom{2}}}  \\
   {{\framebox{$\begin{array}{*{20}l}
   {\rho_{5,5}}  \\[0.3ex]
\end{array}$}}_{\,2}}  \\
   {{\makebox{$\begin{array}{*{20}l}
   {\ZeroRhoab{6}{5} }  \\[0.3ex]
\end{array}$}}_{\,\phantom{2}}}  \\
\end{array}} &\!\!\!\!\!\!
   {\begin{array}{*{20}l}
   {{\framebox{$\begin{array}{*{20}l}
   {\rho_{1,6}}  \\[0.3ex]
\end{array}$}}_{\,1}}  \\
   {{\makebox{$\begin{array}{*{20}l}
   {\ZeroRhoab{2}{6} }  \\[0.3ex]
\end{array}$}}_{\,\phantom{2}}}  \\
   {{\framebox{$\begin{array}{*{20}l}
   {\rho _{3,6}}  \\
   {\rho _{4,6}}  \\[0.3ex]
\end{array}$}}_{\,1}}  \\
   {{\makebox{$\begin{array}{*{20}l}
   {\ZeroRhoab{5}{6} }  \\[0.3ex]
\end{array}$}}_{\,\phantom{2}}}  \\
   {{\framebox{$\begin{array}{*{20}l}
   {\rho_{6,6}}  \\[0.3ex]
\end{array}$}}_{\,1}}  \\
\end{array}}  \\
\end{array}}\!\!\! \right)\!,}
\end{Equation}
{~\vspace{-16pt}\\}
where the boxes' subscripts label the subspaces and show which elements belong to which subspace.
\item[\textbf{7.}]\hypertarget{minTGXIconcurrenceFact:7}{}By \hyperlink{minTGXIconcurrenceFact:6}{Fact 6}, internally spacewise orthogonal matrices are always a \textit{direct sum} of their spacewise orthogonal subspaces. For example, \Eq{27} can be rewritten [using \smash{$\rho\equiv\rhoMinSGX$} and the subspace notation after \Eq{5}] as
\begin{Equation}                      {28}
\rhoMinSGX  = \rho ^{\{ 1,3,4,6\} }  \oplus \rho ^{\{ 2,5\} }.
\end{Equation}
{~\vspace{-34pt}\\}
\item[\textbf{8.}]\hypertarget{minTGXIconcurrenceFact:8}{}From \hyperlink{minTGXIconcurrenceFact:6}{Facts 6} \hyperlink{minTGXIconcurrenceFact:7}{and 7}, eigenvectors for each spacewise-orthogonal subspace of an internally spacewise orthogonal matrix always exist such that those eigenvectors only have nonzero entries within their respective subspaces. (It is well-known that the set of eigenvectors of the full matrix of a direct sum is the union of the eigenvectors of its direct summands.)
\item[\textbf{9.}]\hypertarget{minTGXIconcurrenceFact:9}{}From \hyperlink{minTGXIconcurrenceFact:8}{Fact 8}, we see that none of the eigenvectors from different spacewise orthogonal subspaces are needed to decompose a given spacewise orthogonal subspace.  For example, all decompositions of the $\{1,3,4,6\}$ subspace of $\rho\equiv\rhoMinSGX$ from \Eq{27} only require eigenvectors of $\rho ^{\{1,3,4,6\} } $ embedded in the full space, and those have zeros in all elements except those in $\{1,3,4,6\}$. Therefore, by \hyperlink{minTGXIconcurrenceFact:7}{Fact 7}, decompositions of $\rho\equiv\rhoMinSGX$ into convex sums of pure states can be constructed as direct sums of decompositions of each spacewise orthogonal part.  For example,
\begin{Equation}                      {29}
\begin{array}{*{20}l}
   {\rhoMinSGX} &\!\! { = (\sum\nolimits_{j_1 } {p_{j_1 }^{\{ 1,3,4,6\} } \rho _{j_1 }^{\{ 1,3,4,6\} } } ) \oplus (\sum\nolimits_{j_2 } {p_{j_2 }^{\{ 2,5\} } \rho _{j_2 }^{\{ 2,5\} } } ),}  \\
\end{array}
\end{Equation}
where for instance, \smash{$\rho _{j_1 }^{\{ 1,3,4,6\} } $} are pure decomposition states of \smash{$\rho ^{\{ 1,3,4,6\} } $} constructed from its own eigenstates and eigenvalues.
\item[\textbf{10.}]\hypertarget{minTGXIconcurrenceFact:10}{}The \textit{Lewenstein-Sanpera (LS) decomposition} \cite{LS} of any state $\rho$ is one for which all of the entanglement is concentrated in a single part $\rho_E$ (generally mixed, but pure in $2\times 2$, and as we will show, pure for minimal SGX states in $2\times 3$), and the other part is separable $\rho_S$, where $p_{E}E(\rho _E )$ (which is the entanglement of the unnormalized entangled part $p_{E}\rho_E$) is equal to $E(\rho)$, the minimum average entanglement over all decompositions. Thus, the LS decomposition is
{~\vspace{-16pt}\\}
\begin{Equation}                      {30}
\rho  = p_{E}\rho _E  \!+\! (1 \!-\! p_{E})\rho _S \;\;\;\text{s.t.}\;\; E(\rho ) = p_{E}E(\rho _E ),
\end{Equation}
{~\vspace{-18pt}\\}
which happens when $p_{S}\equiv 1 - p_{E}$ is maximized over all such decompositions (in general, not all decompositions of $\rho$ into entangled and separable parts are optimal, but there always exist ones that \textit{are} optimal, and we call those \textit{the} LS decompositions).
\item[\textbf{11.}]\hypertarget{minTGXIconcurrenceFact:11}{}Since by \hyperlink{minTGXIconcurrenceFact:5}{Fact 5}, all of the  entanglement of $\rhoMinSGX$ comes from a \textit{single} spacewise-orthogonal subspace quartet, and by \hyperlink{minTGXIconcurrenceFact:9}{Fact 9}, only the eigenstates of that subspace matter in the general pure-state decompositions of that subspace, and since the subspace spacewise orthogonal to that which is $\{2,5\}$ is always separable, then \textit{the LS entanglement-minimizing decomposition of $\rhoMinSGX$ gets all of its entanglement from the entangled part of the LS decomposition of a single spacewise-orthogonal quartet subspace}.

$\phantom{\text{in}}$For example, in \Eq{27}, the LS-entangled part of $\rho\equiv\rhoMinSGX$ gets all of its entanglement from the LS-entangled part of $\rho ^{\{ 1,3,4,6\} }  \oplus 0^{\{ 2,5\} } $ because by \hyperlink{minTGXIconcurrenceFact:5}{Fact 5}, \smash{$C^{\{1,3,4,6\}}$} is the only nonzero concurrence, and the part of $\rhoMinSGX$ \textit{outside of} $\{ 1,3,4,6\}$ (which is $\rho^{\{2,5\}}$) is separable, so it can be added to the LS-separable part of $\rho ^{\{ 1,3,4,6\} }  \oplus 0^{\{ 2,5\} } $ and that sum will still be separable by the definition of separability. Thus, the $2\times 3$ LS decomposition of this $\rho\equiv\rhoMinSGX$ is
\begin{Equation}                      {31}
\begin{array}{*{20}l}
   {\rhoMinSGX = } &\!\! { p_{E}^{\{1,3,4,6 \}}(\rho^{\{ 1,3,4,6\} })_E \oplus 0^{\{ 2,5\} } }  \\[0.5ex]
   {} &\!\! {+ [ \tr ( \rho^{\{1,3,4,6 \}} ) - p_{E}^{\{1,3,4,6 \}}](\rho^{\{ 1,3,4,6\} })_S  \oplus \rho^{\{2,5 \}}.}  \\
\end{array}
\end{Equation}
{~\vspace{-30pt}\\}
\item[\textbf{12.}]\hypertarget{minTGXIconcurrenceFact:12}{}By \hyperlink{minTGXIconcurrenceFact:11}{Fact 11}, the LS decomposition of $\rhoMinSGX$ always has a \textit{pure} entangled part (because the LS decomposition of its single entangled quartet is a $2\times 2$ system whose entangled part is always pure), and that entangled part always exists in a \textit{single} quartet's subspace, and therefore since the LS decomposition is an optimal decomposition, \textit{the convex-roof extension of the I-concurrence for a minimal SGX state in $2\times 3$ is always equal to a single subspace concurrence}.  (In other words, the existence of the LS decomposition for all SGX states in $2\times 3$ [proved in \hyperlink{minTGXIconcurrenceFact:11}{Fact 11}] means we do not have to consider optimal decompositions in which multiple pure states simplify to single subspace concurrences in different subspaces.)  Thus, for minimal SGX states in $2\times 3$, if we use an optimal decomposition of the LS form, we get
\begin{Equation}                      {32}
\begin{array}{*{20}l}
   {E(\rhoMinSGX)} &\!\! {=\mathop {\min }\limits_{\forall \{ p_j ,\rho _j \} } \sum\nolimits_j {p_j \|\mathbf{C}(\rho _j )\|_2 } }  \\
   {} &\!\! { = \sum\nolimits_j {p_j C([\rho _j ]^{\{ \mathbf{q}_k \} } )};\;{k\!\in\!\{1,2,3\}} }  \\[0.5ex]
   {} &\!\! { = C([\rhoMinSGX]^{\{ \mathbf{q}_k \} } );\;{k\!\in\!\{1,2,3\}}}  \\
   {} &\!\! { = \max \{ C(\rho^{\{ \mathbf{q}_1 \} } ),C(\rho^{\{ \mathbf{q}_2 \} } ),C(\rho^{\{ \mathbf{q}_3 \} } )\} } \\
   {} &\!\! { = \|\mathbf{C}(\rhoMinSGX)\|_\infty , }  \\
\end{array}
\end{Equation}
{~\vspace{-10pt}\\}
where $\rho _j$ are pure decomposition states of \smash{$\rho\equiv\rhoMinSGX$}, and in line 2 of \Eq{32}, we used \hyperlink{minTGXIconcurrenceFact:5}{Facts 5} \hyperlink{minTGXIconcurrenceFact:11}{and 11} that the pure entangled state of the LS decomposition of \smash{$\rhoMinSGX$} only exists in the spacewise orthogonal subspace of a \textit{single quartet} in $2\times 3$ to simplify the $2$-norm to the concurrence of a single subspace as in \Eq{26}. Then, in line 3 of \Eq{32}, we used the fact that since this LS decomposition is optimal, the average concurrence is equal to the actual concurrence of that same subspace of the full state. Finally, since minimal SGX states in $2\times 3$ only have one nonzero subspace concurrence, lines 4 and 5 of \Eq{32} express that as the infinity norm (for reasons we explain next). 

$\phantom{\text{in}}$Thus, \Eq{32} completes the proof of \Eq{21}.
\item[\textbf{13.}]\hypertarget{minTGXIconcurrenceFact:13}{}For all TGX states in all systems (including multipartite systems), the subspace states of all quartets appearing in the I-concurrence are always X states. Therefore, this is also true for minimal TGX states, as seen in \Eq{24}, and also for their subsets such as EPU-minimal TGX states.  [Then, since minimal TGX states are subsets of minimal SGX states, we can use \Eq{32} to compute I-concurrence of minimal TGX states, and since in $2\times 3$ that yields a $p$-norm of subspace concurrences on the whole state, those subspace concurrences simplify to the form of the X-concurrence formula of \Eq{3}, since those subspaces all have X form for minimal TGX states.]

$\phantom{\text{in}}$Specifically, for \textit{any} TGX state $\rho  \equiv \rhoTGX$, if we abbreviate $C^{\{ a,b,c,d\} }  \equiv C(\rho ^{\{ a,b,c,d\} } )$, then
\begin{Equation}                      {33}
C^{\{ a,b,c,d\} }  \!=\! 2\max \{ 0,|\rho _{a,d} | - \sqrt {\rho _{b,b} \rho _{c,c} } ,|\rho _{b,c} | - \sqrt {\rho _{a,a} \rho _{d,d} } \} ,
\end{Equation}
for valid quartets $\{ a,b,c,d\}$ (see \App{B}), where these matrix elements' indices reference the TGX parent state itself, not its subspace state $\rho ^{\{ a,b,c,d\} } $ even though this is the concurrence of $\rho ^{\{ a,b,c,d\} } $.  Note that \Eq{33} holds for minimal TGX states \smash{$\rhoMinTGX$} since they are subsets of TGX states \smash{$\rhoTGX$}.
\item[\textbf{14.}]\hypertarget{minTGXIconcurrenceFact:14}{}By \hyperlink{minTGXIconcurrenceFact:12}{Facts 12} \hyperlink{minTGXIconcurrenceFact:13}{and 13}, applying \Eq{32} to \smash{$\rhoMinTGX$} yields
\begin{Equation}                      {34}
\scalemath{0.96}{
\begin{array}{*{20}l}
   {E(\rho _{\TGX}^{\min } )} &\!\!\! { =\! \max \{ C^{\{ 1,2,4,5\} } ,C^{\{ 1,3,4,6\} } ,C^{\{ 2,3,5,6\} } \} }  \\
   {} &\!\!\! { =\!\max \{ }  \\
   {} &\!\!\! {\!\!\hsp{0.3} \begin{array}{*{20}l}
   { \phantom{=} } &\!\!\! {2\max \{ 0,|\rho _{1,5} | \!- \hsp{-2.0} \sqrt {\rho _{2,2} \rho _{4,4} } ,|\rho _{2,4} | \!- \hsp{-2.0} \sqrt {\rho _{1,1} \rho _{5,5} } \} ,}  \\
   {} &\!\!\! {2\max \{ 0,|\rho _{1,6} | \!- \hsp{-2.0} \sqrt {\rho _{3,3} \rho _{4,4} } ,|\rho _{3,4} | \!- \hsp{-2.0} \sqrt {\rho _{1,1} \rho _{6,6} } \} ,}  \\
   {} &\!\!\! {2\max \{ 0,|\rho _{2,6} | \!- \hsp{-2.0} \sqrt {\rho _{3,3} \rho _{5,5} } ,|\rho _{3,5} | \!- \hsp{-2.0} \sqrt {\rho _{2,2} \rho _{6,6} } \} \} }  \\
\end{array}}  \\
   {} &\!\!\! {\!\!\hsp{0.3} \begin{array}{*{20}l}
   { =\! 2\max \{ 0,} &\!\!\! {|\rho _{1,5} | \!- \hsp{-2.0} \sqrt {\rho _{2,2} \rho _{4,4} } ,|\rho _{2,4} | \!- \hsp{-2.0} \sqrt {\rho _{1,1} \rho _{5,5} } \} ,}  \\
   {} &\!\!\! {|\rho _{1,6} | \!- \hsp{-2.0} \sqrt {\rho _{3,3} \rho _{4,4} } ,|\rho _{3,4} | \!- \hsp{-2.0} \sqrt {\rho _{1,1} \rho _{6,6} } \} ,}  \\
   {} &\!\!\! {|\rho _{2,6} | \!- \hsp{-2.0} \sqrt {\rho _{3,3} \rho _{5,5} } ,|\rho _{3,5} | \!- \hsp{-2.0} \sqrt {\rho _{2,2} \rho _{6,6} } \} \}, }  \\
\end{array}}  \\
\end{array}}\!\!
\end{Equation}
where again \smash{$C^{\{\mathbf{q}\}}\equiv C(\rho^{\{\mathbf{q}\}})$}, which proves \Eq{7}, and in the last step we used the fact that $\max \{ \max \{ a,b\} ,\max \{ c,d\} \}  = \max \{ a,b,c,d\} $, which generalizes to any number of arguments.  Also, \Eq{34} shows why we used the infinity norm; when simplifying the minimal SGX I-concurrence formula from \Eq{32} for input of minimal TGX states, the fact that the $2\times 2$ X-concurrence formula of \Eq{33} already contains a max function allows the max function in the infinity norm to merge with it, reducing the overall expression to a single max function.
\end{itemize}

Now that we have proven \Eq{7} and \Eq{21}, we are justified in using them in the proofs that follow.  Note that \Eq{7} and \Eq{21} are LPU invariant (since LPUs preserve entanglement as well as TGX form and also minimal SGX form, and thus they preserve minimal TGX form as well). Furthermore, since EPU-minimal TGX states \smash{$\rhoEPUminTGX$} such as \Eq{8} and their LPU variations are subsets of the minimal TGX states \smash{$\rhoMinTGX$} of \Eq{16}, \Eq{7} and \Eq{21} apply to all EPU-minimal TGX states as well.
\subsection{\label{sec:III.B}Proof that EPU-Minimal TGX States \smash{$\rhoEPUminTGX$} are EPU-Equivalent to All States}
To prove that \smash{$\rhoEPUminTGX$} of \Eq{8} is EPU-equivalent to all states in $2\times 3$, we adopt the following strategy. First, \Sec{III.B.1} lists some known facts and conditions to motivate what follows.  Then, similarly to \cite{HeXU}, \Sec{III.B.2} proves that \smash{$\rhoEPUminTGX$} has the proper spectrum, \Sec{III.B.3} proves that \smash{$\rhoEPUminTGX$} has the proper I-concurrence, and \Sec{III.B.4} proves that all physical spectrum-entanglement combinations are achievable by \raisebox{1.0pt}{\smash{$\rhoEPUminTGX$}}. In analogy to \cite{HeXU}, we start by breaking \Eq{8} into $Q$ cases as
\begin{Equation}                      {35}
\begin{array}{*{20}l}
   {\rhoEPUminTGX  = }  \\[1.3mm]
   {\left\{\!\! {\begin{array}{*{20}l}
   {\left( {\begin{array}{*{20}c}
   {\frac{{\lambda _1  + \lambda _5  + \sqrt Q }}{2}} &  \cdot  &  \cdot  &  \cdot  &  \cdot  & {\frac{{E + 2\sqrt {\lambda _4 \lambda _6 } }}{2}}  \\
    \cdot  & {\lambda _2 } &  \cdot  &  \cdot  &  \cdot  &  \cdot   \\
    \cdot  &  \cdot  & {\lambda _4 } &  \cdot  &  \cdot  &  \cdot   \\
    \cdot  &  \cdot  &  \cdot  & {\lambda _6 } &  \cdot  &  \cdot   \\
    \cdot  &  \cdot  &  \cdot  &  \cdot  & {\lambda _3 } &  \cdot   \\
   {\frac{{E + 2\sqrt {\lambda _4 \lambda _6 } }}{2}} &  \cdot  &  \cdot  &  \cdot  &  \cdot  & {\frac{{\lambda _1  + \lambda _5  - \sqrt Q }}{2}}  \\
\end{array}} \right)\!;}  & {Q \geq 0}   \\
   {\left( {\begin{array}{*{20}c}
   {\frac{{\lambda _1  + \lambda _5 }}{2}} &  \cdot  &  \cdot  &  \cdot  &  \cdot  & {\frac{{\lambda _1  - \lambda _5 }}{2}}  \\
    \cdot  & {\lambda _2 } &  \cdot  &  \cdot  &  \cdot  &  \cdot   \\
    \cdot  &  \cdot  & {\lambda _4 } &  \cdot  &  \cdot  &  \cdot   \\
    \cdot  &  \cdot  &  \cdot  & {\lambda _6 } &  \cdot  &  \cdot   \\
    \cdot  &  \cdot  &  \cdot  &  \cdot  & {\lambda _3 } &  \cdot   \\
   {\frac{{\lambda _1  - \lambda _5 }}{2}} &  \cdot  &  \cdot  &  \cdot  &  \cdot  & {\frac{{\lambda _1  + \lambda _5 }}{2}}  \\
\end{array}} \right)\!;}  & {Q < 0,}   \\
\end{array}} \right.}  \\
\end{array}
\end{Equation}
where we used the fact from \Eq{8} that
\begin{Equation}                      {36}
\Omega  \equiv \max \{ 0,Q\} ;\;\;Q \equiv (\lambda _1  - \lambda _5 )^2  - (E + 2\sqrt {\lambda _4 \lambda _6 } )^2 .
\end{Equation}
Note also that we could have defined the $Q<0$ case to be a \textit{diagonal} state, but the form here permits the unification of cases to the compact from in \Eq{8}.

The eigenvalues in \Eq{8} and \Eq{35} are those of some generally nonTGX state $\rho$, and the entanglement $E$ is the I-concurrence of that same $\rho$. Although a computable formula for I-concurrence of any general mixed $\rho$ is not yet known, that will not hinder us here; we will prove that \textit{if} we knew the value of $E$, we could make \raisebox{1.0pt}{\smash{$\rho _{\EPU_{\TGX}^{\min}}$}} in \Eq{8} and it would be EPU-equivalent to $\rho$. Then we will show that \Eq{8} lets us explicitly create a state of any physical spectrum-entanglement combination in $2\times 3$, which makes it extremely useful.  Furthermore, we can create a large collection of nonTGX states with parametric entanglement and spectrum by using LU variations of \Eq{8}.

In what follows, we use $\Lambda$ to mean spectrum in the form of the diagonal eigenvalue matrix, and to refer to all combinations of quantities, we will use the notations $\Lambda E$ and $\Lambda EQ$, analogously to \cite{HeXU}.
\subsubsection{\label{sec:III.B.1}Facts and Conditions for $2\times 3$ Entanglement}
To begin our proofs for $2\times 3$ entanglement, we list some known facts and conditions that motivate what follows.
\begin{itemize}[leftmargin=*,labelindent=4pt]\setlength\itemsep{0pt}
\item[\textbf{1.}]\hypertarget{Condition2x3:1}{}From \cite{Pere,Vida}, positive partial transpose (PPT) is N\&S for separability in $2\times 3$. Also, adapted from \cite{Hild},
\begin{Equation}                      {37}
\lambda _1  - \lambda _5  - 2\sqrt {\lambda _4 \lambda _6 }  \leq 0 \;\;\Rightarrow\;\; \rho  \in \mathbb{S},
\end{Equation}
where $\lambda _1  \geq  \cdots  \geq \lambda _6 $ are eigenvalues of $\rho$, and $\mathbb{S}$ is the set of separable states. This is analogous to the $2\times 2$ case where $(\lambda _1  - \lambda _3  - 2\sqrt {\lambda _2 \lambda _4 }  \le 0) \;\Rightarrow\; \rho  \in \mathbb{S}$ for $\lambda _1  \geq  \cdots  \geq \lambda _4 $.  Note that the opposite condition of \Eq{37}, $\lambda _1  - \lambda _5  - 2\sqrt {\lambda _4 \lambda _6 }>0$, does \textit{not} imply entanglement, but is merely \textit{necessary} for entanglement, so it means $\rho$ \textit{might} be entangled \textit{or} it might be separable. (For example, computational basis states are separable but have $\lambda _1  - \lambda _5  - 2\sqrt {\lambda _4 \lambda _6 }=1>0$. This proves by counterexample that the ``iff'' part of the claim in \cite{Hild} is too strong; it should have said that $\lambda _1  - \lambda _5  - 2\sqrt {\lambda _4 \lambda _6 }  \leq 0$ is \textit{sufficient} for separability in $2\times 3$, and similarly for its $2\times 2$ result, with corresponding ammendments to its PPT claims as well.)
\item[\textbf{2.}]\hypertarget{Condition2x3:2}{}Any EM should give $1$ for ME states. In $2\times 3$, the simplest ME states are the ME TGX states,
\begin{Equation}                      {38}
\begin{array}{*{20}l}
   {|\Phi _{1,5}^ \pm  \rangle } &\!\! { \equiv \frac{1}{{\sqrt 2 }}(|1\rangle  \pm |5\rangle ),} &\;\, {|\Phi _{1,6}^ \pm  \rangle } &\!\! { \equiv \frac{1}{{\sqrt 2 }}(|1\rangle  \pm |6\rangle ),}  \\
   {|\Phi _{2,6}^ \pm  \rangle } &\!\! { \equiv \frac{1}{{\sqrt 2 }}(|2\rangle  \pm |6\rangle ),} &\;\, {|\Phi _{2,4}^ \pm  \rangle } &\!\! { \equiv \frac{1}{{\sqrt 2 }}(|2\rangle  \pm |4\rangle ),}  \\
   {|\Phi _{3,4}^ \pm  \rangle } &\!\! { \equiv \frac{1}{{\sqrt 2 }}(|3\rangle  \pm |4\rangle ),} &\;\, {|\Phi _{3,5}^ \pm  \rangle } &\!\! { \equiv \frac{1}{{\sqrt 2 }}(|3\rangle  \pm |5\rangle ),}  \\
\end{array}
\end{Equation}
\cite{HedX,HedE}, which each have the property that both reductions are as simultaneously mixed as they can be given that the parent state is pure, as explained in \cite{HedX,HedE} (see \App{D} for more details about TGX states).

~~~Note that local-permutation unitaries (LPUs) preserve \textit{both} TGX form and entanglement, so all states in \Eq{38} are related to each other by an LPU.  Also, \Eq{17} shows that in $2\times 3$, not all X states can host ME; X states must be in TGX space to have ME (since X state $\rho^{\{2,5\}}$ is always separable since $\{|2\rangle,|5\rangle\}=\{|1,2\rangle,|2,2\rangle\}$, no entangled state can be EPU converted to it).

~~~Note also that the nonzero levels of the ME TGX states of \Eq{38} are the ME TGX tuples of \Eq{23}.
\item[\textbf{3.}]\hypertarget{Condition2x3:3}{}Any EM in $2\times 3$ should give (or appropriately relate to) the MEMS \textit{minimum average pre-entanglement} (which we just call the pre-entanglement) value,
\begin{Equation}                      {39}
e_{\MEMS}  \equiv \lambda _1  - \lambda _5  - 2\sqrt {\lambda _4 \lambda _6 } ,
\end{Equation}
for maximally entangled mixed states (MEMS) wrt spectrum (which we just call MEMS here).  As we will see, $e_{\MEMS}$ can be negative for certain separable states and its relation to the MEMS I-concurrence is
\begin{Equation}                      {40}
E(\rho _{\MEMS} ) = \max \{ 0,e_{\MEMS} \} ,
\end{Equation}
proven by applying the minimal TGX I-concurrence of \Eq{7} (proved in \Sec{III.A}) to MEMS, which are, as proven in \cite{MeMH}, any $2\times 3$ states EPU-equivalent to
\begin{Equation}                      {41}
\rho _{\MEMS}  =\! \left( {\begin{array}{*{20}c}
   {\frac{{\lambda _1  + \lambda _5 }}{2}} &  \cdot  &  \cdot  &  \cdot  &  \cdot  & {\frac{{\lambda _1  - \lambda _5 }}{2}}  \\
    \cdot  & {\lambda _2 } &  \cdot  &  \cdot  &  \cdot  &  \cdot   \\
    \cdot  &  \cdot  & {\lambda _4 } &  \cdot  &  \cdot  &  \cdot   \\
    \cdot  &  \cdot  &  \cdot  & {\lambda _6 } &  \cdot  &  \cdot   \\
    \cdot  &  \cdot  &  \cdot  &  \cdot  & {\lambda _3 } &  \cdot   \\
   {\frac{{\lambda _1  - \lambda _5 }}{2}} &  \cdot  &  \cdot  &  \cdot  &  \cdot  & {\frac{{\lambda _1  + \lambda _5 }}{2}}  \\
\end{array}} \right)\!,
\end{Equation}
where \smash{$\lambda _1  \geq  \cdots  \geq \lambda _6 $}, and where we adapted the state \smash{$\rho _{\MEMS_{\Lambda}}$} from \cite{MeMH} as \smash{$\rho _{\MEMS}  \equiv U_{\LPU} \rho _{\MEMS_\Lambda  } U_{\LPU}^\dag $} where $U_{\LPU}  \equiv I^{(1)}  \otimes (|3\rangle \langle 1| + |2\rangle \langle 2| + |1\rangle \langle 3|)$ to conform to the convention of \cite{HeXU}.
\item[\textbf{4.}]\hypertarget{Condition2x3:4}{}The \textit{generalized concurrence} $C_G$ \cite{Uhl1} in $2\times 3$ has a maximal value wrt spectrum over all states of
\begin{Equation}                      {42}
\max (C_G ) = \max \{0,\lambda _1  - \lambda _4  - 2\sqrt {\lambda _2 \lambda _6 }  - 2\sqrt {\lambda _3 \lambda _5 }\} ,
\end{Equation}
as proved in \App{E}, and so generally does \textit{not} give the proper $e_{\MEMS}$ of \Eq{39}, so it is not equal to I-concurrence. Therefore, it is not yet clear whether $C_G$ is an EM in $2\times 3$, despite being a minimum average value over all decompositions; it simply does not seem relate to entanglement in a direct way. (Shifting $C_G$ by either eigenvalue functions or singular values in the concurrence calculation does not help except for MEMS, and may cause misdiagnosis of separability and entanglement.)
\item[\textbf{5.}]\hypertarget{Condition2x3:5}{}Any EM should be a convex-roof extension of its action on pure states, so it yields a minimum average value of its pure-state form over all decompositions as
\begin{Equation}                      {43}
E(\rho ) = \mathop {\min }\limits_{\forall \{p_{j}\rho_{j}\}} \left[ {\sum\nolimits_j {p_j E(\rho _j )} } \right],
\end{Equation}
where $\rho _j$ is a pure decomposition state of $\rho$ with probability $p_j$ as explained in \App{C}.
\item[\textbf{6.}]\hypertarget{Condition2x3:6}{}Any valid EM should be LU invariant, as
\begin{Equation}                      {44}
E(U_{\LU} \rho U_{\LU}^\dag  ) = E(\rho );\;\;\forall U_{\LU}  \equiv U^{(1)}  \otimes  \cdots  \otimes U^{(N)} .
\end{Equation}
\end{itemize}
Note that the I-concurrence of \Eq{5} satisfes \Eq{43} by definition, and since it is LU invariant on pure states, it retains this LU invariance on mixed states by \Eq{43} as well, thus satisfying \Eq{44}.  Next, we will use the above facts in the remaining proofs.
\subsubsection{\label{sec:III.B.2}Proof that $\rhoEPUminTGX$ Has Proper Spectrum}
In both cases of \Eq{35},
\begin{Equation}                      {45}
\det (\lambda I - \rhoEPUminTGX ) = \prod\nolimits_{k = 1}^6 {(\lambda  - \lambda _k )}  = 0,
\end{Equation}
which proves that \smash{$\rhoEPUminTGX$} has the spectrum of the $\rho$ whose eigenvalues $\{ \lambda _k \} $ are used to create it, and that \smash{$\rhoEPUminTGX$} and $\rho$ are unitarily equivalent.
\subsubsection{\label{sec:III.B.3}Proof that $\rhoEPUminTGX$ Has Proper Entanglement}
In the $Q\geq 0$ case of \Eq{35}, putting \Eq{35} into \Eq{7} gives (abbreviating as $\rho '_{a,b}  \equiv \langle a|\rhoEPUminTGX |b\rangle $),
\begin{Equation}                      {46}
\begin{array}{*{20}l}
   {E(\rhoEPUminTGX)} &\!\! {= 2\max \{ 0,} &\!\!\! {\phantom{|\rho '_{1,6} |} - \sqrt {\smash{\rho '_{2,2} \rho '_{4,4} }\rule{0pt}{8pt}\vphantom{\rho _{4,4}}} , - \sqrt {\smash{\rho '_{1,1} \rho '_{5,5}}\rule{0pt}{8pt}\vphantom{\rho _{5,5}} } ,}  \\
   {} &\!\! {} &\!\!\! {|\rho '_{1,6} | - \sqrt {\smash{\rho '_{3,3} \rho '_{4,4}}\rule{0pt}{8pt}\vphantom{\rho _{4,4}} } , - \sqrt {\smash{\rho '_{1,1} \rho '_{6,6}}\rule{0pt}{8pt}\vphantom{\rho _{6,6}} } ,}  \\
   {} &\!\! {} &\!\!\! {\phantom{|\rho '_{1,6} |} - \sqrt {\smash{\rho '_{3,3} \rho '_{5,5}}\rule{0pt}{8pt}\vphantom{\rho _{5,5}} } , - \sqrt {\smash{\rho '_{2,2} \rho '_{6,6}}\rule{0pt}{8pt}\vphantom{\rho _{6,6}} } \} }  \\
   {} &\!\! { = 2\max \{ 0,} &\!\!\! {\frac{{E + 2\sqrt {\lambda _4 \lambda _6 } }}{2} - \sqrt {\lambda _4 \lambda _6 } \} }  \\
   {} &\!\! { = 2\max \{ 0,} &\!\!\! {\frac{E}{2}\} }  \\
   {} &\!\! { = E,} &\!\!\! {}  \\
\end{array}\!
\end{Equation}
which, with \Eq{45}, proves that\hsp{0.5} \smash{$\rhoEPUminTGX$}\hsp{-0.5} is EPU-equivalent to $\rho$ when $Q\geq 0$, since $E$ is the\hsp{-1} I-concurrence\hsp{1} of the $\rho$ used to compute \raisebox{1.0pt}{\smash{$\rhoEPUminTGX$}} in \Eq{35}, and we are justified in using \Eq{7} to show this since we proved it in \Sec{III.A}.

In analogy to \cite{HeXU}, when $Q<0$, $\rho$ is always separable (which we will prove next as part of \Sec{III.B.4}), so in this case, $E=0$ and putting \Eq{35} into \Eq{7} gives
\begin{Equation}                      {47}
\begin{array}{*{20}l}
   {E(\rhoEPUminTGX)} &\!\! {= 2\max \{ 0,} &\!\!\! {\phantom{|\rho '_{1,6} |} - \sqrt {\smash{\rho '_{2,2} \rho '_{4,4} }\rule{0pt}{8pt}\vphantom{\rho _{4,4}}} , - \sqrt {\smash{\rho '_{1,1} \rho '_{5,5}}\rule{0pt}{8pt}\vphantom{\rho _{5,5}} } ,}  \\
   {} &\!\! {} &\!\!\! {|\rho '_{1,6} | - \sqrt {\smash{\rho '_{3,3} \rho '_{4,4}}\rule{0pt}{8pt}\vphantom{\rho _{4,4}} } , - \sqrt {\smash{\rho '_{1,1} \rho '_{6,6}}\rule{0pt}{8pt}\vphantom{\rho _{6,6}} } ,}  \\
   {} &\!\! {} &\!\!\! {\phantom{|\rho '_{1,6} |} - \sqrt {\smash{\rho '_{3,3} \rho '_{5,5}}\rule{0pt}{8pt}\vphantom{\rho _{5,5}} } , - \sqrt {\smash{\rho '_{2,2} \rho '_{6,6}}\rule{0pt}{8pt}\vphantom{\rho _{6,6}} } \} }  \\
   {} &\!\! { = 2\max \{ 0,} &\!\!\! {\frac{{\lambda _1  - \lambda _5 }}{2} - \sqrt {\lambda _4 \lambda _6 }\}}  \\
   {} &\!\! { = \max \{ 0,} &\!\!\! {\!\!\!\!\lambda _1  - \lambda _5  - 2\sqrt {\lambda _4 \lambda _6 } \}}  \\
   {} &\!\! { = 0} &\!\!\! {}  \\
   {} &\!\! { = E,} &\!\!\! {}  \\
\end{array}\!
\end{Equation}
where we used the fact that $Q<0$ implies that $\lambda _1  - \lambda _5  - 2\sqrt {\lambda _4 \lambda _6 }  < 0$, which we will also prove in \Sec{III.B.4}.

Therefore, we have proven that both $Q$ cases of \smash{$\rhoEPUminTGX$} preserve both the spectrum and entanglement of any general $\rho$ used to construct it, but as in \cite{HeXU}, this proof also requires that we prove that $E=0$ when $Q<0$ and that both $Q$ cases cover all physically possible spectrum-entanglement combinations, which we do next in \Sec{III.B.4}.
\subsubsection{\label{sec:III.B.4}Proof that All $\Lambda E$ Combinations are Achievable by \smash{$\rhoEPUminTGX$}}
Here we follow the proof from \cite{HeXU} fairly closely with a only few simple changes, but it is important enough to merit writing it in detail. First we show that (i) all states $\rho$ can only qualify as one of the two $Q$ cases from \Eq{35}.  Then we show that (ii) both of those forms of \raisebox{1.0pt}{\smash{$\rhoEPUminTGX$}} admit all possible $\Lambda E$ combinations for each $Q$. Here, we use the notation that $\Lambda\!\equiv\!\text{diag}\{\lambda_k\}$ where $\{\lambda_k\}\!\equiv\!\{\lambda_k\}|_{k=1}^{k=6}\!\equiv\!\{\lambda_1,\ldots,\lambda_6\}$.
\begin{itemize}[leftmargin=*,labelindent=8pt]\setlength\itemsep{0pt}
\item[\textbf{i.}]\hypertarget{LambdaCExhaustive:i}{}\textbf{Proof~that~All~$\rho$~Qualify~as~One~of~the~$Q$ Cases:} Noting the dependence of $Q$ as
\begin{Equation}                      {48}
\begin{array}{*{20}l}
   Q &\!\! { \equiv (\lambda _1  - \lambda _5 )^2  - (E + 2\sqrt {\lambda _4 \lambda _6 } )^2 }  \\
   {} &\!\! { \equiv Q(\lambda _1 ,\lambda _4 ,\lambda _5 ,\lambda _6 ,E),}  \\
\end{array}
\end{Equation}
\vspace{-12pt}\\
(which does not depend on \textit{all} the eigenvalues; just a subset of them), and its case splitting from \Eq{35} as
\begin{Equation}                      {49}
\left\{ {\begin{array}{*{20}l}
   {\text{$Q$ Case 1:}} & {Q \geq 0}  \\
   {\text{$Q$ Case 2:}} & {Q < 0,}  \\
\end{array}} \right.
\end{Equation}
\vspace{-12pt}\\
we can make the following conclusions;
    \begin{itemize}[leftmargin=*,labelindent=4pt]\setlength\itemsep{0pt}
    \item[\textbf{a.}]\hypertarget{LambdaEexhaustive:i.a}{}The fact that $Q$ is \textit{real} for every combination of $\Lambda$ and $E$, proves that the two mutually exclusive $Q$ cases of \Eq{49} cover all possible values of $Q$ [i.e., given the possible values of the arguments in \Eq{48}, we are not missing any $Q$ values by partitioning its values as in \Eq{49}].
    \item[\textbf{b.}]\hypertarget{LambdaEexhaustive:i.b}{}Since every $\rho$ has a spectrum $\{\lambda_k\}$ and entanglement $E$ (even if zero), then for every $\rho$, there exists a value of $Q$ [i.e., there are no states that somehow do not have a $Q$ value].
    \item[\textbf{c.}]\hypertarget{LambdaEexhaustive:i.c}{}By \hyperlink{LambdaEexhaustive:i.a}{Conclusion a} and \hyperlink{LambdaEexhaustive:i.b}{Conclusion b}, every $\rho$ falls into exactly one of the $Q$ cases in \Eq{49}.
    \end{itemize}
Now that we have established that every $\rho$ has a $Q$ value qualifying as one of the two cases in \Eq{49}, we must show that the $Q$ cases do not limit the $\Lambda E$ combinations [i.e., \textit{does the form of} \raisebox{1.0pt}{\smash{$\rhoEPUminTGX$}} \textit{for each $Q$ case in \Eq{35} permit all possible $E$ values for a given spectrum?}].
\item[\textbf{ii.}]\hypertarget{LambdaEexhaustive:ii}{}\textbf{Proof that Each $Q$ Case of \smash{$\rhoEPUminTGX$} Admits All $\Lambda E$ Combinations:~~}
    \begin{itemize}[leftmargin=*,labelindent=4pt]\setlength\itemsep{0pt}
    \item[\textbf{a.}]\hypertarget{LambdaEexhaustive:ii.a}{}For $Q\!\geq\! 0$, the result in \Eq{46} that \smash{$E(\protect\shiftmath{1.0pt}{\rhoEPUminTGX})\!=\!E$} $\forall\,\{\lambda_k\}$ proves that this $Q$-case \smash{$\rhoEPUminTGX$} admits\rule{0pt}{9pt} all $\Lambda E$ combinations.
    \item[\textbf{b.}]\hypertarget{LambdaEexhaustive:ii.b}{}For $Q<0$, we use the following facts:
        \begin{itemize}[leftmargin=*,labelindent=4pt]\setlength\itemsep{0pt}
        \item[\textbf{1.}]\hypertarget{LambdaEexhaustive:ii.b.1}{}At the edge of the $Q\geq 0$ case where $Q=0$, then by \Eq{48}, $E + 2\sqrt {\lambda _4 \lambda _6 }  = \lambda _1  - \lambda _5 $, so then \smash{$\rhoEPUminTGX$} becomes (still in the $Q\geq 0$ case),
\begin{Equation}                      {50}
\rhoEPUminTGX  =\! \left( {\begin{array}{*{20}c}
   {\frac{{\lambda _1  + \lambda _5 }}{2}} &  \cdot  &  \cdot  &  \cdot  &  \cdot  & {\frac{{\lambda _1  - \lambda _5 }}{2}}  \\
    \cdot  & {\lambda _2 } &  \cdot  &  \cdot  &  \cdot  &  \cdot   \\
    \cdot  &  \cdot  & {\lambda _4 } &  \cdot  &  \cdot  &  \cdot   \\
    \cdot  &  \cdot  &  \cdot  & {\lambda _6 } &  \cdot  &  \cdot   \\
    \cdot  &  \cdot  &  \cdot  &  \cdot  & {\lambda _3 } &  \cdot   \\
   {\frac{{\lambda _1  - \lambda _5 }}{2}} &  \cdot  &  \cdot  &  \cdot  &  \cdot  & {\frac{{\lambda _1  + \lambda _5 }}{2}}  \\
\end{array}} \right)\!,
\end{Equation}
which, by \Eq{41}, is a \textit{maximally entangled mixed state} (MEMS) \cite{IsHi,ZiBu,HoBM,VeAM,WNGK} with respect to (wrt) a given spectrum as proven in \cite{MeMH} in which \smash{$\rho_{\MEMS_\Lambda}$} is LPU equivalent to \raisebox{1.2pt}{\smash{$\rhoEPUminTGX$}} as \smash{$\protect\shiftmath{1.5pt}{\rhoEPUminTGX}  \equiv U_{\LPU} \rho _{\MEMS_\Lambda  } U_{\LPU}^\dag $} where $U_{\LPU}  \equiv I^{(1)}  \otimes (|3\rangle \langle 1| + |2\rangle \langle 2| + |1\rangle \langle 3|)$. [Also note that by \Eq{41}, \Eq{50} is a MEMS for \textit{all} $Q$ values, regardless of the sign of $Q$.]
        \item[\textbf{2.}]\hypertarget{LambdaEexhaustive:ii.b.2}{}Since \Eq{50} qualifies as minimal TGX form and holds for all $Q$, we can use \Eq{7} to get the I-concurrence of any MEMS wrt spectrum as
\begin{Equation}                      {51}
\hsp{35}E_{\MEMS} \equiv\max\{0,\lambda _1  - \lambda _5 -2\sqrt {\lambda _4 \lambda _6 }\},
\end{Equation}
        (also valid for all $Q$), which proves \Eq{40} with \Eq{39} as input, and we can identify
\begin{Equation}                      {52}
\hsp{35}\epre\equiv e_{\MEMS}\equiv\lambda _1  - \lambda _5 -2\sqrt {\lambda _4 \lambda _6 },
\end{Equation}
[which is \Eq{39}] as the analogue of minimum average preconcurrence for MEMS, which can be negative, for example if all eigenvalues are $\frac{1}{6}$.
        \item[\textbf{3.}]\hypertarget{LambdaEexhaustive:ii.b.3}{}No state can have a larger $E$ than $E_{\MEMS}$;
\begin{Equation}                      {53}
\hsp{35}0\leq E(\rho)\leq E_{\MEMS}\;\;\forall\,\rho, \;\forall Q.
\end{Equation}
        \item[\textbf{4.}]\hypertarget{LambdaEexhaustive:ii.b.4}{}Focusing only on the $Q<0$ case, we now solve for the conditions that $Q<0$ implies for $E$:
\begin{Equation}                      {54}
\hsp{35}\begin{array}{*{20}l}
   {\hsp{109.5}Q} &\!\!\!  <  &\!\!\! 0  \\
   {(\lambda _1  \hsp{-1}-\hsp{-1} \lambda _5 )^2  \hsp{-1}-\hsp{-1} (E \hsp{-1}+\hsp{-1} 2\sqrt {\lambda _4 \lambda _6 } )^2} &\!\!\!  <  &\!\!\! 0  \\
   {\hsp{65}E \hsp{-1}+\hsp{-1} 2\sqrt {\lambda _4 \lambda _6 } } &\!\!\!  >  &\!\!\! {\lambda _1  \hsp{-1}-\hsp{-1} \lambda _5 }  \\
   {\hsp{109.5}E} &\!\!\!  >  &\!\!\! {\lambda _1  \hsp{-1}-\hsp{-1} \lambda _5  \hsp{-1}-\hsp{-1} 2\sqrt {\lambda _4 \lambda _6 } }  \\
   {\hsp{109.5}E} &\!\!\!  >  &\!\!\! {\,\epre,}  \\
\end{array}
\end{Equation}
        which shows that
\begin{Equation}                      {55}
\hsp{40}(Q < 0) \Rightarrow (E > \epre) \Rightarrow \left\{ {\begin{array}{*{20}l}
   {E > E_{\MEMS} ;} & {\epre \geq 0}  \\
   {E > \epre;} & {\epre < 0,}  \\
\end{array}} \right.
\end{Equation}
since $\epre = E_{\MEMS}$ when $\epre \geq 0$ by \Eqs{51}{52}.
        \item[\textbf{5.}]\hypertarget{LambdaEexhaustive:ii.b.5}{}In \Eq{55}, the $\epre \geq 0$ case would mean that $Q<0$ implies that $E>E_{\MEMS}$, \textit{which is never possible} because $\max(E)=E_{\MEMS}$ from \Eq{53}.  Therefore, only the $\epre < 0$ case of \Eq{55} can apply to physical states, so \Eq{55} becomes
\begin{Equation}                      {56}
\hsp{40}(Q < 0) \Rightarrow (E > \epre) \Rightarrow (\epre < 0) \;\;\forall\,\rho.
\end{Equation}
        \item[\textbf{6.}]\hypertarget{LambdaEexhaustive:ii.b.6}{}From \Eqs{51}{52}, we know that 
\begin{Equation}                      {57}
\hsp{35}(\epre < 0)\Rightarrow (E_{\MEMS}=0),
\end{Equation}
which agrees with part of the fact from \Eq{37}.
        \item[\textbf{7.}]\hypertarget{LambdaEexhaustive:ii.b.7}{}So putting $E_{\MEMS}=0$ from \Eq{57} into \Eq{53},
\begin{Equation}                      {58}
\hsp{35}(\epre < 0)\Rightarrow [0\leq E(\rho)\leq 0\;\;\forall\,\rho]\Rightarrow [E(\rho)=0\;\;\forall\,\rho].
\end{Equation}
        \item[\textbf{8.}]\hypertarget{LambdaEexhaustive:ii.b.8}{}Putting \Eq{58} into \Eq{56},
\begin{Equation}                      {59}
\hsp{35}(Q<0)\Rightarrow [E(\rho)=0\;\;\forall\,\rho].
\end{Equation}
        \item[\textbf{9.}]\hypertarget{LambdaEexhaustive:ii.b.9}{}Since by \Eq{56}, $(Q<0)\Rightarrow(\epre<0) \;\forall\,\rho$, then by \Eq{52} we also have $\lambda _1  - \lambda _5 -2\sqrt {\lambda _4 \lambda _6 }<0 \;\forall\,\rho$ when $Q<0$, which \textit{proves} the claim in \Eq{47} that $E(\rhoEPUminTGX)=0$ for all spectra $\Lambda$ for which $Q<0$. Thus, comparing the now-proven result in \Eq{47} and the result of \Eq{59} regarding \textit{general} states $\rho$ proves that: 
\begin{Equation}                      {60}
\hsp{40}\begin{array}{*{20}l}
   {\parbox{2.0in}{The $Q<0$ case of \protect\raisebox{1.0pt}{\smash{$\rhoEPUminTGX$}} in \Eq{35} \textit{does} exhaust all possible $\Lambda E$ combinations when $Q<0$.}}  \\
\end{array}
\end{Equation}
        \end{itemize}
    \end{itemize}
    Therefore, since \hyperlink{LambdaEexhaustive:ii.a}{(a)} and \hyperlink{LambdaEexhaustive:ii.b}{(b)} prove that \raisebox{1.2pt}{\smash{$\rhoEPUminTGX$}} contains all possible $\Lambda E$ combinations for all possible $Q$ cases, the claim of \hyperlink{LambdaEexhaustive:ii}{(ii)} is proven true.
\end{itemize}
Thus, since Claims \hyperlink{LambdaEexhaustive:i}{(i)} and \hyperlink{LambdaEexhaustive:ii}{(ii)} have been proven true, this completes the proof that \raisebox{1.2pt}{\smash{$\rhoEPUminTGX$}} of \Eq{35} and \Eq{8} is fully EPU-equivalent to the set of all states $\rho$ in $2\times 3$, since the set of all possible $\Lambda EQ$ combinations achievable by $\rho$ is also achievable by \raisebox{1.2pt}{\smash{$\rhoEPUminTGX$}}.

Furthermore, note that the fact we proved from \Eq{56} that $(Q < 0) \Rightarrow (\epre < 0) \;\;\forall\,\rho$ where $\epre\equiv e_{\MEMS}\equiv\lambda _1  - \lambda _5 -2\sqrt {\lambda _4 \lambda _6 }$ by \Eq{52} is fully consistent with \Eq{37}, where the $Q=0$ case corresponds to the ``edge case'' from \hyperlink{LambdaEexhaustive:ii.b.1}{(ii.b.1)}, for which $E=\lambda _1  - \lambda _5 -2\sqrt {\lambda _4 \lambda _6 }=\epre$. In that case, the facts that $E\geq 0$ and $E=\epre$ mean that $(Q=0)\Rightarrow (\epre\geq 0)$ so when $Q=0$, the entanglement is only $0$ when $\lambda _1  - \lambda _5 -2\sqrt {\lambda _4 \lambda _6 }$ itself is $0$ [again consistent with \Eq{37}], but if $\lambda _1  - \lambda _5 -2\sqrt {\lambda _4 \lambda _6 }> 0$, the state is entangled. [This does not contradict the discussion afer \Eq{37}; in this state family, $Q=0$ means the state is a MEMS with $E=e_{\MEMS}$, whereas for general $Q$ values, the state is not necessarily a MEMS, so its $E$ is not necessarily equal to $e_{\MEMS}$. For instance, a computational basis state has $E=0$, $Q=1$, but $e_{\MEMS}=1\neq E$, so having $e_{\MEMS}>0$ does not guarantee entanglement in general. Also note that all states have an $e_{\MEMS}$ value whether or not they are MEMS.] Thus, $Q=0$ does not necessarily mean the state is separable, whereas $Q<0$ \textit{does} mean the state is separable, and is why we made $Q<0$ its own case and grouped $Q=0$ with the positive $Q$ case.

Also, due to \Eq{53}, throughout this paper, by ``all $E$ values,'' we mean \textit{physical} $E$, which are given by 
\begin{Equation}                      {61}
E \in [0,\max \{ 0,\lambda _1  - \lambda _5  - 2\sqrt {\lambda _4 \lambda _6 } \} ],
\end{Equation}
or simply $E \in [0,\max \{ 0,e_{\MEMS} \} ]=[0,E_{\MEMS} ]$.
\subsection{\label{sec:III.C}Derivation of the EPU-Equivalent Minimal TGX Family \smash{$\rhoEPUminTGX$}}
Our next task is to use the minimal TGX I-concurrence formula of \Eq{7}, which we proved in \Sec{III.A}, to derive the family of \textit{EPU-minimal TGX states} \smash{$\rhoEPUminTGX$} of \Eq{8} that has EPU equivalence to the set of\hsp{-1} general states.

To achieve EPU equivalence, this family must include MEMS as a special case (since otherwise it would be missing some $\Lambda E$ combinations), and so we choose the $\rho_\MEMS$ of \Eq{41} as the special case of our EPU family since it has minimal TGX form, which will let us use \Eq{7}.

Immediately, we get its eigenstates in order of its eigenvalues  $\lambda _1  \geq  \cdots  \geq \lambda _6 $ as
\begin{Equation}                      {62}
\left( {\begin{array}{*{20}c}
   {\frac{1}{{\sqrt 2 }}}  \\
    \cdot   \\
    \cdot   \\
    \cdot   \\
    \cdot   \\
   {\frac{1}{{\sqrt 2 }}}  \\
\end{array}} \right)\!,\!\left( {\begin{array}{*{20}c}
    \cdot\vphantom{\frac{1}{{\sqrt 2 }}}   \\
   1  \\
    \cdot   \\
    \cdot   \\
    \cdot   \\
    \cdot\vphantom{\frac{1}{{\sqrt 2 }}}   \\
\end{array}} \right)\!,\!\left( {\begin{array}{*{20}c}
    \cdot\vphantom{\frac{1}{{\sqrt 2 }}}   \\
    \cdot   \\
    \cdot   \\
    \cdot   \\
   1  \\
    \cdot\vphantom{\frac{1}{{\sqrt 2 }}}   \\
\end{array}} \right)\!,\!\left( {\begin{array}{*{20}c}
    \cdot\vphantom{\frac{1}{{\sqrt 2 }}}   \\
    \cdot   \\
   1  \\
    \cdot   \\
    \cdot   \\
    \cdot\vphantom{\frac{1}{{\sqrt 2 }}}   \\
\end{array}} \right)\!,\!\left( {\begin{array}{*{20}c}
   {\frac{1}{{\sqrt 2 }}}  \\
    \cdot   \\
    \cdot   \\
    \cdot   \\
    \cdot   \\
   {\frac{{ - 1}}{{\sqrt 2 }}}  \\
\end{array}} \right)\!,\!\left( {\begin{array}{*{20}c}
    \cdot\vphantom{\frac{1}{{\sqrt 2 }}}   \\
    \cdot   \\
    \cdot   \\
   1  \\
    \cdot   \\
    \cdot\vphantom{\frac{1}{{\sqrt 2 }}}   \\
\end{array}} \right)\!.
\end{Equation}

Our next goal is to find a mixed state related to $\rho_\MEMS$ of \Eq{41} that includes as many \textit{lower} entanglement values as possible for any given spectrum, while still reaching that MEMS form for certain values of its parameters.  The need to keep spectrum general means we must look for ways to generalize only the \textit{eigenstates} of \Eq{62}.

There are many ways we could generalize \Eq{62}, but since the MEMS of \Eq{41} (which is a TGX state) only has superposition in the $\{1,6\}$ subspace, and since the only quartet that contains $\{1,6\}$ in an inseparable qubit is $\{1,3,4,6\}$, and since all quartets of TGX states have X form which would let us use \Eq{7} since its superposition is in just \textit{one} quartet, then the most general way we might need to adapt \Eq{62} is by defining orthogonal states of general superposition ($\theta$-\textit{states} in the parlance of \cite{HedE}) that involve the nonzero levels of the ME TGX states of \Eq{38} that live exclusively in  $\{1,3,4,6\}$.

Thus, we generalize \smash{$\frac{1}{{\sqrt 2 }}(|1\rangle  + |6\rangle )$} as $c_\alpha  |1\rangle  + s_\alpha  |6\rangle $, and \smash{$\frac{1}{{\sqrt 2 }}(|1\rangle  - |6\rangle )$} as $s_\alpha  |1\rangle  - c_\alpha  |6\rangle $   and generalize $\{|3\rangle,|4\rangle\}$ as $c_\beta  |3\rangle  + s_\beta  |4\rangle $ and $-s_\beta  |3\rangle  + c_\beta  |4\rangle $, where $c_\theta   \equiv \cos (\theta )$, $s_\theta   \equiv \sin (\theta )$, and  $\alpha,\beta  \in [0,\frac{\pi }{2}]$, and phases were chosen so that \Eq{63} becomes exactly \Eq{62} when $(\alpha,\beta)=(\frac{\pi}{4},0)$, all of which, ordering by eigenvalue index as $|\epsilon _1 \rangle , \ldots ,|\epsilon _6 \rangle $ gives
\begin{Equation}                      {63}
\left( {\begin{array}{*{20}c}
   {c_\alpha  }  \\
   \cdot  \\
   \cdot  \\
   \cdot  \\
   \cdot  \\
   {s_\alpha  }  \\
\end{array}} \right)\!,\!\left( {\begin{array}{*{20}c}
   \cdot  \\
   {1  }  \\
   \cdot  \\
   \cdot  \\
   \cdot  \\
   \cdot  \\
\end{array}} \right)\!,\!\left( {\begin{array}{*{20}c}
   \cdot  \\
   \cdot  \\
   \cdot  \\
   \cdot  \\
   {1  }  \\
   \cdot  \\
\end{array}} \right)\!,\!\left(\!\! {\begin{array}{*{20}c}
   \cdot  \\
   \cdot  \\
   {c_\beta}  \\
   {s_\beta}  \\
   \cdot  \\
   \cdot  \\
\end{array}}\!\! \right)\!,\!\left(\!\! {\begin{array}{*{20}c}
   {s_\alpha  }  \\
   \cdot  \\
   \cdot  \\
   \cdot  \\
   \cdot  \\
   { - c_\alpha  }  \\
\end{array}}\!\! \right)\!,\!\left(\!\! {\begin{array}{*{20}c}
   \cdot  \\
   \cdot  \\
   {-s_\beta}  \\
   {c_\beta}  \\
   \cdot  \\
   \cdot  \\
\end{array}}\!\! \right)\!.
\end{Equation}
These states are then combined as
\begin{Equation}                      {64}
\rhoMinTGX  \equiv \sum\limits_{k = 1}^6 {\lambda _k |\epsilon _k \rangle \langle \epsilon _k |} ,
\end{Equation}
which expands as
\begin{Equation}                      {65}
\begin{array}{*{20}l}
   {\rhoMinTGX =}  \\
   {\left(\!\! {\begin{array}{*{20}c}
   {\lambda _1 c_\alpha ^2  \!\!+\!\! \lambda _5 s_\alpha ^2 } &\!\!  \cdot  &\!\!  \cdot  &\!\!  \cdot  &\!\!  \cdot  &\!\! {\frac{{\lambda _1  - \lambda _5 }}{2}s_{2\alpha } }  \\
    \cdot  &\!\! {\lambda _2 } &\!\!  \cdot  &\!\!  \cdot  &\!\!  \cdot  &\!\!  \cdot   \\
    \cdot  &\!\!  \cdot  &\!\! {\lambda _4 c_\beta ^2  \!\!+\!\! \lambda _6 s_\beta ^2 } &\!\! {\frac{{\lambda _4  - \lambda _6 }}{2}s_{2\beta } } &\!\!  \cdot  &\!\!  \cdot   \\
    \cdot  &\!\!  \cdot  &\!\! {\frac{{\lambda _4  - \lambda _6 }}{2}s_{2\beta } } &\!\! {\lambda _4 s_\beta ^2  \!\!+\!\! \lambda _6 c_\beta ^2 } &\!\!  \cdot  &\!\!  \cdot   \\
    \cdot  &\!\!  \cdot  &\!\!  \cdot  &\!\!  \cdot  &\!\! {\lambda _3 } &\!\!  \cdot   \\
   {\frac{{\lambda _1  - \lambda _5 }}{2}s_{2\alpha } } &\!\!  \cdot  &\!\!  \cdot  &\!\!  \cdot  &\!\!  \cdot  &\!\! {\lambda _1 s_\alpha ^2  \!\!+\!\! \lambda _5 c_\alpha ^2 }  \\
\end{array}}\!\! \right)\!.}  \\
\end{array}
\end{Equation}

Then, since \Eq{65} has minimal TGX form, \Eq{7} gives its entanglement via I-concurrence as
\begin{Equation}                      {66}
\begin{array}{*{20}l}
   {E(\alpha ,\beta ) \equiv } &\!\! {2\max \left\{ {\rule{0pt}{9pt}} \right.\hsp{-2}0,}  \\
   {} &\!\! {\frac{{\lambda _1  - \lambda _5 }}{2}s_{2\alpha }  -\hsp{-1} \sqrt {(\lambda _4 c_{\smash{\beta}\vphantom{\alpha}} ^2  + \lambda _6 s_{\smash{\beta}\vphantom{\alpha}} ^2 )(\lambda _4 s_{\smash{\beta}\vphantom{\alpha}} ^2  + \lambda _6 c_{\smash{\beta}\vphantom{\alpha}} ^2 )\rule{0pt}{8.0pt}} ,}  \\
   {} &\!\! {\frac{{\lambda _4  - \lambda _6 }}{2}s_{2\beta }  -\hsp{-1} \sqrt {(\lambda _1 c_{\alpha} ^2  + \lambda _5 s_{\alpha} ^2 )(\lambda _1 s_{\alpha} ^2  + \lambda _5 c_{\alpha} ^2 )\rule{0pt}{8.0pt}} \left. {\rule{0pt}{9pt}} \right\}\hsp{-1}.}  \\
\end{array}\!
\end{Equation}

Now we wish to determine the parameters $\alpha$ and $\beta$ that will allow us to achieve EPU equivalence to general states by preserving the entanglement (spectrum is already preserved since we are building the state via orthonormal eigenstates weighted by a general spectrum).  

However, in contrast to $2\times 2$, since we do not have a computable I-concurrence formula for general $2\times 3$ states, here we must make physical parameter sets more methodically.  For this, we simply use this set of normalized nonnegative eigenvalues $\lambda _1  \geq  \cdots  \geq \lambda _6 $ to determine a physical entanglement from \Eq{61} as 
\begin{Equation}                      {67}
E\equiv E_\eta\equiv\eta\max \{ 0,\lambda _1  - \lambda _5  - 2\sqrt {\lambda _4 \lambda _6 } \},
\end{Equation}
for $\eta\in[0,1]$.  Then this spectrum $\{\lambda _k \}$ and its associated physical entanglement $E$ corresponds to some general physical state whose values of these parameters exist.

We now need to show that we can always find $\alpha$ and $\beta$ such that $E(\alpha,\beta)=E$.  Explicitly, we can do this by observing that in \Eq{66}, since $\rhoMinTGX$ should coincide with $\rho _{\MEMS}$ of \Eq{41}, the entanglement of which involves the term $\lambda _1  - \lambda _5 $, then we want to keep only line 2 of \Eq{66}.  Furthermore, since only \textit{one} of the arguments will win in the maximization, we can force it to be the line-2 argument by choosing 
\begin{Equation}                      {68}
\beta  = 0
\end{Equation}
to get rid of the other argument, which, when put into \Eq{66}, and setting $E(\alpha ,\beta )=E$ (since that is what we want to achieve) gives
\begin{Equation}                      {69}
\begin{array}{*{20}l}
   {E} &\!\! {= 2\max \{ 0,\frac{{\lambda _1  - \lambda _5 }}{2}s_{2\alpha }  - \sqrt {\lambda _4 \lambda _6 } \} .}  \\
\end{array}
\end{Equation}
Now that we have motivated $\beta$, \Eq{69} will determine the value of $\alpha$ and our only job is to verify that $\alpha$ has enough freedom to handle all physical values it needs to without limiting the spectrum-entanglement combinations.  

The remainder of the derivation of the desired family of states follows the derivation in \cite{HeXU} almost exactly with $\lambda _3  \to \lambda _5 $, $\lambda _4  \to \lambda _6 $, $\lambda _2  \to \lambda _4 $, and $C\to E$ since the math problem is otherwise identical at this stage. To highlight an important part, when solving for $\alpha$ when $(\lambda _1  - \lambda _5 )s_{2\alpha }  - 2\sqrt {\lambda _4 \lambda _6 }  \geq 0$ which gives
\begin{Equation}                      {70}
\alpha  = \left\{ {\begin{array}{*{20}l}
   {\frac{1}{2}\sin ^{ - 1} (\frac{{E + 2\sqrt {\lambda _4 \lambda _6 } }}{{\lambda _1  - \lambda _5 }});} & {\lambda _1  \ne \lambda _5 }  \\
   {\frac{\pi }{4};} & {\lambda _1  = \lambda _5 ,}  \\
\end{array}} \right.
\end{Equation}
here the $\lambda _1  = \lambda _5$ case arises because given $\lambda _1  \geq  \cdots  \geq \lambda _6 $ and $\sum\nolimits_{k=1}^{6}\lambda_k =1$, then the facts that $\lambda _1  = \lambda _5$ and $(\lambda _1  - \lambda _5 )s_{2\alpha }  - 2\sqrt {\lambda _4 \lambda _6 }  \geq 0$ imply that $\lambda_{1}=\cdots=\lambda_{5}=\frac{1}{5}$ \textit{and} $\lambda_6 =0$, so then the MEMS limit $\lambda _1  - \lambda _5   - 2\sqrt {\lambda _4 \lambda _6 } =0$, which by \Eq{37} means $E=0$, and since $\alpha$ is free when $\lambda _1  = \lambda _5$, we can put $E=0$ into the $\lambda _1  \neq \lambda _5$ case and use $\lambda _1  - \lambda _5  = 2\sqrt {\lambda _4 \lambda _6 } $ to get the $\lambda _1  = \lambda _5$ case result for continuity (which is justified since $(\alpha,\beta)=(\frac{\pi}{4},0)$ gives a MEMS with $E=0$ in this case of \Eq{70}.

\begin{figure}[H]
\centering
\includegraphics[width=1.00\linewidth]{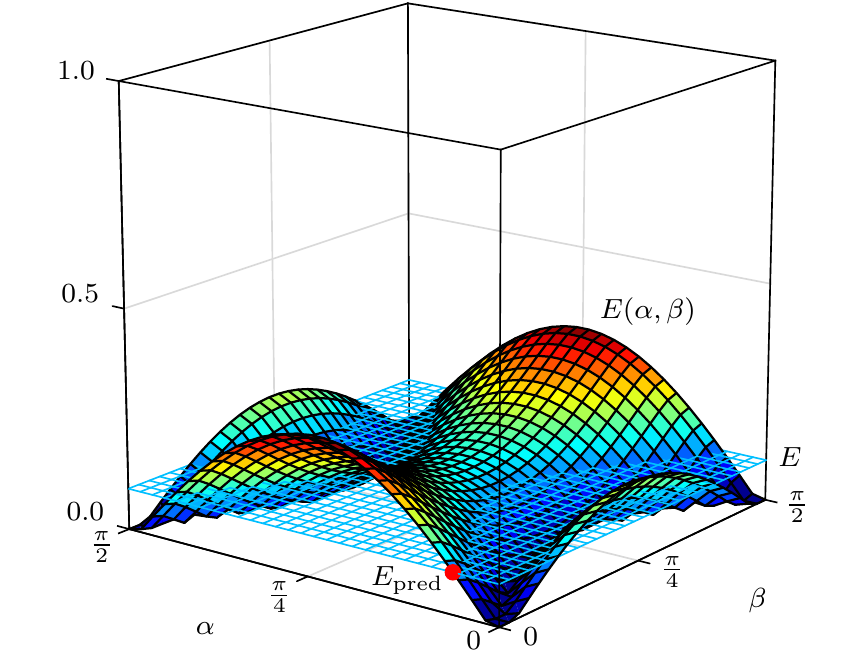}%
\vspace{-4pt}
\caption[]{(color online) Example of $E(\alpha,\beta)$ from \Eq{66} for a particular arbitrary rank-$6$ spectrum and target I-concurrence $E$ from \Eq{67}. $E_{\text{pred}}$, shown as the red dot, is the right side of \Eq{69}, which is $E(\alpha,\beta)$ with $\alpha$ from \Eq{70} and $\beta$ from \Eq{68}, while the planar surface shows the target $E$. If the red dot always lies on an intersection of the plane $E$ and the surface $E(\alpha,\beta)$, then $E_{\text{pred}}$ is correct.  Repeating this test $10^3$ times for arbitrary input spectra and random physical $E$ showed no failures, which provided strong motivation to use \Eq{68} and \Eq{70} as part of the ansatz for \smash{$\rhoEPUminTGX$} in \Eq{8} (which was then \textit{proven} to be a full solution, as shown in \Sec{III.B}). The color scheme here is not related to the other figures, but is chosen to match that of \cite{HeXU} for sake of comparison.}
\label{fig:4}
\end{figure}

Formally, the state we are deriving here can function as an ansatz, and the proofs in \Sec{III.B} verify that ansatz to have EPU equivalence, so that \textit{proves} it causes no limitations. To get a visual sense that this ansatz does not cause limitations, \Fig{4} shows an example of numerical checks that the choice of $\beta=0$ does indeed always permit $E(\alpha ,\beta )$ to equal $E$ when $\alpha$ is given by \Eq{70}.
Thus, closely following \cite{HeXU} with the above changes, we get the EPU-minimal TGX family for $2\times 3$ as
\begin{Equation}                      {71}
\rhoEPUminTGX \! =\! \!\left(\!\! {\begin{array}{*{20}c}
   {\frac{{\lambda _1  + \lambda _5  + \sqrt \Omega  }}{2}} &\!  \cdot  &\!  \cdot  &\!  \cdot  &\!  \cdot  &\! {\frac{{\sqrt {(\lambda _1  - \lambda _5 )^2  - \Omega } }}{2}}  \\[-0.5mm]%
    \cdot  &\! {\lambda _2 } &\!  \cdot  &\!  \cdot  &\!  \cdot  &\!  \cdot   \\[-0.5mm]%
    \cdot  &\!  \cdot  &\! {\lambda _4 } &\!  \cdot  &\!  \cdot  &\!  \cdot   \\[-0.5mm]%
    \cdot  &\!  \cdot  &\!  \cdot  &\! {\lambda _6 } &\!  \cdot  &\!  \cdot   \\[-0.5mm]%
    \cdot  &\!  \cdot  &\!  \cdot  &\!  \cdot  &\! {\lambda _3 } &\!  \cdot   \\[-0.5mm]%
   {\frac{{\sqrt {(\lambda _1  - \lambda _5 )^2  - \Omega } }}{2}} &\!  \cdot  &\!  \cdot  &\!  \cdot  &\!  \cdot  &\! {\frac{{\lambda _1  + \lambda _5  - \sqrt \Omega  }}{2}}  \\
\end{array}} \right)\!,
\end{Equation}
which agrees with \Eq{8}, where $\Omega  \equiv \max \{ 0,Q\} $, and
\begin{Equation}                      {72}
Q \equiv (\lambda _1  - \lambda _5 )^2  - (E + 2\sqrt {\lambda _4 \lambda _6 } )^2 ,
\end{Equation}
where this $Q$ has the same general properties as $Q$ in \cite{HeXU}; by \Eq{56}, any input state $\rho$ with $Q<0$ implies $e_\MEMS <0$ and therefore $E(\rho)\!=\! 0$. Also, the physical limits of $E$ are those in \Eq{61} and \Eq{67}. Thus, \Eqs{71}{72} and \Eq{67} can be used to parameterize a state of any physical spectrum-entanglement combination in $2\times 3$, and any $2\times 3$ state $\rho$ can be transformed to an EPU-minimal TGX state of this form (or LPU variation) with the same $E$ and spectrum as $\rho$, in analogy to the $2\times 2$ case in \cite{HeXU}.

Again, the \textit{proof} that \raisebox{1.0pt}{\smash{$\rhoEPUminTGX $}} is EPU-equivalent to all states is in \Sec{III.B}, while the result in \Eq{71} can be taken as an ansatz which is verified by that proof.
\vspace{0pt}\\
\subsection{\label{sec:III.D}Derivation of the LS Decomposition of \smash{$\rhoEPUminTGX$}}
In \cite{LSD}, an explicit LS decomposition for $2\times 2$ was given based on Wootters's decomposition states from \cite{Woot}.  Here we show how to apply this to the EPU-minimal TGX states of \Eq{8} to derive \Eqs{9}{15}.

First, in \Sec{III.D.1} we derive the overall LS decomposition in terms of Wootters x kets, then in \Sec{III.D.2}, we derive the explicit forms of the x kets.
\subsubsection{\label{sec:III.D.1}LS Decomposition in Terms of Wootters x Kets}
Since the LS decomposition in \cite{LSD} is for $2\times 2$, and since minimal TGX states only have one quartet with potential entanglement, such as $\mathbf{e} = \{ 1,3,4,6\} $ in \Eq{8}, then that is the $2\times 2$ system which gets the LS decomposition, and we call $\mathbf{e}$ \textit{the entanglement quartet} of this EPU-minimal TGX state.  Since $\mathbf{e}$ is spacewise orthogonal to the rest of the state, rather than decomposing \raisebox{1.3pt}{\smash{$\rho_{\EPU_{\TGX}^{\min}}^{\protect\shiftmath{-0.1pt}{\{\mathbf{e}\}}}$}}, which makes for messy notation, we can use the fact that its eigenstates are also eigenstates of the full state, with zeros in the parts outside $\mathbf{e}$ (see \hyperlink{minTGXIconcurrenceFact:6}{Facts 6}\hyperlink{minTGXIconcurrenceFact:11}{--11} in \Sec{III.A}, which apply here because EPU-minimal TGX states are subsets of minimal SGX states).

Thus, if the normalized eigenstates of \raisebox{1.3pt}{\smash{$\rho_{\EPU_{\TGX}^{\min}}^{\{\mathbf{e}\}}$}} are \smash{$|\epsilon _h ^{\{ \mathbf{e}\} } \rangle $} for $h \in 1,2,3,4$ where the indices match eigenvalues such that \smash{$\lambda _1 ^{\{ \mathbf{e}\} }  \ge \lambda _2 ^{\{ \mathbf{e}\} }  \ge \lambda _3 ^{\{ \mathbf{e}\} }  \ge \lambda _4 ^{\{ \mathbf{e}\} } $} [notice here that, in contrast to \cite{Woot}, we use $\lambda$ for eigenvalues of states, and reserve $\xi$ for eigenvalues of spin-flip products as in \Eq{1}], then its \textit{subnormalized} eigenstates are
\begin{Equation}                      {73}
|v_h ^{\{ \mathbf{e}\} } \rangle  \equiv \sqrt {\lambda _h ^{\{ \mathbf{e}\} } } |\epsilon _h ^{\{ \mathbf{e}\} } \rangle ;\;\;\;h \in 1,2,3,4.
\end{Equation}
From \Eq{8}, these eigenvalues in terms of those of the full state are
\begin{Equation}                      {74}
\begin{array}{l}
 \lambda _1 ^{\{\mathbf{e}\} }  = \lambda _1  \\ 
 \lambda _2 ^{\{\mathbf{e}\} }  = \lambda _4  \\ 
 \lambda _3 ^{\{\mathbf{e}\} }  = \lambda _5  \\ 
 \lambda _4 ^{\{\mathbf{e}\} }  = \lambda _6 , \\ 
 \end{array}
\end{Equation}
and rewriting \Eq{73} embedded into the $\mathbf{e}$ subspace of zero vectors in the full space, using \Eq{74} to relabel, we get the full subnormalized eigenvectors of \Eq{8} as
\begin{Equation}                      {75}
\begin{array}{*{20}c}
   {\begin{array}{*{20}l}
   {|u_1 \rangle  \equiv |v_1 \rangle  \equiv \sqrt {\lambda _1 } |\epsilon _1 \rangle ,}  \\
   {|u_2 \rangle  \equiv |v_4 \rangle  \equiv \sqrt {\lambda _4 } |\epsilon _4 \rangle ,}  \\
   {|u_3 \rangle  \equiv |v_5 \rangle  \equiv \sqrt {\lambda _5 } |\epsilon _5 \rangle ,}  \\
   {|u_4 \rangle  \equiv |v_6 \rangle  \equiv \sqrt {\lambda _6 } |\epsilon _6 \rangle ,}  \\
\end{array}} & {\begin{array}{*{20}l}
   {|v_2 \rangle  \equiv \sqrt {\lambda _2 } |\epsilon _2 \rangle ,}  \\
   {|v_3 \rangle  \equiv \sqrt {\lambda _3 } |\epsilon _3 \rangle ,}  \\
\end{array}}  \\
\end{array}
\end{Equation}
where $|\epsilon _h \rangle $ are eigenvectors of the full state, and the left column in \Eq{75} are those whose only nonzero entries are in $\mathbf{e} = \{ 1,3,4,6\} $, while the right column are states whose only nonzero entries are \textit{outside} of $\mathbf{e}$. We relabel the $\mathbf{e}$-related states as $|u_k \rangle $ for $k \in 1,2,3,4$ to simplify equations focusing on $\mathbf{e}$.  Furthermore, in this section, we abbreviate as \smash{$\rho\equiv\rhoEPUminTGX$}.

We then make symmetric matrix $\tau$ with elements\rule{0pt}{12.5pt}
\begin{Equation}                      {76}
\tau _{k,l}  \equiv \langle u_k |\widetilde{u}_l \rangle  \equiv \langle u_k |S|u_l ^* \rangle ,
\end{Equation}
for $k,l \in 1,2,3,4$ [which is reducible to \smash{$k,l \in 1, \ldots ,r'$} where \smash{$r'\! \equiv\! r^{\{ \mathbf{e}\} }  \!\equiv\! \text{rank}(\rho ^{\{ \mathbf{e}\} } )\!=\!\text{rank}(\text{diag}\{\lambda_{1},\lambda_{4},\lambda_{5},\lambda_{6}\})$}, which is generally \textit{not} equal to $r''\equiv\text{rank}(\tau )$ as we will see], where \smash{$|\widetilde{\psi} \rangle  \equiv S|\psi ^* \rangle $} where \smash{$|\psi ^* \rangle $} is the complex conjugate of $|\psi \rangle $, and we use the $0$-embedded spin-flip operator,
\begin{Equation}                      {77}
S \equiv (\sigma _2  \otimes \sigma _2 )  \oplus 0^{\{ \overline{\mathbf{e}}\} } ,
\end{Equation}
where $\overline{\mathbf{e}}$ means ``not $\mathbf{e}$,'' meaning the subspace spacewise orthogonal to $\mathbf{e}$.  Again, we can ignore the eigenstates in $\overline{\mathbf{e}}$ for now because we are simply doing the LS decomposition for \smash{$\rho ^{\{ \mathbf{e}\} } $} $0$-embedded in the full space.

Then get the Autonne-Takagi factorization of $\tau$ as
\begin{Equation}                      {78}
\tau  = UDU^T ,
\end{Equation}
where $U$ is unitary, and $D$ is real nonnegative diagonal with diagonal values in descending order of value, as
\begin{Equation}                      {79}
D_{j,j}  = \xi _j ;\;\;\;j\in 1,\ldots,r'';\;\;\;r''\equiv\text{rank}(\tau ),
\end{Equation}
which are concurrence singular values, as in \Eq{1}. More importantly, we need $U$ to make Wootters's subnormalized decomposition x kets (a term we use based on Wootters's labels, not because of their shape as states) as
\begin{Equation}                      {80}
|x_a \rangle  \equiv \sum\limits_{j = 1}^{r'} {U_{j,a} |u_j \rangle } ,
\end{Equation}
for $a \in 1,2,3,4$ (or $a\in 1,\ldots,r'$), with the property that
\begin{Equation}                      {81}
\langle x_a |\widetilde{x}_b \rangle  = (U^\dag  \tau U^* )_{a,b}  = \xi _a \delta _{a,b} .
\end{Equation}
[Notice our $r'$ is Wootters's $n$ and our $U$ is Wootters's $U^\dag  $ because we use the standard notation in \Eq{78} of putting the ``bare'' operator on the left of the diagonal part.]

Then, following \cite{LSD} with some modifications, we define a variation (not a renormalization) of the $\{|x_a \rangle\}$ as
\begin{Equation}                      {82}
|x'_a \rangle  \equiv \frac{1}{{\sqrt {\xi _a +\delta_{{\xi_a},0}} }}|x_a \rangle ;\;\;\;a \in 1, \ldots ,r',
\end{Equation}
where the Kronecker deltas protect against division by $0$, while keeping the number of states at $r'$ ensures that we maintain a proper decomposition, so the $0$-embedded subspace decomposition is
\begin{Equation}                      {83}
\begin{array}{*{20}l}
   {\rho ^{\{ \mathbf{e}\}  \oplus \mathbf{0}}  \equiv \rho ^{\{ \mathbf{e}\} }  \oplus 0^{\{ \overline{\mathbf{e}}\} }} &\!\! {=\sum\limits_{a = 1}^{r'} {|x_a \rangle \langle x_a |}}  \\
   {} &\!\! {= \sum\limits_{a = 1}^{r'} {(\xi _a +\delta_{{\xi_a},0}) |x'_a \rangle \langle x'_a |},}  \\
\end{array}
\end{Equation}
where the notation on the far left means to extract the $\mathbf{e}$ subspace of $\rho$ and then re-install it in that same subspace of a zero matrix in the full system, an operation we call ``$0$-embedding.''

Now that we have a decomposition where the weights are the concurrence singular values, a simple process of adding zeros creatively can let us find a new decomposition with the actual concurrence as the first weight as
\begin{Equation}                      {84}
\begin{array}{*{20}l}
   {\rho ^{\{ \mathbf{e}\}  \oplus \mathbf{0}} } &\!\!  =  &\!\! {\sum\limits_{a = 1}^{r'} {\!(\xi _a +\delta_{{\xi_a},0}) |x'_a \rangle \langle x'_a |} }  \\
   {} &\!\!  =  &\!\! {(\xi _1  - \xi _2  - \xi _3  - \xi _4 )|x'_1 \rangle \langle x'_1 |}  \\
   {} &\!\! {} &\!\! { + (\delta_{{\xi_1},0}+\xi _2  + \xi _3  + \xi _4 )|x'_1 \rangle \langle x'_1 | }  \\
   {} &\!\! {} &\!\! {+ \sum\limits_{a = 2}^{r'} {\!(\xi _a +\delta_{{\xi_a},0}) |x'_a \rangle \langle x'_a |}} \\
   {} &\!\!  =  &\!\! {\max \{ 0,\xi _1  - \xi _2  - \xi _3  - \xi _4 \} |x'_1 \rangle \langle x'_1 |}  \\
   {} &\!\! {} &\!\! { +\! (\delta_{{\xi_1},0}+\min \{ \xi _1 ,\xi _2  \!+\! \xi _3  \!+\! \xi _4 \}) |x'_1 \rangle \langle x'_1 | }  \\
   {} &\!\! {} &\!\! {+ \sum\limits_{a = 2}^{r'} {\!(\xi _a +\delta_{{\xi_a},0}) |x'_a \rangle \langle x'_a |,}} \\
\end{array}
\end{Equation}
where we inserted the max function to protect against negative probability in the first term (and in the process, making that weight exactly the concurrence of the $\mathbf{e}$ subspace of $\rho$), while the min function compensates for the max function's presence when its output is $0$ (in that case the whole $0$-embedded subspace state is separable).  Thus, the first term in the last equality of \Eq{84} is the entangled part of the LS decomposition (the term with the max function) and the remaining terms are the separable part. The $\delta_{{\xi_1},0}$ is grouped with the separable part because when $\xi_1 =0$, then $\xi_2 =\xi_3 =\xi_4 =0$ and \smash{$\rho ^{\{ \mathbf{e}\}  \oplus \mathbf{0}}$} is both pure and separable, so it is appropriate that the entangled part is $0$, while the separable part then becomes $|x'_1 \rangle \langle x'_1 |$. Note also that when ``adding zeros creatively'' in the second equality of \Eq{84}, we did not include extra deltas since those new terms sum to $0$ anyway.

Next, we rewrite \Eq{84} in terms of the original Wootters x kets using \Eq{82} as
\begin{Equation}                      {85}
\begin{array}{*{20}l}
   {\rho ^{\{ \mathbf{e}\}  \oplus \mathbf{0}} } &\!\!  =  &\!\! {\frac{{\max \{ 0,\xi _1  - \xi _2  - \xi _3  - \xi _4 \} }}{{\xi _1 +\delta_{{\xi_1},0}}}|x_1 \rangle \langle x_1 |}  \\
   {} &\!\! {} &\!\! { + \frac{{\min \{ \xi _1 ,\xi _2  + \xi _3  + \xi _4 \} +\delta_{{\xi_1},0}}}{{\xi _1 +\delta_{{\xi_1},0}}}|x_1 \rangle \langle x_1 | +\! \sum\limits_{a = 2}^{r'} {|x_a \rangle \langle x_a |}, }  \\
\end{array}
\end{Equation}
and then, normalizing each of the Wootters x kets,
\begin{Equation}                      {86}
\begin{array}{*{20}l}
   {\rho ^{\{ \mathbf{e}\}  \oplus \mathbf{0}}  = }  &\!\! {\frac{{\max \{ 0,\xi _1  - \xi _2  - \xi _3  - \xi _4 \} }}{{\xi _1 +\delta_{{\xi_1},0}}}\langle x_1 |x_1 \rangle \frac{{|x_1 \rangle \langle x_1 |}}{{\langle x_1 |x_1 \rangle }}}   \\
   {}  &\!\! { + \frac{{\min \{ \xi _1 ,\xi _2  + \xi _3  + \xi _4 \}+\delta_{{\xi_1},0} }}{{\xi _1 +\delta_{{\xi_1},0}}}\langle x_1 |x_1 \rangle \frac{{|x_1 \rangle \langle x_1 |}}{{\langle x_1 |x_1 \rangle }}}   \\
   {}  &\!\! { + \sum\limits_{a = 2}^{r'} {\langle x_a |x_a \rangle \frac{{|x_a \rangle \langle x_a |}}{{\langle x_a |x_a \rangle }}}, }   \\
\end{array}
\end{Equation}
where we note that it is only the partial ``tilde overlap'' of the x kets that gives the $\xi _a$ values, and in general their regular self overlaps do not give those same values. 

From \Eq{86}, the entangled part's normalized state is
\begin{Equation}                      {87}
\rho _E^{\{ \mathbf{e}\}  \oplus \mathbf{0}}  \equiv \frac{{|x_1 \rangle \langle x_1 |}}{{\langle x_1 |x_1 \rangle }},
\end{Equation}
with decomposition probability [noting that for minimal TGX states, $ E(\rho ^{\{ \mathbf{e}\}  \oplus \mathbf{0}})=C(\rho ^{\{ \mathbf{e}\}})\equiv C^{\{ \mathbf{e}\}}$],
\begin{Equation}                      {88}
p_E ^{\{ \mathbf{e}\} }  \!\equiv\! \frac{{\max \{ 0,\xi _1  - \xi _2  - \xi _3  - \xi _4 \} }}{{\xi _1 +\delta_{{\xi_1},0}}}\langle x_1 |x_1 \rangle \! =\! \frac{{C^{\{ \mathbf{e}\} } \langle x_1 |x_1 \rangle }}{{\xi _1 +\delta_{{\xi_1},0}}},
\end{Equation}
and the normalized separable part is
\begin{Equation}                      {89}
\begin{array}{*{20}l}
   {\rho _S^{\{ \mathbf{e}\}  \oplus \mathbf{0}}  \equiv } &\!\! {\frac{1}{{p_S ^{\{ \mathbf{e}\} } +\delta_{p_S ^{\{ \mathbf{e}\} },0}}}\left( {\rule{0pt}{10pt}} \right.\!\frac{{\min \{ \xi _1 ,\xi _2  + \xi _3  + \xi _4 \}+\delta_{{\xi_1},0} }}{{\xi _1 +\delta_{{\xi_1},0}}}|x_1 \rangle \langle x_1 |}  \\
   {} &\!\! { + \sum\limits_{a = 2}^{r'} {|x_a \rangle \langle x_a |} \left. {\rule{0pt}{10pt}}\! \right),}  \\
\end{array}
\end{Equation}
with probability
\begin{Equation}                      {90}
\begin{array}{*{20}l}
   {p_S ^{\{ \mathbf{e}\} } } &\!\! { \equiv \frac{{\min \{ \xi _1 ,\xi _2  + \xi _3  + \xi _4 \}+\delta_{{\xi_1},0} }}{{\xi _1 +\delta_{{\xi_1},0}}}\langle x_1 |x_1 \rangle  + \sum\limits_{a = 2}^{r'} {\langle x_a |x_a \rangle}  }  \\
   {} &\!\! { = \tr(\rho ^{\{ \mathbf{e}\}  \oplus \mathbf{0}} ) - p_E ^{\{ \mathbf{e}\} } }  \\
   {} &\!\! { = \lambda _1  \!+\! \lambda _4  \!+\! \lambda _5  \!+\! \lambda _6  \!-\! \frac{{\max \{ 0,\xi _1  - \xi _2  - \xi _3  - \xi _4 \} }}{{\xi _1 +\delta_{{\xi_1},0} }}\langle x_1 |x_1 \rangle ,}  \\
\end{array}
\end{Equation}
where we used the fact that the $\mathbf{e}$ subspace state is unnormalized so that $p_E ^{\{ \mathbf{e}\} }+p_S ^{\{ \mathbf{e}\} }=\tr(\rho ^{\{ \mathbf{e}\}  \oplus \mathbf{0}} )=\lambda _1  + \lambda _4  + \lambda _5  + \lambda _6 $, so we can simply write \Eq{89} as
\begin{Equation}                      {91}
\begin{array}{*{20}l}
   {\rho _S^{\{ \mathbf{e}\}  \oplus \mathbf{0}}  \!\!= } &\!\!\! {\frac{1}{{\tr(\rho ^{\{ \mathbf{e}\}  \oplus \mathbf{0}} ) - p_E ^{\{ \mathbf{e}\} } +\delta_{p_S ^{\{ \mathbf{e}\} },0}}}}  \\
   {} &\!\!\! {\times\!\left( {\rule{0pt}{10pt}} \right.\!\!\frac{{\min \{ \xi _1 ,\xi _2  + \xi _3  + \xi _4 \}+\delta_{{\xi_1},0} }}{{\xi _1 +\delta_{{\xi_1},0}}}|x_1 \rangle \langle x_1 |\!+\!\!\sum\limits_{a = 2}^{r'} {\!|x_a \rangle \langle x_a |} \left. {\rule{0pt}{10pt}}\!\! \right)\!,} \\
\end{array}\!
\end{Equation}
or even more simply, as
\begin{Equation}                      {92}
\rho _S^{\{ \mathbf{e}\}  \oplus \mathbf{0}}  = \frac{{\rho ^{\{ \mathbf{e}\}  \oplus \mathbf{0}}  - p_E^{\{ \mathbf{e}\} } \rho _E^{\{ \mathbf{e}\}  \oplus \mathbf{0}}}}{{\tr(\rho ^{\{ \mathbf{e}\}  \oplus \mathbf{0}} ) - p_E ^{\{ \mathbf{e}\} } +\delta_{p_S ^{\{ \mathbf{e}\} },0}}}.
\end{Equation}
Thus, the LS decomposition of \smash{$\rho ^{\{ \mathbf{e}\}  \oplus \mathbf{0}}$} is
\begin{Equation}                      {93}
\rho ^{\{ \mathbf{e}\}  \oplus \mathbf{0}}  = p_E ^{\{ \mathbf{e}\} } \rho _E^{\{ \mathbf{e}\}  \oplus \mathbf{0}}  + [\tr(\rho ^{\{ \mathbf{e}\}  \oplus \mathbf{0}} ) - p_E ^{\{ \mathbf{e}\} } ]\rho _S^{\{ \mathbf{e}\}  \oplus \mathbf{0}},
\end{Equation}
the average I-concurrence of which is
\begin{Equation}                      {94}
\begin{array}{*{20}l}
   {\langle E(\rho ^{\{ \mathbf{e}\}  \oplus \mathbf{0}} )\rangle } &\!\! { = p_E ^{\{ \mathbf{e}\} } E(\rho _E^{\{ \mathbf{e}\}  \oplus \mathbf{0}} ) }  \\
   {} &\!\! {\phantom{=}+ [\tr(\rho ^{\{ \mathbf{e}\}  \oplus \mathbf{0}} ) - p_E ^{\{ \mathbf{e}\} } ]E(\rho _S^{\{ \mathbf{e}\}  \oplus \mathbf{0}} )} \\
   {} &\!\! {= p_E ^{\{ \mathbf{e}\} } C(\rho _E^{\{ \mathbf{e}\} } )} \\
   {} &\!\! {\phantom{=}+ [\tr(\rho ^{\{ \mathbf{e}\}  \oplus \mathbf{0}} ) - p_E ^{\{ \mathbf{e}\} } ]C(\rho _S^{\{ \mathbf{e}\}} )} \\
   {} &\!\! { = p_E ^{\{ \mathbf{e}\} } \frac{{|\langle x_1 |\widetilde{x}_1 \rangle |}}{{\langle x_1 |x_1 \rangle }} + 0}  \\
   {} &\!\! { = \frac{{\max \{ 0,\xi _1  - \xi _2  - \xi _3  - \xi _4 \} }}{{\xi _1 +\delta_{{\xi_1},0}}}\langle x_1 |x_1 \rangle \frac{{\xi _1 }}{{\langle x_1 |x_1 \rangle }}}  \\
   {} &\!\! { = \max \{ 0,\xi _1  - \xi _2  - \xi _3  - \xi _4 \} }  \\
   {} &\!\! { = E(\rho ^{\{ \mathbf{e}\}  \oplus \mathbf{0}} ),}  \\
\end{array}
\end{Equation}
which shows that this is truly an optimal decomposition of $\rho ^{\{ \mathbf{e}\}  \oplus \mathbf{0}}$ since its \textit{average} concurrence is the \textit{actual} concurrence.  Again see \cite{LSD} for the full proof of this result. Note that we used the fact that the I-concurrence of \smash{$\rho ^{\{ \mathbf{e}\}  \oplus \mathbf{0}}$} is just the regular concurrence of \smash{$\rho^{\{\mathbf{e}\}}$} due to the fact that \smash{$\rho ^{\{ \mathbf{e}\}  \oplus \mathbf{0}}$} is $0$ outside of $\mathbf{e}$, and any nonzero parts of other quartet subspaces are always separable.

Now that we found the LS decomposition of the 0-embedded entanglement subspace of \Eq{8}, we can get a decomposition of the \textit{full} state of \Eq{8} as
\begin{Equation}                      {95}
\begin{array}{*{20}l}
   \rho  &\!\!  =  &\!\! {\rho ^{\{ \mathbf{e}\}  \oplus \mathbf{0}}  + \rho ^{\{ \overline{\mathbf{e}}\}  \oplus \mathbf{0}} }  \\
   {} &\!\!  =  &\!\! {p_E ^{\{ \mathbf{e}\} } \rho _E^{\{ \mathbf{e}\}  \oplus \mathbf{0}}  + [\tr(\rho ^{\{ \mathbf{e}\}  \oplus \mathbf{0}} ) - p_E ^{\{ \mathbf{e}\} } ]\rho _S^{\{ \mathbf{e}\}  \oplus \mathbf{0}} }  \\
   {} &\!\! {} &\!\! { + \lambda _2 \rho _{|2\rangle }  + \lambda _3 \rho _{|5\rangle } ,}  \\
\end{array}
\end{Equation}
where $\rho _{|\psi \rangle }  \equiv |\psi \rangle \langle \psi |$.  But here, since these last two terms are \textit{always} separable since they are computational basis states $|2\rangle  \equiv |1\rangle  \otimes |2\rangle $ and $|5\rangle  \equiv |2\rangle  \otimes |2\rangle $, then we can add them to any other separable state and the sum is still separable by the definition of separable states (specifically note that their I-concurrence is $0$ in the \textit{full space}, and furthermore since the separable part in $\mathbf{e}$ in \Eq{95} [from \Eq{91}] has zeros in all other elements, its I-concurrence in the full space is $0$ as well). Therefore, the \textit{total} separable part of the LS decomposition of \Eq{95} is
\begin{Equation}                      {96}
\begin{array}{*{20}l}
   {\rho _S } &\!\! { \equiv \frac{{[\tr(\rho ^{\{ \mathbf{e}\}  \oplus \mathbf{0}} ) - p_E ^{\{ \mathbf{e}\} } ]\rho _S^{\{ \mathbf{e}\}  \oplus \mathbf{0}}  + \lambda _2 \rho _{|2\rangle }  + \lambda _3 \rho _{|5\rangle } }}{{[\tr(\rho ^{\{ \mathbf{e}\}  \oplus \mathbf{0}} ) - p_E ^{\{ \mathbf{e}\} } ]   + \lambda _2  + \lambda _3 +\delta_{p_E ^{\{ \mathbf{e}\} },1}}}}  \\
   {} &\!\! { = \frac{{[\tr(\rho ^{\{ \mathbf{e}\}  \oplus \mathbf{0}} ) - p_E ^{\{ \mathbf{e}\} } ]\rho _S^{\{ \mathbf{e}\}  \oplus \mathbf{0}}  + \lambda _2 \rho _{|2\rangle }  + \lambda _3 \rho _{|5\rangle } }}{{\lambda _1  + \lambda _4  + \lambda _5  + \lambda _6  - p_E ^{\{ \mathbf{e}\} }  + \lambda _2  + \lambda _3 +\delta_{p_E ^{\{ \mathbf{e}\} },1}}}}  \\
   {} &\!\! { = \frac{{(\lambda _1  + \lambda _4  + \lambda _5  + \lambda _6  - p_E ^{\{ \mathbf{e}\} } )\rho _S^{\{ \mathbf{e}\}  \oplus \mathbf{0}}  + \lambda _2 \rho _{|2\rangle }  + \lambda _3 \rho _{|5\rangle } }}{{1 - p_E ^{\{ \mathbf{e}\} } +\delta_{p_E ^{\{ \mathbf{e}\} },1}}},}  \\
\end{array}
\end{Equation}
so the probability of the separable part is \smash{$1 - p_E ^{\{ \mathbf{e}\} } $} which means that the entangled-part probability of $\rho$ is the same as for its entanglement subspace, so
\begin{Equation}                      {97}
p_E  = p_E ^{\{ \mathbf{e}\} } ,
\end{Equation}
and we get the explicit LS decomposition for the full state in \Eq{8} as (recalling our abbreviation that \smash{$\rho\equiv \rhoEPUminTGX $}),
\begin{Equation}                      {98}
\rhoEPUminTGX = p_E \rho _E  + (1 - p_E )\rho _S ,
\end{Equation}
where from the above argument, the entangled part of the full state is the same as for the $\mathbf{e}$ subspace, so
\begin{Equation}                      {99}
E= C^{\{ \mathbf{e}\} },
\end{Equation}
where $E$ is the I-concurrence of the full state $\rho$ [keeping in mind that $C^{\{ \mathbf{e}\} }$ is \textit{not} the same thing as $E(\rho_{E})$], so putting \Eq{88} into \Eq{97}, and \Eq{99} into that gives
\begin{Equation}                      {100}
p_E  \!=\!\frac{{\max \{ 0,\xi _1  - \xi _2  - \xi _3  - \xi _4 \} }}{{\xi _1 +\delta_{{\xi_1},0} }}\langle x_1 |x_1 \rangle \! =\!\frac{{E \langle x_1 |x_1 \rangle }}{{\xi _1 +\delta_{{\xi_1},0}}},
\end{Equation}
and from \Eq{87} and \Eq{95},
\begin{Equation}                      {101}
\rho _E  = \frac{{|x_1 \rangle \langle x_1 |}}{{\langle x_1 |x_1 \rangle }},
\end{Equation}
while putting \Eq{91} with $\tr(\rho ^{\{ \mathbf{e}\}  \oplus \mathbf{0}} )=\lambda _1  + \lambda _4  + \lambda _5  + \lambda _6 $ into line 3 of \Eq{96} and using \Eq{97} gives
\begin{Equation}                      {102}
\begin{array}{*{20}l}
   {\rho _S  = } &\!\! {\frac{1}{{1 - p_E +\delta_{p_{E},1}}}\left( {\rule{0pt}{10pt}}\! \right.{\lambda _2 \rho _{|2\rangle }  + \lambda _3 \rho _{|5\rangle }}}  \\
   {} &\!\! {{+ \frac{{\min \{ \xi _1 ,\xi _2  + \xi _3  + \xi _4 \}+\delta_{{\xi_1},0} }}{{\xi _1 +\delta_{{\xi_1},0}}}|x_1 \rangle \langle x_1 | +\sum\limits_{a = 2}^{r'} {|x_a \rangle \langle x_a |}}\left. {\rule{0pt}{10pt}}\! \right)\!.}  \\
\end{array}
\end{Equation}
Thus we have derived the LS decomposition of \Eq{8} in \Eqs{9}{12}, where it is helpful to note that here, 
\begin{Equation}                      {103}
E(\rho_E)=C(\rho_E^{\{\mathbf{e}\}})=\frac{{|\langle x_1 |\widetilde{x}_1 \rangle |}}{{\langle x_1 |x_1 \rangle }}=\frac{\xi_{1}}{{\langle x_1 |x_1 \rangle }},
\end{Equation}
[noting that in general, $E(\rho_E)\neq E$] which shows that 
\begin{Equation}                      {104}
p_E E(\rho_E)=E=\max \{ 0,\xi _1  - \xi _2  - \xi _3  - \xi _4 \},
\end{Equation}
which is what makes \Eq{98} an optimal LS decomposition since its average I-concurrence is the actual I-concurrence of the state.

Thus, this decomposition concentrates the minimum average I-concurrence into a single weighted decomposition state (which happens to be pure for EPU-minimal TGX states in $2\times 3$, as we have just proven).  This is valid because the entangled part is a pure minimal TGX state so its I-concurrence always simplifies to a single subspace concurrence, which is what allowed us to get this LS decomposition. Thus, this agrees with several of the claims used in the proofs in \Sec{III} as well.

Note that the entire derivation above applies equally well to the middle minimal SGX state \smash{$\rhoMinSGX$} in \Eq{19} (and to all minimal SGX states by LPU variation), with the only difference being that $\lambda _2 \rho _{|2\rangle }  + \lambda _3 \rho _{|5\rangle }$ must be replaced by \rule{0pt}{9.5pt}\smash{$\lambda _2 \rho _{\protect\rule{0pt}{7pt}|\epsilon _{\hsp{-0.5}1}^{\protect\shiftmath{0.2pt}{\{ 2,\!5\}} } \rangle }^{\protect\shiftmath{0pt}{\{ 2,\!5\}  \oplus \mathbf{0}}}  + \lambda _3 \rho _{\protect\rule{0pt}{7pt}|\epsilon _2^{\protect\shiftmath{0.2pt}{\{ 2,\!5\}} } \rangle }^{\protect\shiftmath{0pt}{\{ 2,\!5\}  \oplus \mathbf{0}}}$} where \smash{$|\epsilon _1^{\hsp{0.2}\protect\shiftmath{0.5pt}{\{ 2,\!5\}}} \rangle $} and \smash{$|\epsilon _2^{\hsp{0.2}\protect\shiftmath{0.5pt}{\{ 2,\!5\}}} \rangle$} are\hsp{-0.5} eigenstates\hsp{-0.7} of\hsp{-1.3} \rule{0pt}{12pt}\smash{$\rho^{\protect\shiftmath{-0.9pt}{\{2,\!5\hsp{-0.3}\}}}$} with eigenvalues \smash{$\lambda _1^{\{ 2,5\} }  = \lambda _2$} and \smash{$\lambda _2^{\hsp{0.2}\protect\shiftmath{0.5pt}{\{ 2,\!5\}}}  = \lambda _3$}. But since those states are always separable as explained in \hyperlink{minTGXIconcurrenceFact:5}{Fact 5} of \Sec{III.A}, that changes none of these results about the entanglement. We simply focused on EPU-minimal TGX states to make the discussion simpler and to provide additional proof of the EPU equivalence of \Eq{8}.  Thus, with the above modification, \Eqs{98}{104} hold in general for all minimal SGX states in $2\times 3$, and can be used with the $r'$ limits as shown in numerical calculations, while our results in the next section require us to sacrifice that specificity and use upper limits of $4$ as seen in \Eqs{9}{15} to gain an overall explicit form.

Also note that these methods are adaptable to the $2\times 2$ states of \Eq{4} by simply setting $\lambda_2 \to 0 $, $\lambda_3 \to 0 $, $\lambda_4 \to \lambda_2 $, $\lambda_5 \to \lambda_3 $, $\lambda_6 \to \lambda_4 $, and transplanting only the $\mathbf{e}$ subspace of our EPU-minimal TGX family to the $2\times 2$ space in order, including for all eigenstates, etc.  Thus, even though the LS decomposition was not given in \cite{HeXU}, the one given here can be converted to that system with ease (and this applies to the next section here as well).

Implicit in all of the above is that care must be taken to ensure that each subnormalized eigenvector receives the appropriate index to ensure valid decomposition of each spacewise orthogonal subspace, which may require explicit definitions beyond standard routines, particularly in cases with degenerate eigenvalues.
\subsubsection{\label{sec:III.D.2}Explicit Forms of the Wootters x Kets}
Here, we will use the fact that the EPU-minimal TGX family of \Eq{8} has real TGX eigenvectors, which means that its $\tau$ also has real eigenvectors, to easily convert from a spectral decomposition of $\tau$ to a Takagi decomposition.

First, the eigenvector matrices of the EPU-minimal TGX state of \Eq{8} are
\begin{Equation}                      {105}
\begin{array}{*{20}c}
   {|\epsilon _1 \rangle  = \left( {\begin{array}{*{20}c}
   {\sqrt {\frac{{\Delta  + \sqrt \Omega  }}{{2\Delta }}} }  \\
    \cdot   \\
    \cdot   \\
    \cdot   \\
    \cdot   \\
   {\sqrt {\frac{{\Delta  - \sqrt \Omega  }}{{2\Delta }}} }  \\
\end{array}} \right)\!,\;|\epsilon _2 \rangle  = \left( {\begin{array}{*{20}c}
    \cdot   \\
   1  \\
    \cdot   \\
    \cdot   \\
    \cdot   \\
    \cdot   \\
\end{array}} \right)\!,\;|\epsilon _3 \rangle  = \left( {\begin{array}{*{20}c}
    \cdot   \\
    \cdot   \\
    \cdot   \\
    \cdot   \\
   { 1}  \\
    \cdot   \\
\end{array}} \right)\!,}  \\
   {|\epsilon _4 \rangle  = \left( {\begin{array}{*{20}c}
    \cdot   \\
    \cdot   \\
   1  \\
    \cdot   \\
    \cdot   \\
    \cdot   \\
\end{array}} \right)\!,\;|\epsilon _5 \rangle  = \left( {\begin{array}{*{20}c}
   {\sqrt {\frac{{\Delta  - \sqrt \Omega  }}{{2\Delta }}} }  \\
    \cdot   \\
    \cdot   \\
    \cdot   \\
    \cdot   \\
   { - \sqrt {\frac{{\Delta  + \sqrt \Omega  }}{{2\Delta }}} }  \\
\end{array}} \right)\!,\;|\epsilon _6 \rangle  = \left( {\begin{array}{*{20}c}
    \cdot   \\
    \cdot   \\
    \cdot   \\
   { 1}  \\
    \cdot   \\
    \cdot   \\
\end{array}} \right)\!,}  \\
\end{array}
\end{Equation}
where $\Delta$ and $\Omega$ are given in \Eq{15}, so \Eq{105} gives columns of \protect\raisebox{1.5pt}{\smash{$\epsilon _{\rhoEPUminTGX}$}} from \Eq{18}.  Then,\hsp{-1} from\hsp{-1} \Eq{75},\hsp{-1} define\hsp{-1} the\hsp{-1} relabeled subnormalized eigenvectors of the $0$-embedded entanglement subspace as
\begin{Equation}                      {106}
\begin{array}{*{20}l}
   {|u_1 \rangle  \equiv \left(\! {\begin{array}{*{20}c}
   {\sqrt {\lambda _1 } \sqrt {\frac{{\Delta  + \sqrt \Omega  }}{{2\Delta }}} }  \\
    \cdot   \\
    \cdot   \\
    \cdot   \\
    \cdot   \\
   {\sqrt {\lambda _1 } \sqrt {\frac{{\Delta  - \sqrt \Omega  }}{{2\Delta }}} }  \\
\end{array}}\! \right)\!,} &\!\! {|u_2 \rangle \equiv \left(\! {\begin{array}{*{20}c}
    \cdot   \\
    \cdot   \\
   {\sqrt {\lambda _4 } }  \\
    \cdot   \\
    \cdot   \\
    \cdot   \\
\end{array}}\! \right)\!,}  \\
   {|u_3 \rangle  \equiv \left(\! {\begin{array}{*{20}c}
   {\sqrt {\lambda _5 } \sqrt {\frac{{\Delta  - \sqrt \Omega  }}{{2\Delta }}} }  \\
    \cdot   \\
    \cdot   \\
    \cdot   \\
    \cdot   \\
   { - \sqrt {\lambda _5 } \sqrt {\frac{{\Delta  + \sqrt \Omega  }}{{2\Delta }}} }  \\
\end{array}}\! \right)\!,} &\!\! {|u_4 \rangle  \equiv \left(\! {\begin{array}{*{20}c}
    \cdot   \\
    \cdot   \\
    \cdot   \\
   {\sqrt {\lambda _6 } }  \\
    \cdot   \\
    \cdot   \\
\end{array}}\! \right)\!.}  \\
\end{array}
\end{Equation}
From \Eq{77} we have 
\begin{Equation}                      {107}
S = \left( {\begin{array}{*{20}c}
    \cdot  &  \cdot  &  \cdot  &  \cdot  &  \cdot  & { - 1}  \\
    \cdot  &  \cdot  &  \cdot  &  \cdot  &  \cdot  &  \cdot   \\
    \cdot  &  \cdot  &  \cdot  & 1 &  \cdot  &  \cdot   \\
    \cdot  &  \cdot  & 1 &  \cdot  &  \cdot  &  \cdot   \\
    \cdot  &  \cdot  &  \cdot  &  \cdot  &  \cdot  &  \cdot   \\
   { - 1} &  \cdot  &  \cdot  &  \cdot  &  \cdot  &  \cdot   \\
\end{array}} \right)\!,
\end{Equation}
and the elements of $\tau$ from \Eq{78} are given by
\begin{Equation}                      {108}
\tau _{k,l}  \equiv \langle u_k |\widetilde{u}_l \rangle  \equiv \langle u_k |S|u_l ^* \rangle ;\;\;\;k,l \in 1, \ldots ,4,
\end{Equation}
where here we set $\text{dim}(\tau)=4$ instead of $r'$ to get a general symbolic solution for all cases.

Then, the spin-flipped eigenstates of \Eq{106} are
\begin{Equation}                      {109}
\begin{array}{*{20}l}
   {|\widetilde{u}_1 \rangle  \equiv \left(\! {\begin{array}{*{20}c}
   {-\sqrt {\lambda _1 } \sqrt {\frac{{\Delta  - \sqrt \Omega  }}{{2\Delta }}} }  \\
    \cdot   \\
    \cdot   \\
    \cdot   \\
    \cdot   \\
   {-\sqrt {\lambda _1 } \sqrt {\frac{{\Delta  + \sqrt \Omega  }}{{2\Delta }}} }  \\
\end{array}}\! \right)\!,} &\!\! {|\widetilde{u}_2 \rangle \equiv \left(\! {\begin{array}{*{20}c}
    \cdot   \\
    \cdot   \\
    \cdot  \\
    {\sqrt {\lambda _4 } }   \\
    \cdot   \\
    \cdot   \\
\end{array}}\! \right)\!,}  \\
   {|\widetilde{u}_3 \rangle  \equiv \left(\! {\begin{array}{*{20}c}
   {\sqrt {\lambda _5 } \sqrt {\frac{{\Delta  + \sqrt \Omega  }}{{2\Delta }}} }  \\
    \cdot   \\
    \cdot   \\
    \cdot   \\
    \cdot   \\
   { - \sqrt {\lambda _5 } \sqrt {\frac{{\Delta  - \sqrt \Omega  }}{{2\Delta }}} }  \\
\end{array}}\! \right)\!,} &\!\! {|\widetilde{u}_4 \rangle  \equiv \left(\! {\begin{array}{*{20}c}
    \cdot   \\
    \cdot   \\
    {\sqrt {\lambda _6 } }   \\
    \cdot  \\
    \cdot   \\
    \cdot   \\
\end{array}}\! \right)\!,}  \\
\end{array}
\end{Equation}
so putting \Eq{106} and \Eq{109} into \Eq{108} gives
\begin{Equation}                      {110}
\tau  = \left( {\begin{array}{*{20}c}
   { - \frac{{\lambda _1 \sqrt {\Delta ^2  - \Omega } }}{\Delta }} &  \cdot  & {\frac{{\sqrt {\lambda _1 \lambda _5 \Omega}  }}{\Delta }} &  \cdot   \\
    \cdot  &  \cdot  &  \cdot  & {\sqrt {\lambda _4 \lambda _6 } }  \\
   {\frac{{\sqrt {\lambda _1 \lambda _5 \Omega}  }}{\Delta }} &  \cdot  & {\frac{{\lambda _5 \sqrt {\Delta ^2  - \Omega } }}{\Delta }} &  \cdot   \\
    \cdot  & {\sqrt {\lambda _4 \lambda _6 } } &  \cdot  &  \cdot   \\
\end{array}} \right)\!.
\end{Equation}
Notice that $\tau$ is real, symmetric, and has spacewise orthogonal subspaces $\{1,3\}$ and $\{2,4\}$. The eigenvalues of $\tau^{\{1,3\}}$ are (showing several forms for reference)
\begin{Equation}                      {111}
\begin{array}{*{20}l}
   {\zeta _ \pm  ^{\{ 1,3\} } } &\!\! { = \frac{{ - (\lambda _1  - \lambda _5 )\sqrt {\Delta ^2  - \Omega }  \pm \sqrt {(\lambda _1  + \lambda _5 )^2 \Delta ^2  - (\lambda _1  - \lambda _5 )^2 \Omega } }}{{2\Delta }}}  \\[0.5ex]
   {} &\!\! { = \frac{{ - (\lambda _1  - \lambda _5 )\sqrt {\Delta ^2  - \Omega }  \pm \sqrt {[(\lambda _1  + \lambda _5 )\sqrt {\Delta ^2  - \Omega } ]^2  + 4\lambda _1 \lambda _5 \Omega } }}{{2\Delta }}}  \\[0.5ex]
   {} &\!\! { = \frac{{ - (\lambda _1  - \lambda _5 )\sqrt {\Delta ^2  - \Omega }  \pm \sqrt {4\lambda _1 \lambda _5 \Delta ^2  + [(\lambda _1  - \lambda _5 )\sqrt {\Delta ^2  - \Omega } ]^2 } }}{{2\Delta }},}  \\
\end{array}
\end{Equation}
with corresponding subspace eigenstates
\begin{Equation}                      {112}
\begin{array}{*{20}l}
   {|e_ -  ^{\{ 1,3\} } \rangle } &\!\! { = \frac{1}{{N_-  }}\!\left( {\begin{array}{*{20}c}
   {\frac{{\sqrt {\lambda _1 \lambda _5 \Omega } }}{\Delta } + \delta _{\lambda _5 \Omega ,0} }  \\
   {\zeta _ -  ^{\{ 1,3\} }  + \frac{{\lambda _1 \sqrt {\Delta ^2  - \Omega } }}{\Delta }}  \\
\end{array}} \right)\;\text{for}\;\;\zeta _ -  ^{\{ 1,3\} },}  \\[2.5ex]
   {|e_ +  ^{\{ 1,3\} } \rangle } &\!\! { = \frac{1}{{N_+  }}\!\left( {\begin{array}{*{20}c}
   {\zeta _ +  ^{\{ 1,3\} }  - \frac{{\lambda _5 \sqrt {\Delta ^2  - \Omega } }}{\Delta }}  \\
   {\frac{{\sqrt {\lambda _1 \lambda _5 \Omega } }}{\Delta } + \delta _{\lambda _5 \Omega ,0} }  \\
\end{array}} \right)\;\text{for}\;\;\zeta _ +  ^{\{ 1,3\} },}  \\
\end{array}
\end{Equation}
[which were carefully derived so that when off-diagonals of $\tau$ are zero, its remaining diagonal elements are the eigenvalues in \textit{ascending} order since diagonals of $\tau$ are ascending, and \smash{$\zeta _ -  ^{\{ 1,3\} } \leq 0\leq \zeta _ +  ^{\{ 1,3\} }$} by line 3 of \Eq{111}] with normalization factors,  
\begin{Equation}                      {113}
\begin{array}{*{20}l}
   {N_ -  } &\!\! { \equiv \sqrt {(\zeta _ -  ^{\{ 1,3\} }  + \frac{{\lambda _1 \sqrt {\Delta ^2  - \Omega } }}{\Delta })^2  + (\frac{{\sqrt {\lambda _1 \lambda _5 \Omega } }}{\Delta } + \delta _{\lambda _5 \Omega ,0} )^2 } \,,}  \\
   {N_ +  } &\!\! { \equiv \sqrt {(\zeta _ +  ^{\{ 1,3\} }  - \frac{{\lambda _5 \sqrt {\Delta ^2  - \Omega } }}{\Delta })^2  + (\frac{{\sqrt {\lambda _1 \lambda _5 \Omega } }}{\Delta } + \delta _{\lambda _5 \Omega ,0} )^2 } \,.}  \\
\end{array}
\end{Equation}
Eigenvalues of \smash{$\tau^{\{2,4\}}$} are
\begin{Equation}                      {114}
\zeta _ \pm  ^{\{ 2,4\} }  =  \pm \sqrt {\lambda _4 \lambda _6 } \,,
\end{Equation}
with corresponding subspace eigenstates
\begin{Equation}                      {115}
|f_ \pm  ^{\{ 2,4\} } \rangle  = \frac{1}{{\sqrt 2 }}\left( {\begin{array}{*{20}c}
   1  \\
   { \pm 1}  \\
\end{array}} \right)\!.
\end{Equation}

Then, a spectral decomposition of $\tau$ (except for a permutation for descending-order eigenvalues) is
\begin{Equation}                      {116}
\tau  = VdV^\dag  ,
\end{Equation}
with unitary eigenvector matrix
\begin{Equation}                      {117}
V \equiv \left( {\begin{array}{*{20}c}
   {\frac{{\frac{{\sqrt {\lambda _1 \lambda _5 \Omega } }}{\Delta } + \delta _{\lambda _5 \Omega ,0} }}{{N_ -  }}} &  \cdot  & {\frac{{\zeta _ +  ^{\{ 1,3\} }  - \frac{{\lambda _5 \sqrt {\Delta ^2  - \Omega } }}{\Delta }}}{{N_ +  }}} &  \cdot   \\
    \cdot  & {\frac{1}{{\sqrt 2 }}} &  \cdot  & {\frac{1}{{\sqrt 2 }}}  \\
   {\frac{{\zeta _ -  ^{\{ 1,3\} }  + \frac{{\lambda _1 \sqrt {\Delta ^2  - \Omega } }}{\Delta }}}{{N_ -  }}} &  \cdot  & {\frac{{\frac{{\sqrt {\lambda _1 \lambda _5 \Omega } }}{\Delta } + \delta _{\lambda _5 \Omega ,0} }}{{N_ +  }}} &  \cdot   \\
    \cdot  & {\frac{1}{{\sqrt 2 }}} &  \cdot  & {\frac{{ - 1}}{{\sqrt 2 }}}  \\
\end{array}} \right)\!,
\end{Equation}
and eigenvalue matrix (again not in descending order),
\begin{Equation}                      {118}
d \equiv \text{diag}\{ \zeta _ -  ^{\{ 1,3\} } ,\zeta _ +  ^{\{ 2,4\} } ,\zeta _ +  ^{\{ 1,3\} } ,\zeta _ -  ^{\{ 2,4\} } \} .
\end{Equation}

Now that we have a spectral decomposition for $\tau$, we can convert it to a Takagi decomposition. First, peel off the phase factors from $d$ as
\begin{Equation}                      {119}
d = \text{diag}\left( {\begin{array}{*{20}c}
   {\text{sgn}(\zeta _ -  ^{\{ 1,3\} } )|\zeta _ -  ^{\{ 1,3\} } |}  \\
   {\text{sgn}(\zeta _ +  ^{\{ 2,4\} } )|\zeta _ +  ^{\{ 2,4\} } |}  \\
   {\text{sgn}(\zeta _ +  ^{\{ 1,3\} } )|\zeta _ +  ^{\{ 1,3\} } |}  \\
   {\text{sgn}(\zeta _ -  ^{\{ 2,4\} } )|\zeta _ -  ^{\{ 2,4\} } |}  \\
\end{array}} \right)\!.
\end{Equation}
Then, we want to convert these factors to unit-complex exponentials, but that means we have to sacrifice the $0$ case. But that is okay since the magnitude is $0$ anyway in that case.  So we rewrite $d$ as
\begin{Equation}                      {120}
d = \text{diag}\left( {\begin{array}{*{20}c}
   {|\zeta _ -  ^{\{ 1,3\} } |e^{i\frac{{\pi [1 - \text{sgn}_2 (\zeta _ -  ^{\{ 1,3\} } )]}}{2}} }  \\
   {|\zeta _ +  ^{\{ 2,4\} } |e^{i\frac{{\pi [1 - \text{sgn}_2 (\zeta _ +  ^{\{ 2,4\} } )]}}{2}} }  \\
   {|\zeta _ +  ^{\{ 1,3\} } |e^{i\frac{{\pi [1 - \text{sgn}_2 (\zeta _ +  ^{\{ 1,3\} } )]}}{2}} }  \\
   {|\zeta _ -  ^{\{ 2,4\} } |e^{i\frac{{\pi [1 - \text{sgn}_2 (\zeta _ -  ^{\{ 2,4\} } )]}}{2}} }  \\
\end{array}} \right)\!,
\end{Equation}
where we use the two-case sign function
\begin{Equation}                      {121}
\text{sgn}_2 (a) \equiv \left\{ {\begin{array}{*{20}l}
   { + 1;} & {a \ge 0}  \\
   { - 1;} & {a < 0,}  \\
\end{array}} \right.
\end{Equation}
to keep the sign factors real in the $0$ case (even though the magnitudes are $0$ anyway then, just to ensure that near-$0$ eigenvalues stay as true to the actual values of $d$ as possible).  Then, to prepare to peel these factors off on either side, we rewrite \Eq{120} as
\begin{Equation}                      {122}
d = \text{diag}\left( {\begin{array}{*{20}c}
   {e^{i\frac{{\pi [1 - \text{sgn}_2 (\zeta _ -  ^{\{ 1,3\} } )]}}{4}} |\zeta _ -  ^{\{ 1,3\} } |e^{i\frac{{\pi [1 - \text{sgn}_2 (\zeta _ -  ^{\{ 1,3\} } )]}}{4}} }  \\
   {e^{i\frac{{\pi [1 - \text{sgn}_2 (\zeta _ +  ^{\{ 2,4\} } )]}}{4}} |\zeta _ +  ^{\{ 2,4\} } |e^{i\frac{{\pi [1 - \text{sgn}_2 (\zeta _ +  ^{\{ 2,4\} } )]}}{4}} }  \\
   {e^{i\frac{{\pi [1 - \text{sgn}_2 (\zeta _ +  ^{\{ 1,3\} } )]}}{4}} |\zeta _ +  ^{\{ 1,3\} } |e^{i\frac{{\pi [1 - \text{sgn}_2 (\zeta _ +  ^{\{ 1,3\} } )]}}{4}} }  \\
   {e^{i\frac{{\pi [1 - \text{sgn}_2 (\zeta _ -  ^{\{ 2,4\} } )]}}{4}} |\zeta _ -  ^{\{ 2,4\} } |e^{i\frac{{\pi [1 - \text{sgn}_2 (\zeta _ -  ^{\{ 2,4\} } )]}}{4}} }  \\
\end{array}} \right)\!.
\end{Equation}

The \textit{magnitudes} of the eigenvalues of $\tau$ are the Takagi values (up to order), and we can peel off the phase factors and lump them with the eigenvector matrix $V$, after first rewiting the adjoint as transpose (which is possible because $V$ is real), which gives the \textit{almost}-Takagi factorization,
\begin{Equation}                      {123}
\tau  = VdV^\dag   = VdV^T  = U'D'U'^T ,
\end{Equation}
where the almost-Takagi unitary is
\begin{Equation}                      {124}
U' \equiv V\text{diag}\left( {\begin{array}{*{20}c}
   {e^{i\frac{{\pi [1 - \text{sgn}_2 (\zeta _ -  ^{\{ 1,3\} } )]}}{4}} }  \\
   {e^{i\frac{{\pi [1 - \text{sgn}_2 (\zeta _ +  ^{\{ 2,4\} } )]}}{4}} }  \\
   {e^{i\frac{{\pi [1 - \text{sgn}_2 (\zeta _ +  ^{\{ 1,3\} } )]}}{4}} }  \\
   {e^{i\frac{{\pi [1 - \text{sgn}_2 (\zeta _ -  ^{\{ 2,4\} } )]}}{4}} }  \\
\end{array}} \right)\!,
\end{Equation}
and the almost-Takagi-value matrix is
\begin{Equation}                      {125}
D' \equiv \text{diag}\{ |\zeta _ -  ^{\{ 1,3\} } |,|\zeta _ +  ^{\{ 2,4\} } |,|\zeta _ +  ^{\{ 1,3\} } |,|\zeta _ -  ^{\{ 2,4\} } |\} .
\end{Equation}

Now, to get a useful Takagi factorization, we need to determine the descending order of the values in \Eq{125}, and apply the corresponding permutation to $D'$ and $U'$.  

Starting with subspace $\{1,3\}$, although \smash{$\zeta _ +  ^{\{ 1,3\} }\geq \zeta _ -  ^{\{ 1,3\} }$}, since we want to compare \textit{magnitudes}, line 3 of \Eq{111} is most useful since it has common terms inside and outside the larger radical. So the magnitudes are
\begin{Equation}                      {126}
\begin{array}{*{20}l}
   {|\zeta _ \pm  ^{\{ 1,3\} } |=} &\!\! {\frac{1}{2\Delta}\left( {\rule{0pt}{10pt}} \right.\!2(\lambda _1  - \lambda _5 )^2 [\Delta ^2  - \Omega ] + 4\lambda _1 \lambda _5 \Delta ^2 }  \\
   {} &\!\! {\mp 2[(\lambda _1  - \lambda _5 )\sqrt {\Delta ^2  - \Omega } ]}  \\
   {} &\!\! {\times \sqrt {4\lambda _1 \lambda _5 \Delta ^2  + [(\lambda _1  - \lambda _5 )\sqrt {\Delta ^2  - \Omega } ]^2 } \left. {\rule{0pt}{10pt}} \right)^{1/2}\!,}  \\
\end{array}
\end{Equation}
comparison of which shows that
\begin{Equation}                      {127}
 |\zeta _ -  ^{\{ 1,3\} } |\geq |\zeta _ +  ^{\{ 1,3\} } |,
\end{Equation}
in all cases.  Then, since \smash{$|\zeta _ +  ^{\{ 2,4\} } |=|\zeta _ -  ^{\{ 2,4\} } |$}, we have the proper ordering in each subspace separately, so we just need to determine the values of these two sets relative to each other.  For this, we will break the problem $Q$ cases, and also treat any subcases that arise as needed.\\

{\noindent}\textbf{Case 1:} $Q\geq 0$:  Here, by \Eq{15}, $\Omega =Q$ where $Q\equiv (\lambda _1  - \lambda _5 )^2  - (E + 2\sqrt {\lambda _4 \lambda _6 } )^2$, so \Eq{126} becomes
\begin{Equation}                      {128}
\begin{array}{*{20}l}
   {|\zeta _ \pm  ^{\{ 1,3\} } | =} &\!\! {\frac{1}{2\Delta}\left( {\rule{0pt}{10pt}} \right.\!2(\lambda _1  - \lambda _5 )^2 [\Delta ^2  - Q ] + 4\lambda _1 \lambda _5 \Delta ^2 }  \\
   {} &\!\! {\mp 2[(\lambda _1  - \lambda _5 )\sqrt {\Delta ^2  - Q } ]}  \\
   {} &\!\! {\times \sqrt {4\lambda _1 \lambda _5 \Delta ^2  + [(\lambda _1  - \lambda _5 )\sqrt {\Delta ^2  - Q } ]^2 } \left. {\rule{0pt}{10pt}} \right)^{1/2}\!.}  \\
\end{array}
\end{Equation}
Then, since $\Delta\equiv \lambda _1  - \lambda _5  + \delta _{\lambda _1 ,\lambda _5 }$ [where the Kronecker delta comes from the spectral decomposition of \Eq{8}], there are two \textit{subcases}; $\lambda _1  = \lambda _5$ and $\lambda _1  > \lambda _5$. 

In general, regardless of $Q$ case, given eigenvalue constraints of normalization, nonnegativity, and descending order, subcases with $\lambda _1  = \lambda _5$ only admit spectra described by $\lambda _1  =  \cdots  = \lambda _5  = \frac{{1 - \lambda _6 }}{5}$ with $\lambda _6  \in [0,\frac{1}{6}]$.  Then, because $\lambda _1  = \lambda _5$ causes $Q= - (E + 2\sqrt {\lambda _4 \lambda _6 } )^2$, this will split this set of spectra into two subsets based on which $Q$ case pertains.

Since $Q=0$ belongs to the $Q\geq 0$ case, then in the $\lambda _1  = \lambda _5$ subcase, the fact that $Q= - (E + 2\sqrt {\lambda _4 \lambda _6 } )^2$ such that $Q=0$ requires that both terms in the square be $0$ since both are always nonnegative, so we must have $E=0$ and $\lambda _4 \lambda _6 =0$. But due to descending order, $\lambda _1  = \lambda _5$ implies that $\lambda _4 \neq 0$, so $\lambda _6 =0$ is the only sub-subcase that applies when $Q\geq 0$ and $\lambda _1  = \lambda _5$, so we must have $\lambda _1 =\cdots = \lambda _5 =\frac{1}{5}$ with $\lambda _6 =0$ here.

Looking ahead, for the same reason, whenever $Q<0$ and $\lambda _1  = \lambda _5$, the fact that $Q<0$ implies $E=0$ as proven in \Eq{59} means that $Q= - (E + 2\sqrt {\lambda _4 \lambda _6 } )^2 = -4\lambda _4 \lambda _6$, which can only be negative when $\lambda _6 >0$ (since $\lambda _4 \neq 0$ by $\lambda _1  = \lambda _5$ and descending order), so the only spectra that apply when $Q<0$ and $\lambda _1  = \lambda _5$ are $\lambda _1  =  \cdots  = \lambda _5  = \frac{{1 - \lambda _6 }}{5}$ with $\lambda _6  \in (0,\frac{1}{6}]$, which we save for our treatment of the $Q<0$ case. (As we will see, we will not need this, but it was important to rule it out for the present $Q$ case). 

So the only spectrum in the $\lambda _1  = \lambda _5$ subcase of $Q\geq 0$ is $\lambda _1 =\cdots = \lambda _5 =\frac{1}{5}$ with $\lambda _6 =0$, which yields, by \Eq{128}, 
\begin{Equation}                      {129}
{\textstyle |\zeta _ \pm  ^{\{ 1,3\} } | = \sqrt {\lambda _1 \lambda _5 }  = \frac{1}{5},\;\;|\zeta _ \pm  ^{\{ 2,4\} } | = \sqrt {\lambda _4 \lambda _6 }  = 0},
\end{Equation}
where we used $\lambda _1  = \lambda _5 $, $Q = 0$, $\Delta  = 1$, so this yields a descending ordering as [also using \Eq{127}],
\begin{Equation}                      {130}
\begin{array}{*{20}l}
   {Q = 0;\;\;\lambda _1  = \lambda _5 :}  \\
   {\;\;\{ |\zeta _ -  ^{\{ 1,3\} } |\geq |\zeta _ +  ^{\{ 1,3\} } |\geq|\zeta _ +  ^{\{ 2,4\} } |\geq|\zeta _ -  ^{\{ 2,4\} } |\}  = \{ \frac{1}{5},\frac{1}{5},0,0\}. }  \\
\end{array}
\end{Equation}

Then, still in the $Q\geq 0$ case, its other subcase is $\lambda _1  > \lambda _5$, for which $\Delta  = \lambda _1  - \lambda _5 $ and $Q = \Delta ^2  - (E + 2\sqrt {\lambda _4 \lambda _6 } )^2 $, so then \Eq{128} becomes
\begin{Equation}                      {131}
\begin{array}{*{20}l}
   {|\zeta _ \pm  ^{\{ 1,3\} } | =} &\!\! {\frac{1}{2}\left( {\rule{0pt}{10pt}} \right.\!2(E + 2\sqrt {\lambda _4 \lambda _6 } )^2  + 4\lambda _1 \lambda _5 }  \\
   {} &\!\! {\mp 2[(E + 2\sqrt {\lambda _4 \lambda _6 } )]}  \\
   {} &\!\! {\times \sqrt {4\lambda _1 \lambda _5  + (E + 2\sqrt {\lambda _4 \lambda _6 } )^2 } \left. {\rule{0pt}{10pt}} \right)^{1/2}\!.}  \\
\end{array}
\end{Equation}
\vspace{-10pt}\\
Here, it is difficult to get a full ordering of all the Takagi values.  However, since the preconcurrence's form only requires that we know which is the \textit{largest} Takagi value, this simplifies our task.  With this in mind, if we focus on $|\zeta _ -  ^{\{ 1,3\} } |$ from \Eq{131}, comparing it to $|\zeta _ \pm  ^{\{ 2,4\} } |$ as
\begin{Equation}                      {132}
|\zeta _ -  ^{\{ 1,3\} } |\sim |\zeta _ \pm  ^{\{ 2,4\} } |,
\end{Equation}
where we use ``$\sim$'' as an undetermined inequality placeholder, then plugging in \Eq{131} and \Eq{114} into \Eq{132} (being mindful of the absolute values), we get
\begin{Equation}                      {133}
\begin{array}{*{20}l}
   {\left( {\rule{0pt}{10pt}} \right.\!E^2  + 4E\sqrt {\lambda _4 \lambda _6 }  + (E + 2\sqrt {\lambda _4 \lambda _6 } )^2 } &\!\! {}  \\
   { + 4\lambda _1 \lambda _5  + 2[(E + 2\sqrt {\lambda _4 \lambda _6 } )]} & {}  \\
   { \times \sqrt {4\lambda _1 \lambda _5  + (E + 2\sqrt {\lambda _4 \lambda _6 } )^2 } \left. {\rule{0pt}{10pt}} \right)} &\!\! {\sim 0,}  \\
\end{array}
\end{Equation}
\vspace{-10pt}\\
which shows that $\sim\to\geq$ since every term on the left of \Eq{133} is nonnegative, so then \Eq{132} becomes
\begin{Equation}                      {134}
|\zeta _ -  ^{\{ 1,3\} } |\geq |\zeta _ \pm  ^{\{ 2,4\} } |,
\end{Equation}
which, together with \Eq{127} gives us a useful partial ordering for this subcase as
\begin{Equation}                      {135}
\begin{array}{*{20}l}
   {Q \geq 0;\;\;\lambda _1  > \lambda _5 :}  \\
   {\;\;\; |\zeta _ -  ^{\{ 1,3\} } |\geq\{|\zeta _ +  ^{\{ 1,3\} } |,|\zeta _ +  ^{\{ 2,4\} } |,|\zeta _ -  ^{\{ 2,4\} } |\}, }  \\
\end{array}
\end{Equation}
where the order of the set on the right is not yet specified, but we definitely know that \smash{$|\zeta _ -  ^{\{ 1,3\} } |$} is the largest, which is sufficient for our purposes.  Therefore, since the set in \Eq{130} can also be written like the set in \Eq{135}, then both subcases of $Q\geq 0$ can be united as
\begin{Equation}                      {136}
Q \geq 0:\;\;\; |\zeta _ -  ^{\{ 1,3\} } |\geq\{|\zeta _ +  ^{\{ 1,3\} } |,|\zeta _ +  ^{\{ 2,4\} } |,|\zeta _ -  ^{\{ 2,4\} } |\}.
\end{Equation}

{\noindent}\textbf{Case 2:} $Q< 0$:  Here, $\Omega =0$ by \Eq{15}, and $E=0$ by \Eq{59}, so also $Q< 0$ means $(\lambda _1  - \lambda _5 )^2  - (0 + 2\sqrt {\lambda _4 \lambda _6 } )^2 <0$ and therefore $\lambda _1  - \lambda _5 -2\sqrt {\lambda _4 \lambda _6 } <0 $.  So \Eq{126} becomes
\begin{Equation}                      {137}
\begin{array}{*{20}l}
   {|\zeta _ \pm  ^{\{ 1,3\} } |} &\!\! {=\sqrt {\frac{{\lambda _1 ^2  + \lambda _5 ^2  \mp (\lambda _1 ^2  - \lambda _5 ^2 )}}{2}} .}  \\
\end{array}
\end{Equation}
Here, since there is no remaining $\Delta$, we do not need to break this into subcases as we did for the first $Q$ case.  

Furthermore, since we have already settled for simply finding the \textit{largest} Takagi value, then based on \Eq{127}, we note that \smash{$|\zeta _ -  ^{\{ 1,3\} } |$} from \Eq{137} is
\begin{Equation}                      {138}
|\zeta _{-}  ^{\{ 1,3\} } |=\lambda _1,
\end{Equation}
and comparing it to \smash{$|\zeta _ \pm  ^{\{ 2,4\} } |$} gives 
\begin{Equation}                      {139}
\begin{array}{*{20}r}
   {|\zeta _ -  ^{\{ 1,3\} } |\sim} &\!\! {|\zeta _ \pm  ^{\{ 2,4\} } |\phantom{2}}  \\
   {\lambda _1 \sim} &\!\! {\sqrt {\lambda _4 \lambda _6 }\phantom{2} ,}  \\
\end{array}
\end{Equation}
and since $\lambda _1 \geq \lambda _4 \geq \sqrt {\lambda _4 \lambda _6 }  $, \Eq{139} shows that $\sim \to  \geq $, so
\begin{Equation}                      {140}
|\zeta _ -  ^{\{ 1,3\} } | \geq |\zeta _ \pm  ^{\{ 2,4\} } |,
\end{Equation}
which gives the result for this $Q$ case that
\begin{Equation}                      {141}
Q < 0:\;\;\; |\zeta _ -  ^{\{ 1,3\} } |\geq\{|\zeta _ +  ^{\{ 1,3\} } |,|\zeta _ +  ^{\{ 2,4\} } |,|\zeta _ -  ^{\{ 2,4\} } |\},
\end{Equation}
where again, the set on the right is not yet ordered.

Then, since \Eq{136} and \Eq{141} have the same form, we get a unified partial ordering for \textit{all} cases as
\begin{Equation}                      {142}
\forall Q:\;\;\; |\zeta _ -  ^{\{ 1,3\} } |\geq\{|\zeta _ +  ^{\{ 1,3\} } |,|\zeta _ +  ^{\{ 2,4\} } |,|\zeta _ -  ^{\{ 2,4\} } |\},
\end{Equation}
where the order in the set on the right does not matter. (Note that this order can always be obtained numerically for a given set of parameters, but our goal here is to find a \textit{symbolic} solution with little or no case-splitting.) 

Therefore, we can now assign the concurrence singular values to the appropriate Takagi values (where the order of the last three is not necessarily descending, but the first is always the largest) as
\begin{Equation}                      {143}
\xi _1  \equiv |\zeta _ -  ^{\{ 1,3\} } |,\;\;\xi _2  \equiv |\zeta _ +  ^{\{ 1,3\} } |,\;\;\xi _3  \equiv |\zeta _ +  ^{\{ 2,4\} } |,\;\;\xi _4  \equiv |\zeta _ -  ^{\{ 2,4\} } |.
\end{Equation}
Given \Eq{143}, using \Eq{114} and \Eq{126} then yeilds the specific $\{\xi_a\}$ forms in \Eq{14}, which, when compared with \Eq{114} and \Eq{111}, shows that
\begin{Equation}                      {144}
\begin{array}{*{20}l}
   {\zeta _ -  ^{\{ 1,3\} } } &\!\! { =  - \xi _1 }  \\
   {\zeta _ +  ^{\{ 1,3\} } } &\!\! { =  + \xi _2 }  \\
   {\zeta _ +  ^{\{ 2,4\} } } &\!\! { =  + \xi _3 }  \\
   {\zeta _ -  ^{\{ 2,4\} } } &\!\! { =  - \xi _4 .}  \\
\end{array}
\end{Equation}

Then, \Eq{143} gives us a useful ordering for the Takagi decomposition of $\tau$, which we achieve by swapping columns $2$ and $3$ in $U'$ of \Eq{124} and swapping diagonals $2$ and $3$ in $D'$ of \Eq{125}, which yields
\begin{Equation}                      {145}
\tau  = UDU^T ,
\end{Equation}
with Takagi unitary,
\begin{Equation}                      {146}
U \equiv \left(\!\! {\begin{array}{*{20}c}
   {\frac{{(\frac{{\sqrt {\lambda _1 \lambda _5 \Omega } }}{\Delta } + \delta _{\lambda _5 \Omega ,0} )i}}{{N_ 1  }}} & {\frac{{\xi _2  - \frac{{\lambda _5 \sqrt {\Delta ^2  - \Omega } }}{\Delta }}}{{N_ 2  }}} &  \cdot  &  \cdot   \\
    \cdot  &  \cdot  & {\frac{1}{{\sqrt 2 }}} & {\frac{i}{{\sqrt 2 }}}  \\
   {\frac{{ - (\xi _1  - \frac{{\lambda _1 \sqrt {\Delta ^2  - \Omega } }}{\Delta })i}}{{N_ 1  }}} & {\frac{{\frac{{\sqrt {\lambda _1 \lambda _5 \Omega } }}{\Delta } + \delta _{\lambda _5 \Omega ,0} }}{{N_ 2  }}} &  \cdot  &  \cdot   \\
    \cdot  &  \cdot  & {\frac{1}{{\sqrt 2 }}} & {\frac{{ - i}}{{\sqrt 2 }}}  \\
\end{array}}\!\! \right)\!,
\end{Equation}
where, from \Eq{113} and \Eq{144},
\begin{Equation}                      {147}
\begin{array}{*{20}l}
   {N_1  \equiv } &\!\! {N_ -   \equiv \sqrt {(\frac{{\xi _1 \Delta  - \lambda _1 \sqrt {\Delta ^2  - \Omega } }}{\Delta })^2  \!+\! (\frac{{\sqrt {\lambda _1 \lambda _5 \Omega }  + \delta _{\lambda _5 \Omega ,0} \Delta }}{\Delta })^2 } ,}  \\
   {N_2  \equiv } &\!\! {N_ +   \equiv \sqrt {(\frac{{\xi _2 \Delta  - \lambda _5 \sqrt {\Delta ^2  - \Omega } }}{\Delta })^2  \!+\! (\frac{{\sqrt {\lambda _1 \lambda _5 \Omega }  + \delta _{\lambda _5 \Omega ,0} \Delta }}{\Delta })^2 } ,}  \\
\end{array}
\end{Equation}
and we used \Eq{144} in \Eq{124} to get
\begin{Equation}                      {148}
\begin{array}{*{20}l}
   {e^{i\frac{{\pi [1 - {\mathop{\rm sgn}} _2 (\zeta _ -  ^{\{ 1,3\} } )]}}{4}} } &\!\! { = e^{i\frac{{\pi [1 + {\mathop{\rm sgn}} _2 (\xi _1 )]}}{4}} } &\!\! { = i}  \\
   {e^{i\frac{{\pi [1 - {\mathop{\rm sgn}} _2 (\zeta _ +  ^{\{ 1,3\} } )]}}{4}} } &\!\! { = e^{i\frac{{\pi [1 - {\mathop{\rm sgn}} _2 (\xi _2 )]}}{4}} } &\!\! { = 1}  \\
   {e^{i\frac{{\pi [1 - {\mathop{\rm sgn}} _2 (\zeta _ +  ^{\{ 2,4\} } )]}}{4}} } &\!\! { = e^{i\frac{{\pi [1 - {\mathop{\rm sgn}} _2 (\xi _3 )]}}{4}} } &\!\! { = 1}  \\
   {e^{i\frac{{\pi [1 - {\mathop{\rm sgn}} _2 (\zeta _ -  ^{\{ 2,4\} } )]}}{4}} } &\!\! { = e^{i\frac{{\pi [1 + {\mathop{\rm sgn}} _2 (\xi _4 )]}}{4}} } &\!\! { = i,}  \\
\end{array}
\end{Equation}
and the Takagi-value matrix is then
\begin{Equation}                      {149}
D \equiv \text{diag}\{ \xi _1 ,\xi _2 ,\xi _3 ,\xi _4 \} ,
\end{Equation}
where these values can be expressed as seen in \Eq{14}.

Here we pause to prove that these $\{\xi_a\}$ are valid in all cases.  First, putting the $\{\xi_a\}$ from \Eq{14} into \Eq{11} gives
\begin{Equation}                      {150}
\begin{array}{*{20}l}
   E &\!\! { = \max \{ 0,\xi _1  - \xi _2  - \xi _3  - \xi _4 \} }  \\
   {} &\!\! { = \max \{ 0,\frac{{(\lambda _1  - \lambda _5 )\sqrt {\Delta ^2  - \Omega } }}{\Delta } - 2\sqrt {\lambda _4 \lambda _6 } \} ,}  \\
\end{array}
\end{Equation}
which gives us a general formula to check $\{\xi_a\}$.

In the $Q \ge 0$ case when $\lambda _1  = \lambda _5 $, then $\Omega  = Q$, $\Delta  = 1$, and $Q =  - (E + 2\sqrt {\lambda _4 \lambda _6 } )^2  = 0$ since the only spectrum that applies to this subcase is $\lambda _1  =  \cdots  = \lambda _5 $ with $\lambda _6  = 0$, and $E = 0$ as explained in the second new paragraph after \Eq{128}, so then \Eq{150} becomes
\begin{Equation}                      {151}
\begin{array}{*{20}l}
   E &\!\! { = \max \{ 0,\frac{{0\sqrt {1^2  - 0} }}{1} - 2\sqrt {\lambda _4 0} \} }  \\
   {} &\!\! { = 0}  \\
   {} &\!\! { = E.}  \\
\end{array}
\end{Equation}

In the $Q \ge 0$ case when $\lambda _1  > \lambda _5 $, then $\Omega  = Q$ and $\Delta  \!=\! \lambda _1  \!-\! \lambda _5 $, so $Q \!=\! \Delta ^2  \!-\! (E \!+\! 2\sqrt {\lambda _4 \lambda _6 } )^2 $ and \Eq{150} gives
\begin{Equation}                      {152}
\begin{array}{*{20}l}
   E &\!\! { = \max \{ 0,\frac{{(\lambda _1  - \lambda _5 )(E + 2\sqrt {\lambda _4 \lambda _6 } )}}{\Delta } - 2\sqrt {\lambda _4 \lambda _6 } \} }  \\[0.5ex]
   {} &\!\! { = \max \{ 0,E + 2\sqrt {\lambda _4 \lambda _6 }  - 2\sqrt {\lambda _4 \lambda _6 } \} }  \\
   {} &\!\! { = \max \{ 0,E\} }  \\
   {} &\!\! { = E.}  \\
\end{array}
\end{Equation}

In the $Q < 0$ case, $\Omega  = 0$ while $Q < 0$ implies $E = 0$ as proven in \Eq{59}, and $Q < 0$ also implies that $\lambda _1  - \lambda _5  - 2\sqrt {\lambda _4 \lambda _6 }  < 0$ as proved in \Eq{56}, so then \Eq{150} becomes
\begin{Equation}                      {153}
\begin{array}{*{20}l}
   E &\!\! { = \max \{ 0,\lambda _1  - \lambda _5  - 2\sqrt {\lambda _4 \lambda _6 } \} }  \\
   {} &\!\! { = 0}  \\
   {} &\!\! { = E.}  \\
\end{array}
\end{Equation}
Thus, we have proven that the Takagi values $\{\xi_a\}$ from \Eq{14} yield the correct entanglement $E$ for all cases.

Now we have all we need to build the Wootters x kets $\{|x_a\rangle\}$ for this system.  Here, since we do not have a definite symbolic order for the lower three Takagi values, we must use \textit{all four} $|x_a\rangle$ states to ensure that no states are missed, and their own parameters will then take care of causing the necessary zero vectors when appropriate.

Therefore, putting \Eq{146} into \Eq{80} gives
\begin{Equation}                      {154}
\begin{array}{*{20}l}
   {|x_1 \rangle } &\!\! { = \frac{i}{{N_1 }}(\frac{{\sqrt {\lambda _1 \lambda _5 \Omega }  + \delta _{\lambda _5 \Omega ,0} \Delta }}{\Delta }|u_1 \rangle  - \frac{{\xi _1 \Delta  - \lambda _1 \sqrt {\Delta ^2  - \Omega } }}{\Delta }|u_3 \rangle )}  \\
   {|x_2 \rangle } &\!\! { = \frac{1}{{N_2 }}(\frac{{\xi _2 \Delta  - \lambda _5 \sqrt {\Delta ^2  - \Omega } }}{\Delta }|u_1 \rangle  + \frac{{\sqrt {\lambda _1 \lambda _5 \Omega }  + \delta _{\lambda _5 \Omega ,0} \Delta }}{\Delta }|u_3 \rangle )}  \\
   {|x_3 \rangle } &\!\! { = \frac{1}{{\sqrt 2 }}(|u_2 \rangle  + |u_4 \rangle )}  \\
   {|x_4 \rangle } &\!\! { = \frac{i}{{\sqrt 2 }}(|u_2 \rangle  - |u_4 \rangle ),}  \\
\end{array}
\end{Equation}
and then putting \Eq{106} into \Eq{154} gives \Eq{13}, which is the explicit form we sought.

We have now proven our explicit symbolic form of the LS decomposition of the EPU-minimal TGX states of \Eq{8}.  This decomposition, summarized in \Eqs{9}{15}, was also numerically tested on a large number of states and spectra covering all special cases, and found to always give the exact correct entanglement $E$ by both methods shown in \Eq{11}, as well as always being yielding a valid decomposition of the state.  Furthermore, the separable state $\rho_S$ of \Eq{12} constructed from these $\{|x_a\rangle\}$ was also found to always be separable, where we used the minimal TGX formula of \Eq{7} (proved in \Sec{III.A}) to verify this, and the proof of \Sec{III.D.1} also guarantees it.

Note that the results in \Sec{III.D.1} can be done numerically without using \Sec{III.D.2} at all, but the results of \Sec{III.D.2} give us a simple set of parameterized decomposition states allowing symbolic manipulation, which is a valuable tool for proving new results.
\section{\label{sec:IV}Conclusions}
This paper has successfully proven that TGX states are EPU-equivalent to the set of all states in $2\times 3$ systems, where the entanglement is measured by I-concurrence.  Thus, for every general state, there is a TGX state with the same spectrum and I-concurrence.

Specifically, only a very compact subspace of TGX states called \textit{EPU-minimal TGX states} \protect\raisebox{1.0pt}{\smash{$\rhoEPUminTGX$}} are needed for this EPU equivalence, as seen \Eq{8}.  This family is a subset of the somewhat larger family called \textit{minimal TGX states} \smash{$\rhoMinTGX$}, all varieties of which are shown in \Eq{16}.  The general set of TGX states is shown in \Eq{17}.  Each of these sets have their form and entanglement preserved under local-permutation unitary (LPU) operations.

Although a computable formula for I-concurrence of general $2\times 3$ mixed states is not yet known, we have proven that all possible spectrum-entanglement combinations \textit{are} represented in \protect\raisebox{1.0pt}{\smash{$\rhoEPUminTGX$}}.  Therefore if a computable I-concurrence formula for $E$ is ever discovered, then any general state $\rho$ can be converted to an EPU-minimal TGX state \smash{$\rhoEPUminTGX$} simply by harvesting the $E$ and spectrum of $\rho$ to construct \smash{$\rhoEPUminTGX$}, and the unitary relating them will be given by \Eq{18}.

Nevertheless, \protect\raisebox{1.0pt}{\smash{$\rhoEPUminTGX$}} of \Eq{8} can still be used to \textit{parameterize} physical states of all spectrum-entanglement combinations simply by picking any unit-normalized spectrum $\lambda_1 \geq\cdots\geq\lambda_6 \geq 0$ and using \Eq{67} to get any physical I-concurrence $E_\eta\equiv\eta\max \{ 0,\lambda _1  - \lambda _5  - 2\sqrt {\lambda _4 \lambda _6 } \};\;\;\eta\in[0,1]$. By using LU operators, this also allows us to create a very wide range of generally \textit{non}TGX states parameterized by both spectrum and entanglement.

As part of proving the EPU equivalence of \protect\raisebox{1.0pt}{\smash{$\rhoEPUminTGX$}}, we developed an explicit I-concurrence formula in \Eq{7} for all minimal TGX states \smash{$\rhoMinTGX$} [examples of which are in \Eq{16}].  The simple form of \Eq{7} comes from utilizing spacewise orthogonality, LS decomposition, and the fact that quartet subspaces of TGX states always have X form, all of which conspires to let the $2$-norm of the I-concurrence simplify to a single subspace concurrence, as explained in \Sec{III.A}. Note that these LS decompositions are optimal because all quartets outside of the entanglement-containing quartet always have concurrence $0$ since those subspaces are always separable for minimal SGX states and their subsets, such as minimal TGX states and EPU-minimal TGX states.

We also derived the computable I-concurrence formula in \Eq{21} for the more general family of \textit{minimal SGX states} \smash{$\rhoMinSGX$} from \Eq{19}. The formula in \Eq{21} has a relationship to the formula in \Eq{7} that is analogous to that between the $2\times 2$ concurrence $C(\rho)$ of \Eq{1} and the X-concurrence $C(\rho_{\X})$ of \Eq{3}; the subspace concurrences in \Eq{21} generally need to be computed numerically, but they all simplify to an explicit symbolic form for minimal TGX states, as seen in \Eq{7}.  Therefore, since EPU-minimal TGX states \protect\raisebox{1.0pt}{\smash{$\rhoEPUminTGX$}} are a subset of both minimal TGX states and minimal SGX states, this proves that minimal TGX states and minimal SGX states both have EPU equivalence to general states as well.  However, since our goal is to find \textit{the most compact set} that achieves EPU equivalence, the main result here is the EPU equivalence of EPU-minimal TGX states (which are the most compact set since superposition is necessary for entanglement and they only have \textit{one} unique nonzero off-diagonal).

Furthermore, we used the simplicity of our EPU-minimal TGX family of \Eq{8} to derive its \textit{explicit} LS decomposition in \Sec{III.D}, summarized in \Eqs{9}{15}. However, \Eqs{98}{104} can also be used to get LS decompositions numerically for these states, as well as the more general families of minimal TGX states and minimal SGX states with the adjustment mentioned at the end of \Sec{III.D.1}.

All of this generalizes the $2\times 2$ results of \cite{HeXU} to $2\times 3$ (while adding the LS decomposition and explaining how to apply that to $2\times 2$).  Furthermore, we showed that the prediction of \cite{HedX} regarding entanglement universality of TGX states is correct in $2\times 3$; in general, literal X states are not always sufficient to achieve EPU equivalence, since there are X states which are permanently separable in this system, as explained in \hyperlink{ExtraSummaryResults:5}{Item 5} of \Sec{II}. (Indeed, entanglement universality of TGX states was already proven to exclude X states in general, both in \cite{HedE} and \cite{HMME} which discuss example systems such as $3\times 3$ where X states cannot achieve maximal entanglement in the pure case, but TGX states can.) However, in $2\times 3$, \textit{some} X states \textit{can} achieve EPU equivalence, such as the particular family shown in \Eq{8}. But the preservation of its superposition and entanglement properties under LPU operations that cause it wander through TGX space show that TGX form is the more relevant feature for EPU equivalence (particularly since TGX states are known to be necessary for EPU equivalence in larger systems such as $3\times 3$ as mentioned above).

In closing, we have discovered a powerful new family of states that greatly extends our ability to model and explore entanglement in $2\times 3$ systems.  We have also found a satisfying generalization of the X-concurrence formula of $2\times 2$ to the minimal-TGX I-concurrence formula in $2\times 3$, and a further generalization of that to a computable I-concurrence formula for the more general minimal SGX states.  Perhaps most importantly, the methods here have already helped guide similar discoveries in a very wide range of  multipartite quantum systems, even though those involve many other issues that do not arise in $2\times 3$, such as bound entanglement.  At the time of writing, we have prepared another paper extending these ideas to such systems, and will release those results soon.
\vspace{2pt}
\begin{appendix}
\section{\label{sec:App.A}Review of I-Concurrence}
The I-concurrence \cite{AuVD,RBCH,Woo2,AlFe,ZZFL} of bipartite pure states is
\begin{Equation}                      {A.1}
E(\rho _{|\psi \rangle } ) \equiv \sqrt {2[1 - P(\redx{\rho}{m}_{|\psi \rangle})]} ,
\end{Equation}
where \smash{$\rho _{|\psi \rangle }  \equiv |\psi \rangle \langle \psi |$}, \smash{$\redx{\rho}{m}$} is the mode-$m$ reduction of $\rho$, $m\in \{1,2\}$ is a mode label, and \smash{$P(\rho ) \equiv \tr(\rho ^2 )$} is \textit{purity}.

Expressing $|\psi \rangle$ in the coincidence basis as
\begin{Equation}                      {A.2}
|\psi \rangle  \equiv \sum\nolimits_{b_1 ,b_2  = 1,1}^{n_1 ,n_2 } {a_{b_1 ,b_2 } |b_1 ,b_2 \rangle } ,
\end{Equation}
where $a_{b_1 ,b_2 }  \equiv \langle b_1 ,b_2 |\psi \rangle$ causes \Eq{A.1} to expand as
\begin{Equation}                      {A.3}
\begin{array}{*{20}l}
   {E(\rho _{|\psi \rangle } )} &\!\! { = 2\sqrt {\sum\limits_{w < y}^{n_1 } {\sum\limits_{x < z}^{n_2 } {|a_{w,x} a_{y,z} \! -\! a_{w,z} a_{y,x} |^2 } } } }  \\
   {} &\!\! { = 2\sqrt {\sum\limits_{w = 1}^{n_1 } {\sum\limits_{y = w + 1}^{n_1 } {\sum\limits_{x = 1}^{n_2 } {\sum\limits_{z = x + 1}^{n_2 } {\!\! |a_{w,x} a_{y,z}  \!-\! a_{w,z} a_{y,x} |^2 } } } } } .}  \\
\end{array}
\end{Equation}

In $2\times 3$, $n_1  = 2$ and $n_2  = 3$ so the only pair of $\{w,y\}$ in the sums is $\{1,2\}$, while $\{x,z\}\in \{\{1,2\},\{1,3\},\{2,3\}\}$. Thus, the index combinations in terms of the sums are
\begin{Equation}                      {A.4}
\begin{array}{*{20}c}
   {\begin{array}{*{20}c}
   w & y &\vline &  x & z  \\
\hline
   1 & 2 &\vline &  1 & 2  \\
   1 & 2 &\vline &  1 & 3  \\
   1 & 2 &\vline &  2 & 3  \\
\end{array}} &  \to  & {\begin{array}{*{20}c}
   w &\vline &  x &\vline &  y &\vline &  z  \\
\hline
   1 &\vline &  1 &\vline &  2 &\vline &  2  \\
   1 &\vline &  1 &\vline &  2 &\vline &  3  \\
   1 &\vline &  2 &\vline &  2 &\vline &  3  \\
\end{array}}  \\
\end{array},
\end{Equation}
which shows that \Eq{A.3} becomes
\begin{Equation}                      {A.5}
\begin{array}{*{20}l}
   {E(\rho _{|\psi \rangle } ) = 2(} &\!\!\!\! {\phantom{+}|a_{1,1} a_{2,2}  - a_{1,2} a_{2,1} |^2 }  \\
   {} &\!\!\!\! { + |a_{1,1} a_{2,3}  - a_{1,3} a_{2,1} |^2 }  \\
   {} &\!\!\!\! { + |a_{1,2} a_{2,3}  - a_{1,3} a_{2,2} |^2 )^{1/2}. }  \\
\end{array}
\end{Equation}
By \hyperlink{ExtraSummaryResults:5}{Item 5} from \Sec{II}, mapping the coincidence basis to scalar indices converts \Eq{A.5} to
\begin{Equation}                      {A.6}
\begin{array}{*{20}l}
   {E(\rho _{|\psi \rangle } ) = (} &\!\!\!\! {\phantom{+}[2|a_{1} a_{5}  - a_{2} a_{4} |]^2 }  \\
   {} &\!\!\!\! { + [2|a_{1} a_{6}  - a_{3} a_{4} |]^2 }  \\
   {} &\!\!\!\! { + [2|a_{2} a_{6}  - a_{3} a_{5} |]^2 )^{1/2}. }  \\
\end{array}
\end{Equation}
Recalling that in $2\times 2$, pure-state concurrence is
\begin{Equation}                      {A.7}
C(\rho _{|\psi \rangle }^{[2 \times 2]} ) = 2|a_1 a_4  - a_2 a_3 |,
\end{Equation}
this connects the coefficients in $C$ to \textit{levels} $\{1,2,3,4\}$ of the total state. Thus, quantities in \Eq{A.6} are concurrences of \textit{subspaces} of \smash{$\rho _{|\psi \rangle }$} as defined after \Eq{5}, so that
\begin{Equation}                      {A.8}
2|a_{q_1 } a_{q_4 }  - a_{q_2 } a_{q_3 } | = C(\rho _{|\psi \rangle }^{\{ q_1 ,q_2 ,q_3 ,q_4 \} } ) = C(\rho _{|\psi \rangle }^{\{ \mathbf{q}\} } ) \equiv C^{\{ \mathbf{q}\} } ,
\end{Equation}
where\hsp{-1.2} \smash{$\rho _{|\psi \rangle }^{\{ \mathbf{q}\} }$} is the $\mathbf{q}$ subspace of \smash{$\rho _{|\psi \rangle }$} (with no renormalization).\hsp{2.5} Thus, \Eq{A.8} lets us rewrite \Eq{A.6} as
\begin{Equation}                      {A.9}
E(\rho _{|\psi \rangle } ) = \sqrt {[C^{\{ 1,2,4,5\} }]^2  + [C^{\{ 1,3,4,6\} }]^2  + [C^{\{ 2,3,5,6\} }]^2 },
\end{Equation}
which can also be written as the $2$-norm,
\begin{Equation}                      {A.10}
E(\rho _{|\psi \rangle } ) = \|\mathbf{C}(\rho _{|\psi \rangle } )\|_2 \,,
\end{Equation}
of the \textit{subspace concurrence vector},
\begin{Equation}                      {A.11}
\mathbf{C}(\rho_{|\psi \rangle } ) \equiv [C(\rho_{|\psi \rangle } ^{\{ 1,2,4,5\} } ),C(\rho_{|\psi \rangle } ^{\{ 1,3,4,6\} } ),C (\rho_{|\psi \rangle } ^{\{ 2,3,5,6\} } )].
\end{Equation}

The form in \Eqs{A.10}{A.11} is much easier to work with than \Eq{A.3}. To get I-concurrence for mixed states,  use the convex-roof extension as in \Eq{5}.

Note that \textit{not all sets of four indices are used in \Eq{A.9} and \Eq{A.11}}. See \App{B} for \textit{physical meanings} of the special \textit{quartets} that appear in these quantities.
\section{\label{sec:App.B}Quartets}
In \Eq{5} and \Eq{A.9}, we see sets of levels we call \textit{quartets}:
\begin{Equation}                      {B.1}
\mathbf{q}_1  \equiv \{ 1,2,4,5\} ,\;\;\mathbf{q}_2  \equiv \{ 1,3,4,6\} ,\;\;\mathbf{q}_3  \equiv \{ 2,3,5,6\} .
\end{Equation}
Furthermore, in \Eq{5} and \Eq{A.9} we take \textit{concurrences} of subspaces of pure states as \smash{$C(\rho _{|\psi \rangle }^{\{ \mathbf{q}_k\} } )$}.  Yet since there are $\binom{6}{4}=15$ possible sets of $4$ levels in $n=6$ dimensions, the three sets in \Eq{B.1} must have some special significance.

It turns out that the quartets in \Eq{B.1} all have \textit{product form} in the sense that expanding them in the coincidence basis forms an ordered basis factorizable as a tensor product of basis sets in each mode, as we will show.
  
For example, in $2\times 2$, the only quartet possible is
\begin{Equation}                      {B.2}
\mathbf{q}_1 ' \equiv \{ 1,2,3,4\} ,
\end{Equation}
corresponding to coincidence-form computational basis,
\begin{Equation}                      {B.3}
\{ |1\rangle |1\rangle ,|1\rangle |2\rangle ,|2\rangle |1\rangle ,|2\rangle |2\rangle \}  = \{|1\rangle ,|2\rangle \} \otimes \{|1\rangle ,|2\rangle \}.
\end{Equation}
Thus, its quartet is formed from a \textit{tensor product} of two pairs of single-mode basis states, which we call a \textit{product quartet} (or just  \textit{quartet}) which forms a \textit{product subspace}. 

Note also that the outer pair of quartet indices forms the basis of an \textit{inseparable qubit} as does the inner pair;
\begin{Equation}                      {B.4}
\{ 1,4\}  = \{ 11,22\} ,\;\;\{ 2,3\}  = \{ 12,21\} ,
\end{Equation}
where for brevity we omit ket symbols and commas in coincidence strings (such as writing $21$ instead of $\{2,1\}$), and we use ``inseparable'' instead of ``entangled'' since these are \textit{subsets of basis states} rather than actual states.

To see that the $2\times 3$ quartets in \Eq{B.1} have product form, note that mode 1 only has two levels, but mode 2, being a 3-level system, has three pairs of two levels.  Thus the full set of $2\times 2$ product subspaces in $2\times 3$ (including the map from coincidences to single indices) is
\begin{Equation}                      {B.5}
\begin{array}{*{20}c}
   {\left\{ {\begin{array}{*{20}c}
   {11} & {12} & {13} & {21} & {22} & {23}  \\
   1 & 2 & 3 & 4 & 5 & 6  \\
\end{array}} \right\}}  \\[2.0ex]
   {\begin{array}{*{20}l}
   {\{ 1,2\} \{ 1,2\} } &\!\! { = \{ 11,12,21,22\} } &\!\! { = \{ 1,2,4,5\} }  \\
   {\{ 1,2\} \{ 1,3\} } &\!\! { = \{ 11,13,21,23\} } &\!\! { = \{ 1,3,4,6\} }  \\
   {\{ 1,2\} \{ 2,3\} } &\!\! { = \{ 12,13,22,23\} } &\!\! { = \{ 2,3,5,6\},}  \\
\end{array}}  \\
\end{array}
\end{Equation}
which confirms that \textit{the quartets of \Eq{B.1} are exactly the set of} $2\times 2$ \textit{product subspaces in} $2\times 3$. (Note that these are not just any quartets. For example, $\{1,2,3,6\}=\{11,12,13,23\}$ does \textit{not} factor as a tensor product, and also does not have an inseparable qubit in the inner pair.)
\section{\label{sec:App.C}Brief Review of Physical Decompositions of Mixed States into Pure States}
Decomposition of any generally mixed state $\rho$ into a convex sum of pure states is well-known, and already covered in \cite{Woot,HedE}.  Here we briefly review it to explain our parameterization for the sake of reproducibility.

Any $n$-level density matrix $\rho$ of rank $r$ expands as
\begin{Equation}                      {C.1}
{\textstyle \rho  = \sum\nolimits_{j = 1}^{D \ge r} {p_j |w_j \rangle \langle w_j |}  = \sum\nolimits_{j = 1}^{D \ge r} {|\overline{w}_j \rangle \langle \overline{w}_j |,} }
\end{Equation}
where $D\in[r,\infty)$, with normalized decomposition states
\begin{Equation}                      {C.2}
{\textstyle |w_j \rangle  \equiv \frac{1}{{\sqrt {p_j } }}\sum\nolimits_{k = 1}^r {U_{j,k} \sqrt {\lambda _k } |e_k \rangle } ,}
\end{Equation}
where $\lambda _1\geq\cdots\geq\lambda _n$ are eigenvalues of $\rho$ with corresponding eigenstates $|e_k \rangle$, and $U \equiv U^{[D]}$ is any $D$-level unitary matrix. The decomposition probabilities are
\begin{Equation}                      {C.3}
{\textstyle p_j  \equiv \sum\nolimits_{k = 1}^r {|U_{j,k} |^2 \lambda _k } =\langle\overline{w}_j  |\overline{w}_j \rangle,}
\end{Equation}
and \textit{subnormalized} pure decomposition states are
\begin{Equation}                      {C.4}
|\overline{w}_j \rangle  \equiv \sqrt {p_j } |w_j \rangle ,
\end{Equation}
to get the form on the right in \Eq{C.1} which is useful for \textit{degree-$1$ homogeneous} entanglement measures like I-concurrence for which $E(p\rho)=pE(\rho)$.

From \Eq{C.1}, the dependence on $U$ enters as
\begin{Equation}                      {C.5}
{\textstyle |\overline{w}_j \rangle \langle \overline{w}_j | = \sum\nolimits_{k,l = 1,1}^{r,r} {U_{j,k} U_{j,l}^* \sqrt {\lambda _k \lambda _l } |e_k \rangle \langle e_l |} .}
\end{Equation}
so $U$ is special unitary, since global matrix phase would cancel in \Eq{C.5}.  In the simplest case where $D=2$,
\begin{Equation}                      {C.6}
\begin{array}{*{20}c}
   {U = \left( {\begin{array}{*{20}c}
   a & b  \\
   { - b^* } & {a^* }  \\
\end{array}} \right);} & {(a,b) \equiv (c_\theta  e^{i\varphi } ,s_\theta  e^{i\chi } ),}  \\
\end{array}
\end{Equation}
where $\theta  \in [0,\frac{\pi }{2}]$, and $\varphi ,\chi  \in [0,2\pi )$, but $(\upsilon _j )_{k,l}  \equiv U_{j,k} U_{j,l}^* $ act as elements of pure states with density matrices $\upsilon _j$, so to gauge the \textit{effective} degrees of freedom (DOF) of $U$ in the decomposition, we must view $\{\upsilon _j\}$ collectively as
\begin{Equation}                      {C.7}
\begin{array}{*{20}l}
   {\upsilon _1 } &\!\! { = \left( {\begin{array}{*{20}c}
   {c_\theta ^2 } & {s_\theta  c_\theta  e^{ - i(\chi  - \varphi )} }  \\
   {s_\theta  c_\theta  e^{i(\chi  - \varphi )} } & {s_\theta ^2 }  \\
\end{array}} \right)}  \\
   {\upsilon _2 } &\!\! { = \left( {\begin{array}{*{20}c}
   {s_\theta ^2 } & {- s_\theta  c_\theta  e^{ - i(\chi  - \varphi )} }  \\
   { - s_\theta  c_\theta  e^{i(\chi  - \varphi )} } & {c_\theta ^2 }  \\
\end{array}} \right)\!,}  \\
\end{array}
\end{Equation}
which only has \textit{two} DOF which are $\{ \theta ,(\chi  - \varphi )\}$, so we can set $\phi  \equiv \chi  - \varphi  \in [0,2\pi )$ where the range of $\phi$ is limited by the functions in which it appears in \Eq{C.7}.  Thus, noting that the off-diagonals of $\upsilon _1,\upsilon _2$ still reach all the same values if we set $\varphi\equiv 0$, then $\phi=\chi$, so our final parameterization of $U$ for $D=2$ is
\begin{Equation}                      {C.8}
\begin{array}{*{20}c}
   {U = \left( {\begin{array}{*{20}c}
   a & b  \\
   { - b^* } & {a^* }  \\
\end{array}} \right);} & {(a,b) \equiv (c_\theta  ,s_\theta  e^{i\phi } ),}  \\
\end{array}
\end{Equation}
where $\theta  \in [0,\frac{\pi }{2}]$ and $\phi  \in [0,2\pi )$.  Since two DOF can be searched efficiently, this allows useful numerical searches for minimum average entanglement. However, as shown in \cite{Hor1,Cara,Hor6}, separable states can require up to $D=r^2$ to find an optimal decomposition.

For $D\geq 3$, the number of DOF quickly becomes intractable to search deterministically. However, the states we are studying are simple enough (due to spacewise orthogonality) that a random search can give a reasonable approximation, as seen in our plots for $D=3,4$ which show average I-concurrence approaching the theoretical minimum average value.  

Again, we did not rely on numerical searches to reach any of our conclusions; they are merely checks to show that we cannot find any contradictions to our proofs.
\section{\label{sec:App.D}Additional Facts about TGX States}
{~\vspace{-16pt}\\}
From \cite{HedX,HedE}, TGX states also have diagonal reductions (see App.{\kern 2.5pt}H of \cite{HCor} for a brief history of TGX states). ME TGX states always have balanced superposition (nonzero coefficients all with the same magnitude) as seen in \Eq{38} (as proved in \cite{HedE} for all systems), and are also the core states in the multipartite Schmidt decomposition presented in \cite{HedE} and detailed in \cite{HMME}.  

Furthermore, each column of ME TGX states in \Eq{38} forms a maximally entangled basis (MEB) as proposed in \cite{HedX}, the general existence of which was proved for all systems in \cite{HedE}.  As shown in \cite{HedX}, the union of all nonzero elements of ME TGX states gives the full set of $2\times 3$ TGX states as seen in \Eq{17} (which is just one of \textit{many} different characterizations of TGX states).
{~\vspace{-16pt}\\}
\section{\label{sec:App.E}Generalized Concurrence in $2\times 3$}
{~\vspace{-16pt}\\}
Despite the close ties of I-concurrence to $2\times 2$ subspace concurrences, it is \textit{not} equivalent to \textit{generalized concurrence} $C_G$ \cite{Uhl1}. Here we briefly derive the maximal value of $C_G$ wrt spectrum, which is used in the text to show that $C_G$ is not I-concurrence.

From \cite{Uhl1}, \textit{generalized concurrence} of mixed $\rho$ is
\begin{Equation}                      {E.1}
C_G (\rho)  = \max \{ 0,c_G(\rho) \} ,
\end{Equation}
\vspace{-18pt}\\
where $c_G$ is the minimum average generalized preconcurrence over all decompositions of $\rho$, given by
\begin{Equation}                      {E.2}
{\textstyle c_G  \equiv \xi _1  - \sum\limits_{l = 2}^{\text{rank}(R)} {\xi _l } },
\end{Equation}
\vspace{-12pt}\\
where $\xi _1 \geq\xi _2\geq\cdots$, and $R$ is defined as
\begin{Equation}                      {E.3}
{\textstyle R \equiv R(\rho ) \equiv \sqrt {\sqrt \rho  \widetilde{\rho} \sqrt \rho  }} ,
\end{Equation}
\vspace{-18pt}\\
where $\widetilde{\rho}$ is an \textit{antilinearly unitary (antiunitary) conjugation} of $\rho$ (which is defined in \cite{Uhl1}), and
\begin{Equation}                      {E.4}
\{ \xi _l \}  \equiv\text{eig}(R) = \text{sing}(\rho \widetilde{\rho} ).
\end{Equation}
\vspace{-24pt}\\

Since we want to maximize $C_G$ wrt spectrum, we need to find a singular-value inequality involving $c_G$ of \Eq{E.2}.  Following \cite{VeAM}, from \cite{WaXi} (which covers a wider array of situations than we use here), for any complex square matrices $C=AB$, where $A,B,C$ each have $n$ dimensions with $\{\lambda_l\}\equiv\{\lambda_l(C)\}\equiv\text{eig}(C)$ and singular values \smash{$\{\sigma _l \} \equiv\{\sigma _l (C)\} \equiv \{\sqrt {\lambda _l (CC^\dag  )}\}\equiv \text{sing}(C)$}\rule{0pt}{9.5pt} for $l \in 1, \ldots ,n$ such that $\sigma _1  \ge  \cdots  \ge \sigma _n $, then
\begin{Equation}                      {E.5}
{\textstyle \sum\limits_{l = 1}^k {\sigma _l (AB)}  \le \sum\limits_{l = 1}^k {\sigma _l (A)\sigma _l (B)} },
\end{Equation}
for $k \in 1, \ldots ,n$, and also
\begin{Equation}                      {E.6}
{\textstyle \sum\limits_{t = 1}^k {\sigma _{l_t } (AB)}  \ge \sum\limits_{t = 1}^k {\sigma _{l_t } (A)\sigma _{n - t + 1} (B)} },
\end{Equation}
for $k \in 1, \ldots ,n$ and integers $l_t$ s.t. $1 \le l_1  <  \cdots  < l_k  \le n$.

For the desired inequality, first set $k=1$ in \Eq{E.5} to get
\begin{Equation}                      {E.7}
\sigma _1 (AB) \le \sigma _1 (A)\sigma _1 (B).
\end{Equation}
Then, since we want five terms subtracted from $\sigma_1$ [by \Eq{E.2} with $\max\{\text{rank}(R)\}\!=\!n\!=\!6$], set $k\!=\!5$ in \Eq{E.6} and since we want those indices to be $\{2,3,4,5,6\}$, let
\begin{Equation}                      {E.8}
\{l_1 , \ldots ,l_k \} = \{l_1 ,l_2 ,l_3 ,l_4 ,l_5 \} \equiv \{2,3,4,5,6\},
\end{Equation}
which put in \Eq{E.6} and multiplied by $-1$ yields
\begin{Equation}                      {E.9}
\begin{array}{*{20}l}
   { - (\sigma _2  + \sigma _3  + \sigma _4  + \sigma _5  + \sigma _6 )(AB) \le } &\!\! { - [\;\,\sigma _2 (A)\sigma _6 (B)}  \\
   {} &\!\! { \phantom{[}+ \sigma _3 (A)\sigma _5 (B)}  \\
   {} &\!\! { \phantom{[}+ \sigma _4 (A)\sigma _4 (B)}  \\
   {} &\!\! { \phantom{[}+ \sigma _5 (A)\sigma _3 (B)}  \\
   {} &\!\! { \phantom{[}+ \sigma _6 (A)\sigma _2 (B)].}  \\
\end{array}
\end{Equation}
where $(AB)$ on the left represents the argument for that entire quantity. Then, adding \Eq{E.7} and \Eq{E.9} gives
\begin{Equation}                      {E.10}
\begin{array}{*{20}l}
   {(\sigma _1  - \sigma _2  - \sigma _3  - \sigma _4  - \sigma _5  - \sigma _6 )(AB) \le } &\!\! {\phantom{-}\sigma _1 (A)\sigma _1 (B)}  \\
   {} &\!\! { - \sigma _2 (A)\sigma _6 (B)}  \\
   {} &\!\! { - \sigma _3 (A)\sigma _5 (B)}  \\
   {} &\!\! { - \sigma _4 (A)\sigma _4 (B)}  \\
   {} &\!\! { - \sigma _5 (A)\sigma _3 (B)}  \\
   {} &\!\! { - \sigma _6 (A)\sigma _2 (B).}  \\
\end{array}
\end{Equation}

In \cite{VeAM}, it was shown that \Eq{E.4} can be rewritten as
\begin{Equation}                      {E.11}
\{ \xi _l \}  = \text{sing}(\sqrt \Lambda  V\sqrt \Lambda  ),
\end{Equation}
where $V$ is unitary and $\Lambda  \equiv \text{diag}\{\lambda _1 , \ldots ,\lambda _n \}\equiv\text{eig}(\rho)$.  So then, letting
\begin{Equation}                      {E.12}
A \equiv \sqrt \Lambda  ,\;\;\; B \equiv V\sqrt \Lambda  ,
\end{Equation}
that causes the singular values in \Eq{E.10} to become
\begin{Equation}                      {E.13}
\sigma _l(AB)=\xi_l,\;\;\text{and}\;\;\sigma _l (A) = \sigma _l (B) = \sqrt {\lambda _l } ,
\end{Equation}
so then, putting \Eq{E.12} and \Eq{E.13} into \Eq{E.10} yields
\begin{Equation}                      {E.14}
{\textstyle (\sigma _1  - \sum\limits_{l = 2}^6 {\!\sigma _l } )(\sqrt \Lambda  V\sqrt \Lambda  )\! \le\! \lambda _1  - \lambda _4  - 2\sqrt {\lambda _2 \lambda _6 }  - 2\sqrt {\lambda _3 \lambda _5 }}.
\end{Equation}
Then maximizing \Eq{E.14} over all $V$ gives
\begin{Equation}                      {E.15}
\begin{array}{*{20}l}
   {\max (c_G )} &\!\! { \equiv \mathop {\max }\limits_{\forall V} \left[ {(\sigma _1  - \sum\nolimits_{l = 2}^6 {\sigma _l } )(\sqrt \Lambda  V\sqrt \Lambda  )} \right]}  \\
   {} &\!\! { = \lambda _1  - \lambda _4  - 2\sqrt {\lambda _2 \lambda _6 }  - 2\sqrt {\lambda _3 \lambda _5 }, }  \\
\end{array}
\end{Equation}
so putting \Eq{E.15} into \Eq{E.1} maximized over all $V$ gives
\begin{Equation}                      {E.16}
\begin{array}{*{20}l}
   {\max (C_G )} &\!\! { = \max \{ 0,\max (c_G )\} }  \\
   {} &\!\! { = \max \{0,\lambda _1  - \lambda _4  - 2\sqrt {\lambda _2 \lambda _6 }  - 2\sqrt {\lambda _3 \lambda _5 }\}, }  \\
\end{array}
\end{Equation}
which is the maximal value of $C_G$ wrt spectrum in \Eq{42}.
\end{appendix}
%
\end{document}